\documentclass{cernrep} 
\usepackage{texnames}
\usepackage[T1]{fontenc}
\usepackage{cancel}
\usepackage{subfigure}
 \usepackage{units}

\usepackage{color}
\usepackage{epstopdf}
\usepackage{listings}
\usepackage{textcomp}
\usepackage{url}

\graphicspath{{./}{./figures/}}
\def\half{\mbox{\small $\frac{1}{2}$}}

\def\roofit{\texttt{RooFit}}
\def\roostats{\texttt{RooStats}}

\def\vec#1{\ifmmode
\mathchoice{\mbox{\boldmath$\displaystyle\bf#1$}}
{\mbox{\boldmath$\textstyle\bf#1$}}
{\mbox{\boldmath$\scriptstyle\bf#1$}}
{\mbox{\boldmath$\scriptscriptstyle\bf#1$}}\else
{\mbox{\boldmath$\bf#1$}}\fi}

\def\half{\mbox{\small $\frac{1}{2}$}}

\newcommand{\Pois}{{\ensuremath{\rm Pois}}}
\newcommand{\LN}{{\ensuremath{P_{\rm LN}}}}
\newcommand{\Gauss}{{\ensuremath{\rm Gauss}}}
\newcommand{\PGamma}{{\ensuremath{P_{\;\Gamma}}}}
\newcommand{\data}{{\ensuremath{\mathcal{D}}}}
\newcommand{\datasim}{{\ensuremath{\mathcal{D}_{\textrm{sim}}}}}
\newcommand{\F}{{\ensuremath{\mathbf{f}}}}
\newcommand{\f}{{\ensuremath{\mathbf{f}}}}

\newcommand{\globs}{{\ensuremath{\mathcal{G}}}}
\newcommand{\nuisObs}{\ensuremath{\vec\theta_{\rm obs}}}
\newcommand{\hathatthetamu}{\ensuremath{\hat{\hat{\vec\theta}}(\mu)}}

\newcommand{\mh}{\ensuremath{m_H}}

\newcommand{\HF}{\texttt{HistFactory}}

\newcommand{\OS}{\texttt{OverallSys}}
\newcommand{\HS}{\texttt{HistoSys}}

\begin{document}
\title{Practical Statistics for the LHC}
 
\author{Kyle Cranmer}

\institute{Center for Cosmology and Particle Physics, Physics Department, New York University, USA}

\maketitle 

\begin{abstract}
This document is a pedagogical introduction to statistics for particle physics.  
Emphasis is placed on the terminology, concepts, and methods being used at the Large Hadron Collider.  
The document addresses both the statistical tests applied to a model of the data and the modeling itself .
I expect to release updated versions of this document in the future.
\end{abstract}

 \nopagebreak
\tableofcontents\nopagebreak

\newpage

\section{Introduction}

It is often said that the language of science is mathematics.  It could well be said that the language of experimental science is statistics.  It is through statistical concepts that we quantify the correspondence between theoretical predictions and experimental observations.  While the statistical analysis of the data is often treated as a final subsidiary step to an experimental physics result, a more direct approach would be quite the opposite.  In fact, thinking through the requirements for a robust statistical statement is an excellent way to organize an analysis strategy.  

In these lecture notes\footnote{These notes borrow significantly from other documents that I am writing contemporaneously; specifically Ref.\cite{asimov}, documentation for \HF\ ~\cite{histfactory} and the ATLAS Higgs combination. } I will devote significant attention to the strategies used in high-energy physics for developing a statistical model of the data.  This modeling stage is where you inject your understanding of the physics.  I like to think of the modeling stage in terms of a conversation.  When your colleague asks you over lunch to explain your analysis, you tell a story.  It is a story about the signal and the backgrounds -- are they estimated using Monte Carlo simulations, a side-band, or some data-driven technique?    Is the analysis based on counting events or do you use some discriminating variable, like an invariant mass or perhaps the output of a multivariate discriminant?  What are the dominant uncertainties in the rate of signal and background events and how do you estimate them?  What are the dominant uncertainties in the shape of the distributions and how do you estimate them?  The answer to these questions forms a \textit{scientific narrative}; the more convincing this narrative is the more convincing your analysis strategy is.  The statistical model is the mathematical representation of this narrative and you should strive for it to be as faithful a representation as possible.

Once you have constructed a statistical model of the data, the actual statistical procedures should be relatively straight forward.  In particular, the statistical tests can be written for a generic statistical model without knowledge of the physics behind the model.  The goal of the \texttt{RooStats} project was precisely to provide statistical tools based on an arbitrary statistical model implemented with the \texttt{RooFit} modeling language.  While the formalism for the statistical procedures can be somewhat involved, the logical justification for the procedures is based on a number of abstract properties for the statistical procedures. One can follow the logical argument without worrying about the detailed mathematical proofs that the procedures have the required properties.  Within the last five years there has been a significant advance in the field's understanding of certain statistical procedures, which has led to to some commonalities in the statistical recommendations by the major LHC experiments.  I will review some of the most common statistical procedures and their logical justification.

\section{Conceptual building blocks for modeling}

\subsection{Probability densities and the likelihood function}

This section specifies my notations and conventions, which I have chosen with some care.%
\footnote{As in the case of relativity, notational conventions can make some properties of expressions manifest and help identify mistakes.  For example, $g_{\mu\nu}x^\mu y^\nu$ is manifestly Lorentz invariant and $x^\mu + y_\nu$ is manifestly wrong.}  Our statistical claims will be based on the outcome of an experiment.  When discussing frequentist probabilities, one must consider ensembles of experiments, which may either be real, based on computer simulations, or mathematical abstraction.

\begin{figure}[htbp]
\begin{center}
\includegraphics[width=.6\linewidth]{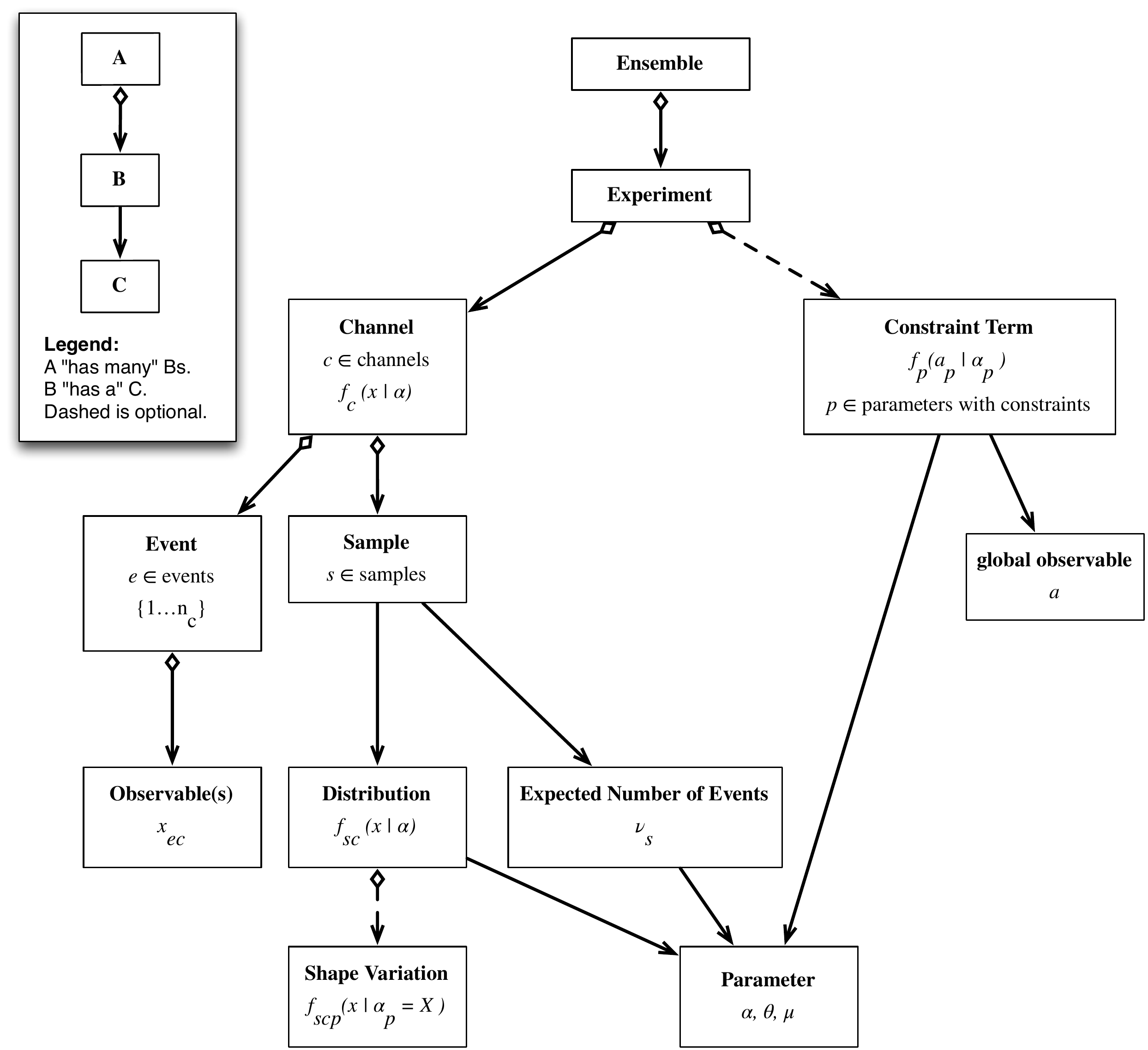}
\caption{A schematic diagram of the logical structure of a typical particle physics probability model and dataset structures.}
\label{fig:hierarchy}
\end{center}
\end{figure}

Figure~\ref{fig:hierarchy} establishes a hierarchy that is fairly general for the context of high-energy physics.  Imagine the search for the Higgs boson, in which the search is composed of several ``channels'' indexed by $c$.  Here a channel is defined by its associated event selection criteria, not an underlying physical process.  In addition to the number of selected events, $n_c$, each channel may make use of some other measured quantity, $x_c$, such as the invariant mass of the candidate Higgs boson. The quantities will be called ``observables'' and will be written in roman letters e.g. $x_c$. The notation is chosen to make manifest that the observable $x$ is  frequentist in nature.  Replication of the experiment many times will result in different values of $x$ and this ensemble gives rise to a \emph{probability density function} (pdf) of $x$, written $f(x)$, which has the important property that it is normalized to unity
\[
\int  f(x) \;dx\;= 1\;.
\]
In the case of discrete quantities, such as the number of events satisfying some event selection, the integral is replaced by a sum.  Often one considers a parametric family of pdfs
\[
f(x | \alpha) \;,
\]
read ``$f$ of $x$ given $\alpha$''  and, henceforth, referred to as a \emph{probability model} or just \emph{model}.  The parameters of the model typically represent parameters of a physical theory or an unknown property of the detector's response.  The parameters are not frequentist in nature, thus any probability statement associated with $\alpha$ is Bayesian.\footnote{Note, one can define a conditional distribution $f(x|y)$ when the joint distribution $f(x,y)$ is defined in a frequentist sense.}  In order to make their lack of frequentist interpretation manifest, model parameters will be written in greek letters, e.g.: $\mu, \theta, \alpha, \nu$.%
 \footnote{While it is common to write $s$ and $b$ for the number of expected signal and background, these are parameters \emph{not} observables, so I will write $\nu_S$ and $\nu_B$.  This is one of few notational differences to Ref.~\cite{asimov}.}
From the full set of parameters, one is typically only interested in a few: the \emph{parameters of interest}.  The remaining parameters are  referred to as \emph{nuisance parameters}, as we must account for them even though we are not interested in them directly.

While $f(x)$ describes the probability density for the observable $x$ for a single event, we also need to describe the probability density for a dataset with many events, $\data = \{x_1,\dots,x_{n}\}$.  If we consider the events as independently drawn from the same underlying distribution, then clearly the probability density is just a product of densities for each event.  However, if we have  a prediction that the total number of events expected, call it $\nu$, then we should also include the overall Poisson probability for observing $n$ events given $\nu$ expected.  Thus, we arrive at what statisticians call a marked Poisson model,
\begin{equation}
\label{Eq:markedPoisson}
\F(\data|\nu,\alpha) = \Pois(n|\nu) \prod_{e=1}^n f(x_e|\alpha) \; ,
\end{equation}
where I use a bold $\F$ to distinguish it from the individual event probability density $f(x)$.  In practice, the expectation is often parametrized as well and some parameters simultaneously modify the expected rate and shape, thus we can write $\nu\rightarrow\nu(\alpha)$.  In \texttt{RooFit} both $f$ and $\F$ are implemented with a \texttt{RooAbsPdf}; where \texttt{RooAbsPdf::getVal(x)} always provides the value of $f(x)$ and depending on \texttt{RooAbsPdf::extendMode()} the value of $\nu$ is accessed via \texttt{RooAbsPdf::expectedEvents()}.

The \emph{likelihood function} $L(\alpha)$ is numerically equivalent to $f(x|\alpha)$ with $x$ fixed -- or $\F(\data|\alpha)$ with \data\ fixed.  The likelihood function should not be interpreted as a probability density for $\alpha$.  In particular, the likelihood function does not have the property that it normalizes to unity
\[
\cancelto{\mathrm{Not ~True!}}{\int L(\alpha) \;d\alpha = 1}\; .
\]
%
It is common to work with the log-likelihood (or negative log-likelihood) function.  In the case of a marked Poisson, we have what is commonly referred to as an extended  likelihood~\cite{Barlow1990496}
\begin{eqnarray}\nonumber
-\ln L( \alpha) &=& \underbrace{\nu(\alpha) - n \ln \nu(\alpha)}_{\rm extended~term} - \sum_{e=1}^n \ln f(x_e)   + \underbrace{~\ln n! ~}_{\mathrm{constant}}\; .
\end{eqnarray}
To reiterate the terminology, \emph{probability density function} refers to the value of $f$ as a function of $x$ given a fixed value of $\alpha$; \emph{likelihood function} refers to the value of $f$ as a function of $\alpha$ given a fixed value of $x$; and \emph{model} refers to the full structure of $f(x|\alpha)$.


Probability models can be constructed to simultaneously describe several channels, that is several disjoint regions of the data defined by the associated selection criteria.  I will use $e$ as the index over events and $c$ as the index over channels.  Thus, the number of events in the $c^{\rm th}$ channel is $n_c$ and the value of the $e^{\rm th}$ event in the $c^{\rm th}$ channel is $x_{ce}$.  In this context, the data is a collection of smaller datasets: \mbox{$\datasim=\{\data_1, \dots, \data_{c_{\rm max}}\}=\{\{x_{c=1,e=1}\dots x_{c=1,e=n_c}\}, \dots \{x_{c=c_{\rm max},e=1}\dots x_{c=c_{\rm max},e=n_{c_{\rm max}}} \}\}$}.  In \texttt{RooFit} the index $c$ is referred to as a \texttt{RooCategory} and it is used to inside the dataset to differentiate events associated to different channels or categories. The class \texttt{RooSimultaneous} associates the dataset $\data_c$ with the corresponding marked Poisson model.  The key point here is that there are now multiple Poisson terms.  Thus we can write the combined (or simultaneous) model 
\begin{equation}
\label{Eq:simultaneous}
\F_{\textrm{sim}}(\datasim|\alpha) = \prod_{c\in\rm channels} \left[ \Pois(n_c|\nu(\alpha)) \prod_{e=1}^{n_c} f(x_{ce}|\alpha) \right] \; ,
\end{equation}
remembering that the symbol product over channels has implications for the structure of the dataset.

\subsection{Auxiliary measurements}\label{S:AuxMeas}

Auxiliary measurements or control regions can be used to estimate or reduce the effect of systematic uncertainties.  The signal region and control region are not fundamentally different.  In the language that we are using here, they are just two different channels.  

A common example is  a simple counting experiment with an uncertain background.  In the frequentist way of thinking, the true, unknown background in the signal region is a nuisance parameter, which I will denote $\nu_B$.\footnote{Note, you can think of a counting experiment in the context of Eq.~\ref{Eq:markedPoisson} with $f(x)=1$, thus it reduces to just the Poisson term.}  If we call the true, unknown signal rate $\nu_S$ and the number of events in the signal region $n_{\rm SR}$ then we can write the model $\Pois(n_{\rm SR} | \nu_S + \nu_B)$.  As long as $\nu_B$ is a free parameter, there is no ability to make any useful inference about $\nu_S$.  Often we have some estimate for the background, which may have come from some control sample with $n_{\rm CR}$ events.  If the control sample has no signal contamination and is populated by the same background processes as the signal region, then we can write $\Pois(n_{\rm CR}|\tau \nu_B)$, where $n_{\rm CR}$ is the number of events in the control region and $\tau$ is a factor used to extrapolate the background from the signal region to the control region.  Thus the total probability model can be written $\F_{\rm sim}(n_{\rm SR},n_{\rm CR} | \nu_S, \nu_B) = \Pois(n_{\rm SR} | \nu_S + \nu_B)\cdot \Pois(n_{\rm CR}|\tau\nu_B)$.  This is a special case of Eq.~\ref{Eq:simultaneous} and is often referred to as the ``on/off' problem~\cite{Cousins:2008zz}.

Based on the control region alone, one would estimate (or `measure') $\nu_B = n_{\rm CR}/\tau$.  Intuitively the estimate comes with an  `uncertainty' of $\sqrt{n_{\rm CR}}/\tau$.   We will make these points more precise in Sec.~\ref{S:estimation}, but the important lesson here is that we can use auxiliary measurements (ie. $n_{\rm CR}$) to describe our uncertainty on the nuisance parameter $\nu_B$ statistically.  Furthermore, we have formed a statistical model that can be treated in a frequentist formalism -- meaning that if we repeat the experiment many times $n_{\rm CR}$ will vary and so will the estimate of $\nu_B$.   It is common to say that auxiliary measurements `constrain' the nuisance parameters.  In principle the auxiliary measurements can be every bit as complex as the main signal region, and there is no formal distinction between the various channels.

The use of auxiliary measurements is not restricted to estimating rates as in the case of the on/off problem above.  One can also use auxiliary measurements to constrain other parameters of the model. To do so, one must relate the effect of some common parameter $\alpha_p$ in multiple channels (ie. the signal region and a control regions).   This is implicit in Eq.~\ref{Eq:simultaneous}.

\subsection{Frequentist and Bayesian reasoning}

The intuitive interpretation of measurement of $\nu_B$ to be $n_{\rm CR}/\tau \pm \sqrt{n_{\rm CR}}/\tau$ is that the parameter $\nu_B$ has a distribution centered around $n_{\rm CR}/\tau$ with a width of $\sqrt{n_{\rm CR}}/\tau$.  With some practice you will be able to immediately identify this type of reasoning as Bayesian.  It is manifestly Bayesian because we are referring to the probability distribution of a parameter.  The frequentist notion of probability of an event is defined as the limit of its relative frequency in a large number of trials.  The large number of trials is referred to as an ensemble.  In particle physics the ensemble is formed conceptually by repeating the experiment many times.  The true values of the parameters, on the other hand, are states of nature, not the outcome of an experiment.  The true mass of the $Z$ boson has no frequentist probability distribution.  The existence or non-existence of the Higgs boson has no frequentist probability associated with it.  There is a sense in which one can talk about the probability of parameters, which follows from Bayes's theorem:
\begin{equation}
\label{Eq:Bayes}
P(A|B) = \frac{P(B|A) P(A)}{P(B)} \; .
\end{equation}
Bayes's theorem is a theorem, so there's no debating it.  It is not the case that Frequentists dispute whether Bayes's theorem is true.  The debate is whether the necessary probabilities exist in the first place.  If one can define the joint probability $P(A,B)$ in a frequentist way, then a Frequentist is perfectly happy using Bayes theorem.   Thus, the debate starts at the very definition of probability.

The Bayesian definition of probability clearly can't be based on relative frequency.  Instead, it is based on a degree of belief.  Formally, the probability needs to satisfy Kolmogorov's axioms for probability, which both the frequentist and Bayesian definitions of probability do.  One can quantify degree of belief through betting odds, thus Bayesian probabilities can be assigned to hypotheses on states of nature.  In practice human's bets are not generally not `coherent' (see `dutch book'), thus this way of quantifying probabilities may not satisfy the Kolmogorov axioms.

Moving past the philosophy and accepting the Bayesian procedure at face value, the practical consequence is that one must supply prior probabilities for various parameter values and/or hypotheses.  In particular, to interpret our example measurement of $n_{\rm CR}$ as implying a probability distribution for $\nu_B$ we would write
\begin{equation}
\pi(\nu_B | n_{\rm CR}) \propto f(n_{\rm CR} | \nu_B) \eta(\nu_B) \; ,
\end{equation}
where $\pi(\nu_B | n_{\rm CR})$ is called the \textit{posterior} probability density, $f(n_{\rm CR} | \nu_B)$ is the likelihood function, and $\eta(\nu_B)$ is the \textit{prior} probability.  Here I have suppressed the somewhat curious term $P(n_{\rm CR})$, which can be thought of as a normalization constant and is also referred to as the \textit{evidence}.  The main point here is that one can only invert `the probability of $n_{\rm CR}$ given $\nu_B$' to be `the probability of $\nu_B$ given $n_{\rm CR}$' if one supplies a prior.  Humans are very susceptible to performing this logical inversion accidentally, typically with a uniform prior on $\nu_B$.  Furthermore, the prior degree of belief cannot be derived in an objective way.  There are several formal rules for providing a prior based on formal rules (see Jefferey's prior and Reference priors), though these are not accurately described as representing a degree of belief.  Thus, that style of Bayesian analysis is often referred to as objective Bayesian analysis.

{\flushleft{Some  useful and amusing quotes on  Bayesian and Frequentist reasoning:}}
\begin{quote}
{\em ``Using Bayes's theorem doesn't make you a Bayesian, \textbf{always} using Bayes's theorem makes you a Bayesian.''} --unknown
\end{quote}
\begin{quote}
{\em
``Bayesians address the questions everyone is interested in by using assumptions that no one believes.
Frequentist use impeccable logic to deal with an issue that is of no interest to anyone.''}- Louis Lyons
\end{quote}


\subsection{Consistent Bayesian and Frequentist modeling of constraint terms}\label{S:Constraint}

Often a detailed probability model for an auxiliary measurement are not included directly into the model.  If the model for the auxiliary measurement were available, it could and should be included as an additional channel as described in Sec.~\ref{S:AuxMeas}.  The more common situation for background  and systematic uncertainties only has an estimate,  ``central value'', or best guess for a parameter $\alpha_p$ and some notion of uncertainty on this estimate.  In this case one typically resorts to including idealized terms into the likelihood function, here referred to as ``constraint terms'', as surrogates for a more detailed model of the auxiliary measurement.   I will denote this estimate for the parameters as $a_p$, to make it manifestly frequentist in nature.  In this case there is a single measurement of $a_p$ per experiment, thus it is referred to as a ``global observable'' in \roostats.  The treatment of constraint terms is somewhat \emph{ad hoc} and discussed in more detail in Section~\ref{S:ConstraintExamples}.  I make it a point to write constraint terms in a manifestly frequentist form $f(a_p | \alpha_p)$.  

Probabilities on parameters are legitimate constructs in a Bayesian setting, though they will always rely on a prior.  In order to distinguish Bayesian pdfs from frequentist ones, greek letters will be used for their distributions.  For instance, a generic Bayesian pdf might be written $\pi(\alpha)$.  In the context of a main measurement, one might have a prior for $\alpha_p$ based on some estimate $a_p$.  In this case, the prior $\pi(\alpha_p )$ is really a posterior from some previous measurement.  It is desirable to write with the help of Bayes theorem
\begin{equation}
\label{eq:urprior}
\pi(\alpha_p | a_p) \propto L( \alpha_p ) \eta(\alpha_p) = f(a_p|\alpha_p) \eta(\alpha_p)\; ,
\end{equation}
where $\eta(\alpha_p)$ is some more fundamental prior.\footnote{Glen Cowan has referred to this more fundamental prior as an 'urprior', which is based on the German use of 'ur' for forming words with the sense of `proto-, primitive, original'.}  By taking the time to undo the Bayesian reasoning into an objective pdf or likelihood and a prior we are able to write a model that can be used in a frequentist context.  Within \roostats, the care is taken to separately track the frequentist component and the prior; this is achieved with the \texttt{ModelConfig} class.

If one can identify what auxiliary measurements were performed to provide the estimate of $\alpha_p$ and its uncertainty, then it is not a logical fallacy to approximate it with a constraint term, it is simply  a convenience.  However, not all uncertainties that we deal result from auxiliary measurements.  In particular, some theoretical uncertainties are not statistical in nature.  For example, uncertainty associated with the choice of renormalization and factorization scales and missing higher-order corrections in a theoretical calculation are not statistical.  Uncertainties from parton density functions are a bit of a hybrid as they are derived from data but require theoretical inputs and make various modeling assumptions.  In a Bayesian setting there is no problem with including a prior on the parameters associated to theoretical uncertainties.  In contrast, in a formal frequentist setting, one should not include constraint terms on theoretical uncertainties that lack a frequentist interpretation.  That leads to a very cumbersome presentation of results, since formally the results should be shown as a function of the uncertain parameter.  In practice, the groups often read Eq.~\ref{eq:urprior} to arrive at an effective frequentist constraint term.

I will denote the set of parameters with constraint terms as $\mathbb{S}$ and the global observables $\mathcal{G}=\{a_p\}$ with $p\in\mathbb{S}$.  By including the constraint terms explicitly (instead of implicitly as an additional channel) we arrive at the total probability model, which we will not need to generalize any further:
\begin{equation}
\label{Eq:ftot}
\F_{\textrm{tot}}(\datasim, \mathcal{G}|\alpha) = \prod_{c\in\rm channels} \left[ \Pois(n_c|\nu_c(\alpha)) \prod_{e=1}^{n_c} f_c(x_{ce}|\alpha) \right] \cdot \prod_{p \in \mathbb{S}} f_p(a_p | \alpha_p)\; .
\end{equation}

\section{Physics questions formulated in statistical language}

\subsection{Measurement as parameter estimation}
\label{S:estimation}

One of the most common tasks of the working physicist is to estimate some model parameter.  We do it so often, that we often don't realize it. For instance, the sample mean $\bar{x} = \sum_{e=1}^n x_e / n$ is an estimate for the mean, $\mu$,  of a Gaussian probability density $f(x|\mu,\sigma) =\Gauss(x|\mu,\sigma)$.  More generally, an \textit{estimator} $\hat{\alpha}(\data)$ is some function of the data and its value is used to estimate the true value of some parameter $\alpha$.  There are various abstract properties such as variance, bias, consistency, efficiency, robustness, etc~\cite{James}.  The bias of an estimator is defined as  $B(\hat\alpha) = E[ \hat\alpha ]-\alpha$, where $E$ means the expectation value of \mbox{$E[ \hat\alpha ]=\int\hat\alpha(x) f(x)dx$} or the probability-weighted average.  Clearly one would like an unbiased estimator. The variance of an estimator is defined as $var[\hat\alpha] = E[ (\alpha - E[\hat{\alpha}] )^2 ]$; and clearly one would like an estimator with the minimum variance.  Unfortunately, there is a tradeoff between bias and variance.  Physicists tend to be allergic to biased estimators, and within the class of unbiased estimators, there is a well defined minimum variance bound referred to as the Cram\'er-Rao bound (that is the inverse of the Fisher information, which we will refer to again later).  

The most widely used estimator in physics is the maximum likelihood estimator (MLE).  It is defined as the value of $\alpha$ which maximizes the likelihood function $L(\alpha)$.  Equivalently this value, $\hat{\alpha}$, maximizes $\log L(\alpha)$ and minimizes $-\log L(\alpha)$.  The most common tool for finding the maximum likelihood estimator is \texttt{Minuit}, which conventionally minimizes $-\log L(\alpha)$ (or any other function)~\cite{James:1975dr}.  The jargon is that one `fits' the function and the maximum likelihood estimate is the `best fit value'.  

When one has a multi-parameter likelihood function $L(\vec{\alpha})$, then the situation is slightly more complicated.  The maximum likelihood estimate for the full parameter list, $\hat{\vec{\alpha}}$, is clearly defined.  The various components $\hat{\alpha}_p$ are referred to as the \textit{unconditional maximum likelihood estimates}.  In the physics jargon, one says all the parameters are `floating'.  One can also ask about maximum likelihood estimate of $\alpha_p$ is with some other parameters $\vec{\alpha}_o$ fixed; this is called the \textit{conditional maximum likelihood estimate} and is denoted $\hat{\hat{\alpha}}_p(\vec{\alpha}_o)$.  These are important quantities for defining the profile likelihood ratio, which we will discuss in more detail later.  The concept of variance of the estimates is also generalized to the covariance matrix $cov[\alpha_p, \alpha_{p'}] = E[(\hat\alpha_p - \alpha_p)(\hat\alpha_{p'}- \alpha_{p'})]$ and is often denoted $\Sigma_{pp'}$.  Note, the diagonal elements of the covariance matrix are the same as the variance for the individual parameters, ie. $cov[\alpha_p, \alpha_{p}] = var[\alpha_p]$.

In the case of a Poisson model $\Pois(n|\nu)$ the maximum likelihood estimate of $\nu$ is simply \mbox{$\hat{\nu}=n$}.  Thus, it follows that the variance of the estimator is $var[\hat{\nu}]=var[n]=\nu$.  Thus if the true rate is $\nu$ one expects to find estimates $\hat{\nu}$ with a characteristic spread around $\nu$; it is in this sense that the measurement has a estimate has some uncertainty or `error' of $\sqrt{n}$.  We will make this statement of uncertainty more precise when we discuss frequentist confidence intervals.

When the number of events is large, the distribution of maximum likelihood estimates approaches a Gaussian or normal distribution.\footnote{There are various conditions that must be met for this to be true, but skip the fine print in these lectures.  There are two conditions that are most often violated in particle physics, which will be addressed later.}  This does not depend on the pdf $f(x)$ having a Gaussian form.  For small samples this isn't the case, but this limiting distribution is often referred to as an \textit{asymptotic distribution}.
Furthermore, under most circumstances in particle physics, the maximum likelihood estimate approaches the minimum variance or Cram\'er-Rao bound. In particular, the inverse of the covariance matrix for the estimates is asymptotically given by
\begin{equation}
\label{Eq:expfisher}
\Sigma_{pp'}^{-1}(\vec\alpha) = E\left[- \frac{\partial^2 \log f(x|\vec{\alpha})}{\partial\alpha_p \partial_{p'}}  \middle| \;\vec\alpha \right ]  \;,
\end{equation}
where I have written explicitly that the expectation, and thus the covariance matrix itself, depend on the true value $\vec\alpha$.  The right side of Eq.~\ref{Eq:expfisher} is called the (expected) Fisher information matrix. Remember that the expectation involves an integral over the observables.  Since that integral is difficult to perform in general, one often uses the observed Fisher information matrix to approximate the variance of the estimator by simply taking the matrix of second derivatives based on the observed data
\begin{equation}
\label{Eq:obsfisher}
\tilde\Sigma_{pp'}^{-1}(\vec\alpha) = - \frac{\partial^2 \log L(\vec{\alpha})}{\partial\alpha_p \partial_{p'}}  \; .
\end{equation}
This is what \texttt{Minuit}'s \texttt{Hesse} algorithm\footnote{The matrix is called the Hessian, hence the name.} calculates to estimate the covariance matrix of the parameters.

\subsection{Discovery as hypothesis tests}\label{S:hypothesis test}

Let us examine the statistical statement associated to the claim of discovery for new physics.  Typically, new physics searches are looking for a signal that is additive on top of the background, though in some cases there are interference effects that need to be taken into account and one cannot really talk about 'signal' and 'background' in any meaningful way.  Discovery is formulated in terms of a hypothesis test where the background-only hypothesis plays the role of the null hypothesis and the signal-plus-background hypothesis plays the roll of the alternative.  Roughly speaking, the claim of discovery is a statement that the data are incompatible with the background-only hypothesis.  Consider the simplest scenario where one is counting events in the signal region, $n_{\rm SR}$ and expects $\nu_B$ events from background and $\nu_S$ events from the putative signal.    Then we have the following hypotheses:
\begin{center}
\begin{tabular}{llll}
symbol & statistical name & physics name & probability model \\ \hline
$H_0$ &  null hypothesis & background-only & $\Pois(n_{SR} | \nu_B)$ \\
$H_1$ &  alternate hypothesis & signal-plus-background & $\Pois(n_{SR} | \nu_S+\nu_B)$ 
\end{tabular}
\end{center}
In this simple example it's fairly obvious that evidence for a signal shows up as an excess of events and a reasonable way to quantify the compatibility of the observed data $n_{CR}^0$ and the null hypothesis is to calculate the probability that the background-only would produce at least this many events; the $p$-value
\begin{equation}
p = \sum_{n=n_{SR}^0}^\infty \Pois(n | \nu_B) \; .
\end{equation}
If this $p$-value is very small, then one might choose to reject the null hypothesis.

Note, the $p$-value is \textit{not} a to be interpreted as the probability of the null hypothesis given the data -- that is a manifestly Bayesian statement.  Instead, the $p$-value is a statement about the probability to have obtained data with a certain property assuming the null hypothesis.

How do we generalize this to more complicated situations?  There were really two ingredients in our simple example.  The first was the proposal that we would reject the null hypothesis based on the probability for it to produce data at least as extreme  as the observed data.  The second ingredient was the prescription for what is meant by more discrepant; in this case the possible observations are ordered according to increasing $n_{SR}$.  One could imagine using difference between observed and expected, $n_{SR}-\nu_B$, as the measure of discrepancy.  In general, a function that maps the data to a single real number is called a \textit{test statistic}: $T(\data)\to\mathbb{R}$.  How does one choose from the infinite number of test statistics?

Neyman and Pearson provided a framework for hypothesis testing that addresses the choice of the test statistic.  This setup treats the null and the alternate hypotheses in an asymmetric way.  First, one defines an \textit{acceptance region} in terms of a test statistic, such that if $T(\data)< k_\alpha$ one accepts the null hypothesis.  One can think of the $T(\data) = k_\alpha$ as defining a contour in the space of the data, which is the boundary of this acceptance region.  Next, one defines the \textit{size of the test}, $\alpha$,\footnote{Note, $\alpha$ is the conventional notation for the size of the test, and has nothing to do with a model parameter in Eq.~\ref{Eq:simultaneous}.} as the probability the null hypothesis will be rejected when it is true (a so-called Type-I error).  This is equivalent to the probability under the null hypothesis that the data will not be found in this acceptance region, ie. $\alpha = P(T(\data) \ge k_\alpha | H_0)$.  Note, it is now clear why there is a subscript on $k_\alpha$, since the contour level is related to the size of the test.  In contrast, if one accepts the null hypothesis when the alternate is true, it is called a Type-II error.  The probability to commit a Type-II error is denoted as $\beta$ and it is given by $\beta=P(T(\data) < k_\alpha|H_1)$.  One calls $1-\beta$ the \textit{power} of the test.  With these definitions in place, one looks for a test statistic that maximizes the power of the test for a fixed test size.  This is a problem for the calculus of variations, and sounds like it might be very difficult for complicated probability models. 

\begin{figure}[h]
\begin{center}
\includegraphics[width=.65\textwidth]{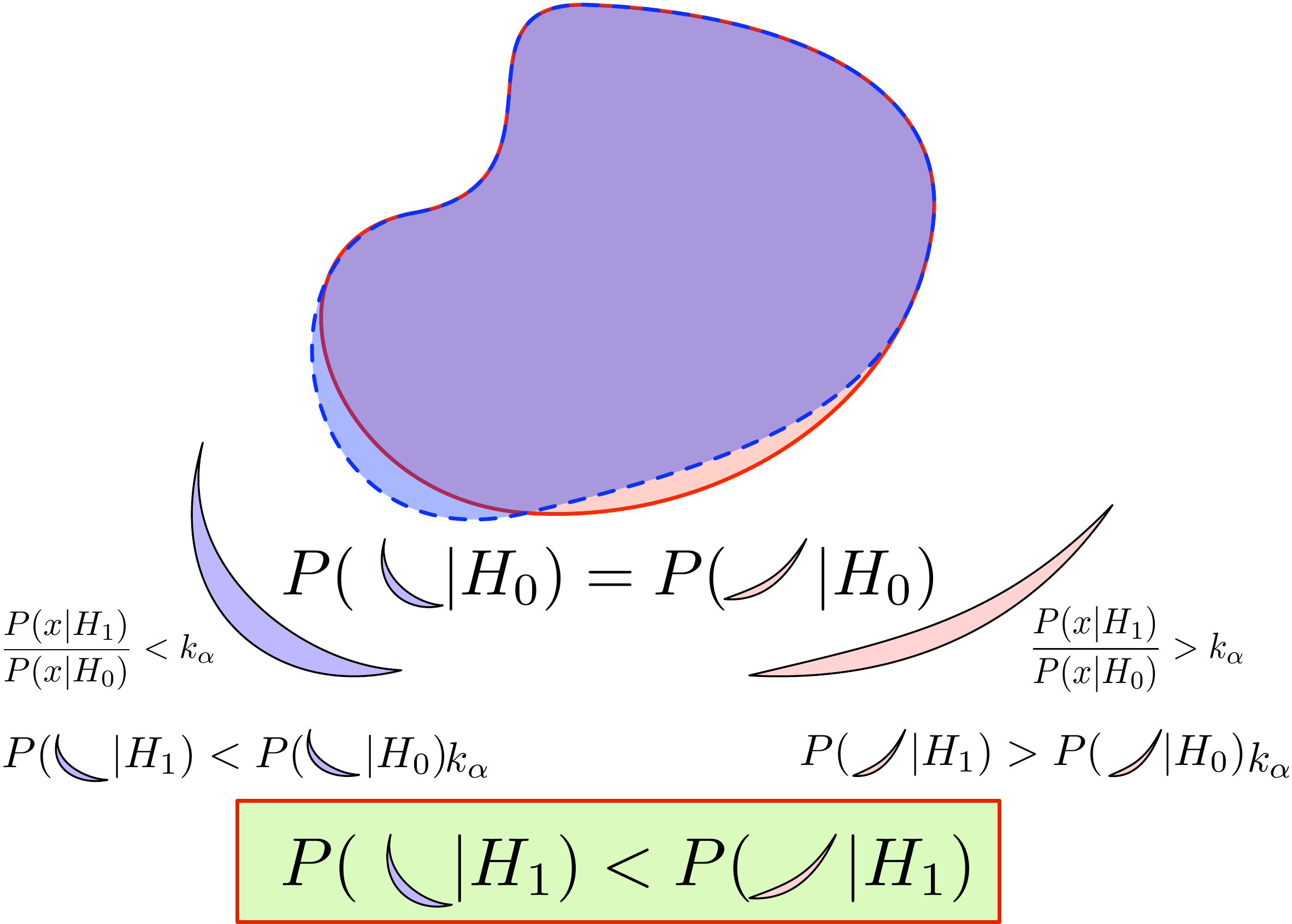}
\caption{A graphical proof of the Neyman-Pearson lemma.}
\label{fig:neymanpearson}
\end{center}
\end{figure}

It turns out that in the case of two simple hypotheses (probability models without any parameters), there is a simple solution!  In particular, the test statistic leading to the most powerful test is given by the likelihood ratio $T_{NP}(\data) = \f(\data|H_1)/\f(\data|H_0)$.  This result is referred to as the Neyman-Pearson lemma, and I will give an informal proof.  We will prove this by considering a small variation to the acceptance region defined by the likelihood ratio.  The solid red contour in Fig.~\ref{fig:neymanpearson} represents the rejection region (the complement to the acceptance region) based on the likelihood ratio and the dashed blue contour represents a small perturbation.  If we can say that any variation to the likelihood ratio has less power, then we will have proved the Neyman-Pearson lemma.  The variation adds (the left, blue wedge) and removes (the right, red wedge) rejection regions.  Because the Neyman-Pearson setup requires that both tests have the same size, we know that the probability for the data to be found in the two wedges must be the same under the null hypothesis.  Because the two regions are on opposite sides of the contour defined by $ \f(\data|H_1)/\f(\data|H_0)$, then we know that the data is less likely to be found in the small region that we added than the small region we subtracted assuming the alternate hypothesis.  In other words, there is less probability to reject the null when the alternate is true; thus the  test based on the new contour is less powerful.

How does this generalize for our most general model in Eq.~\ref{Eq:ftot} with many free parameters?  First one must still define the null and the alternate hypotheses.  Typically is done by saying some parameters -- the parameters of interest $\vec\alpha_{\rm poi}$ --  take on specific values takes on a particular value for the signal-plus-background hypothesis and a different value for the background-only hypothesis.  For instance, the signal production cross-section might be singled out as the \textit{parameter of interest} and it would take on the value of zero for the background-only and some reference value for the signal-plus-background.  The remainder of the parameters are called the \textit{nuisance parameters} $\vec\alpha_{\rm nuis}$.  Unfortunately, there is no equivalent to the Neyman-Pearson lemma for models with several free parameters -- so called, composite models.  Nevertheless, there is a natural generalization based on the profile likelihood ratio.

Remembering that the test statistic $T$ is a real-valued function of the data, then any particular probability model $\f_{\rm tot}(\data|\vec\alpha)$ implies a distribution for the test statistic $f(T|\vec\alpha)$.  Note, the distribution for the test statistic depends on the value of $\vec\alpha$.  Below we will discuss how one constructs this distribution, but lets take it as given for the time being.  Once one has the distribution, then one can calculate the $p$-value is given by
\begin{equation}
p(\vec\alpha) = \int_{T_0}^\infty f(T | \vec\alpha) dT = \int  \f(\data | \vec\alpha )\, \theta(T(\data) - T_0) \,d\data = P(T\ge T_0 | \vec\alpha) \;,
\end{equation}
where $T_0$ is the value of the test statistic based on the observed data and $\theta( \cdot )$ is the Heaviside function.\footnote{The integral $\int d\data$ is a bit unusual for a marked Poisson model, because it involves both a sum over the number of events and an integral over the values of $x_e$ for each of those events.} Usually the $p$-value is just written as $p$, but I have written it as $p(\vec\alpha)$ to make its  $\vec\alpha$-dependence explicit.  

Given that the $p$-value depends on $\vec\alpha$, how does one decide to accept or reject the null hypothesis?  Remembering that $\vec\alpha_{\rm poi}$ takes on a specific value for the null hypothesis, we are worried about how the $p$-value changes as a function of the nuisance parameters.  It is natural to say that one should not reject the null hypothesis if the $p$-value is larger than the size of the test \textit{for any value of the nuisance parameters}.  Thus, in a frequentist approach one should either present $p$-value explicitly as a function of $\vec{\alpha}_{\rm nuis}$ or take its maximal (or supremum) value 
\begin{equation}\label{Eq:psup}
p_{\rm sup}(\vec\alpha_{\rm poi}) = \sup_{ \vec{\alpha}_{\rm nuis}} p(\vec{\alpha}_{\rm nuis}) \; .
\end{equation}

As a final note it is worth mentioning that the size of the test, which serves as the threshold for rejecting the null hypothesis, is purely conventional.  In most sciences conventional choices of the size are 10\%, 5\%, or 1\%.  In particle physics, our conventional threshold for discovery is the infamous $5\sigma$ criterion -- which is a conventional way to refer to $\alpha=2.87 \cdot 10^{-7}$.  This is an incredibly small rate of Type-I error, reflecting that claiming the discovery of new physics would be a monumental statement.  The origin of the $5\sigma$ criterion has its roots in the fact that traditionally we lacked the tools to properly incorporate systematics, we fear that there are systematics that may not be fully under control, and we perform many searches for new physics and thus we have many chances to reject the background-only hypothesis.  We will return to this in the discussion of the look-elsewhere effect.

\subsection{Excluded and allowed regions as confidence intervals}

\begin{figure}[h]
\begin{center}
\subfigure[][]{\includegraphics[width=.57\textwidth]{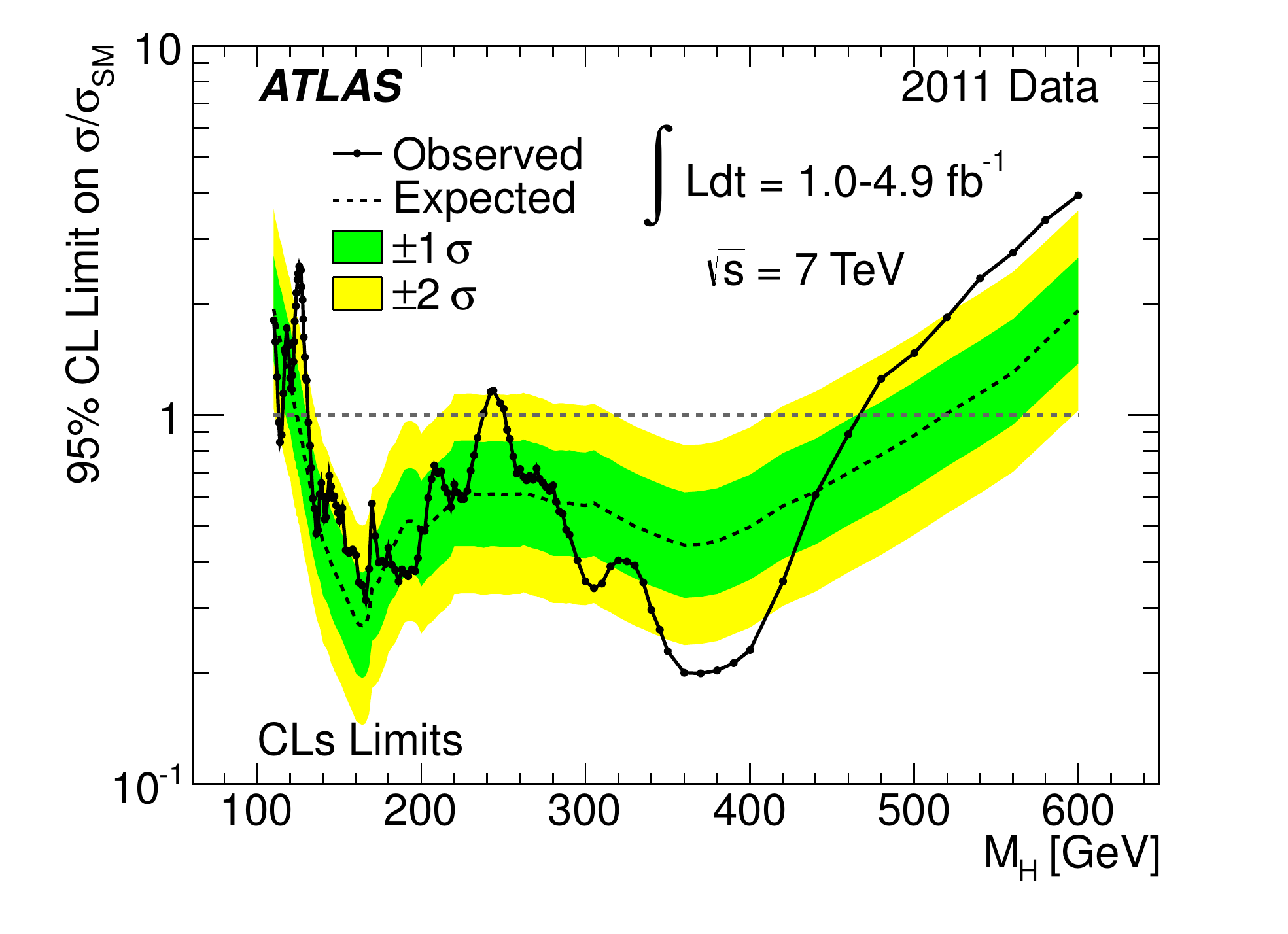}}%
\subfigure[][]{\includegraphics[width=.41\textwidth]{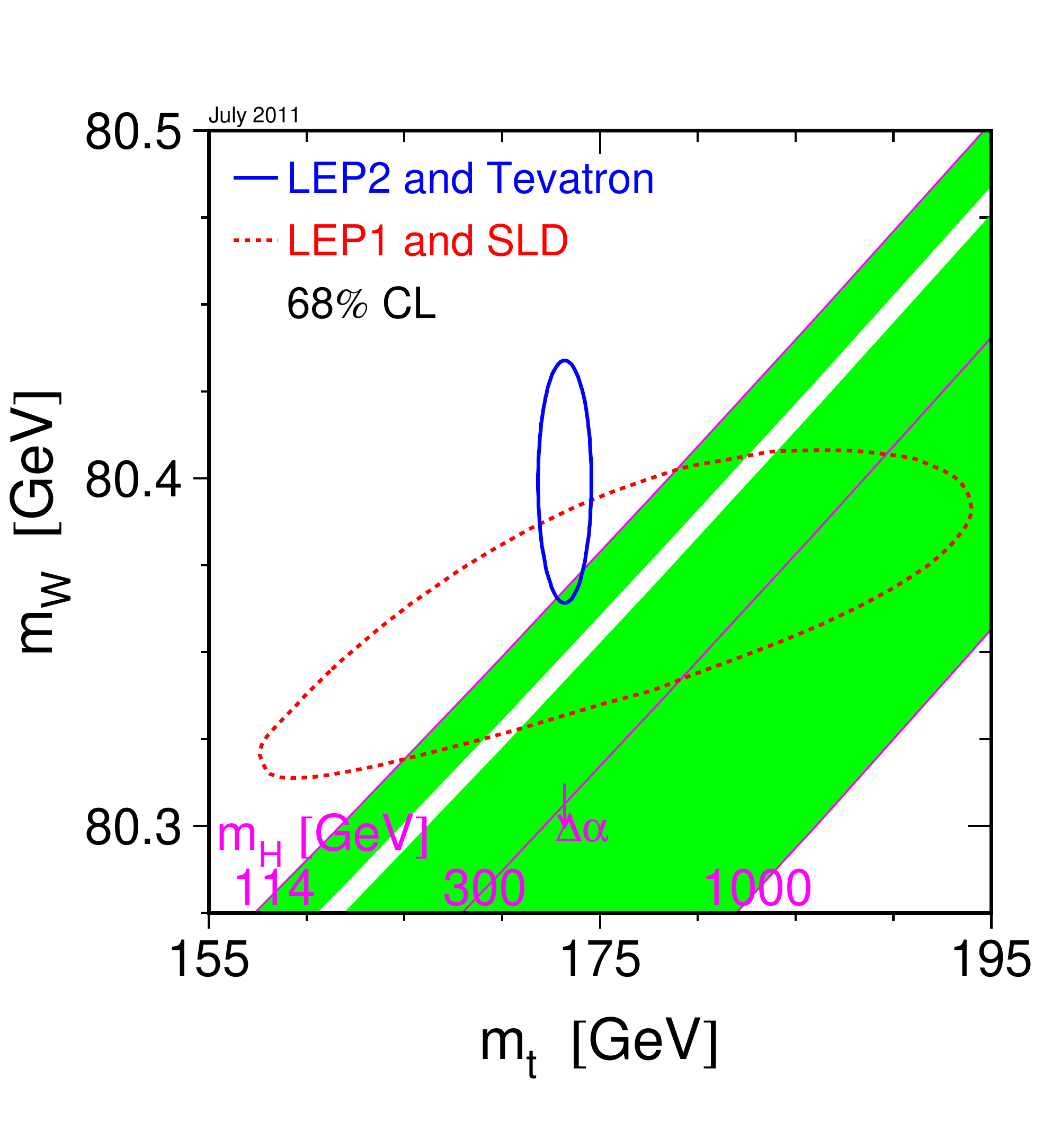}}
\caption{Two examples of confidence intervals.}
\label{fig:confidenceIntervals}
\end{center}
\end{figure}

Often we consider a new physics model that is parametrized by theoretical parameters.  For instance, the mass or coupling of a new particle.  In that case we typically want to ask what values of these theoretical parameters are allowed or excluded given available data.  Figure~\ref{fig:confidenceIntervals} shows two examples.  Figure~\ref{fig:confidenceIntervals}(a) shows an example with $\vec\alpha_{\rm poi} = (\sigma/\sigma_{SM}, M_H)$, where $\sigma/\sigma_{SM}$ is the ratio of the production cross-section for the Higgs boson with respect to its prediction in the standard model and $M_H$ is the unknown Higgs mass parameter in the standard model.  All the parameter points above the solid black curve correspond to scenarios for the Higgs boson that are considered `excluded at the 95\% confidence level'.  Figure~\ref{fig:confidenceIntervals}(b) shows an example with 
$\vec\alpha_{\rm poi} = (m_W,m_t)$ where $m_W$ is the mass of the $W$-boson and $m_t$ is the mass of the top quark. We have discovered the $W$-boson and the top quark and measured their masses.  The blue ellipse `is the 68\% confidence level contour' and all the parameter points inside it are considered `consistent with data at the $1\sigma$ level'.   What is the precise meaning of these statements?

In a frequentist setting, these allowed regions are called \textit{confidence intervals} or \textit{confidence regions}, and the parameter points outside them are considered excluded.  Associated with a confidence interval is a confidence level, i.e. the 95\% and 68\% confidence level in the two examples.  If we repeat the experiments and obtain different data, then these confidence intervals will change.  It is useful to think of the confidence intervals as being random  in the same way the data are random.  The defining property of a 95\% confidence interval is that it \textit{covers} the true value 95\% of the time.  

How can one possibly construct a confidence interval has the desired property, that it \textit{covers} the true value with a specified probability, given that we don't know the true value?  The procedure for building confidence intervals is called the Neyman Construction~\cite{Neyman}, and it is based on `inverting' a series of hypothesis tests (as described in Sec.~\ref{S:hypothesis test}).  In particular, for each value of $\vec\alpha$ in the parameter space one performs a hypothesis test based on some test statistic where the null hypothesis is $\vec\alpha$.  Note, that in this context, the null hypothesis is changing for each test and generally is not the background-only.  If one wants a 95\% confidence interval, then one constructs a series of hypothesis test with a size of 5\%.  The confidence interval $I(\data)$ is constructed by taking the set of parameter points where the null hypothesis is accepted. 
\begin{equation}
I(\data) = \left\{ \vec\alpha \middle |\, P(T(\data)>k_\alpha \,|\, \vec\alpha) < \alpha \right\} \;,
\end{equation}
where the final $\alpha$ and the subscript $k_\alpha$ refer to the size of the test.
Since a hypothesis test with a size of 5\% should accept the null hypothesis 95\% of the time if it is true, confidence intervals constructed in this way satisfy the defining property.  This same property is usually formulated in terms of \textit{coverage}.  Coverage is the probability that the interval will contain (cover) the parameter $\vec\alpha$ when it is true,
\begin{equation}
\textrm{coverage}(\vec\alpha) = P(\vec\alpha \in I\, |\, \vec\alpha) \; .
\end{equation}
The equation above can easily be mis-interpreted as the probability the parameter is in a fixed interval $I$; but one must remember that in evaluating the probability above the data $\data$, and, thus, the corresponding intervals produced by the procedure $I(\data)$, are the random quantities.  
Note, that coverage is a property that can be  quantified for any procedure that produces the confidence intervals $I$.  Intervals produced using the Neyman Construction procedure are said to ``cover by construction''; however, one can consider alternative procedures that may either under-cover or over-cover.  Undercoverage means that \mbox{$P(\vec\alpha \in I\, |\, \vec\alpha)$} is smaller than desired and over-coverage means that $P(\vec\alpha \in I\, |\, \vec\alpha)$ is larger than desired.  Note that in general coverage depends on the assumed true value $\vec\alpha$.

Since one typically is only interested in forming confidence intervals on the parameters of interest, then one could use the supremum $p$-value of Eq.~\ref{Eq:psup}.  This procedure ensures that the coverage is at least the desired level, though for some values of $\vec\alpha$ it may over-cover (perhaps significantly).  This procedure, which I call the `full construction',  is also computationally very intensive when $\vec\alpha$ has many parameters as it require performing many hypothesis tests.  In the naive approach where each $\alpha_p$ is scanned in a regular grid, the number of parameter points tested grows exponentially in the number of parameters.  There is an alternative approach, which I call the `profile construction'~\cite{Feldman,Cranmer:2005hi}
 and which statisticians call an `hybrid resampling technique'~ \cite{Hybrid,Bodhi} that is approximate to the full construction, but typically has good coverage properties.  We return to the procedures and properties for the different types of Neyman Constructions later.

\begin{figure}[htbp]
\begin{center}
\includegraphics[width=.9\textwidth]{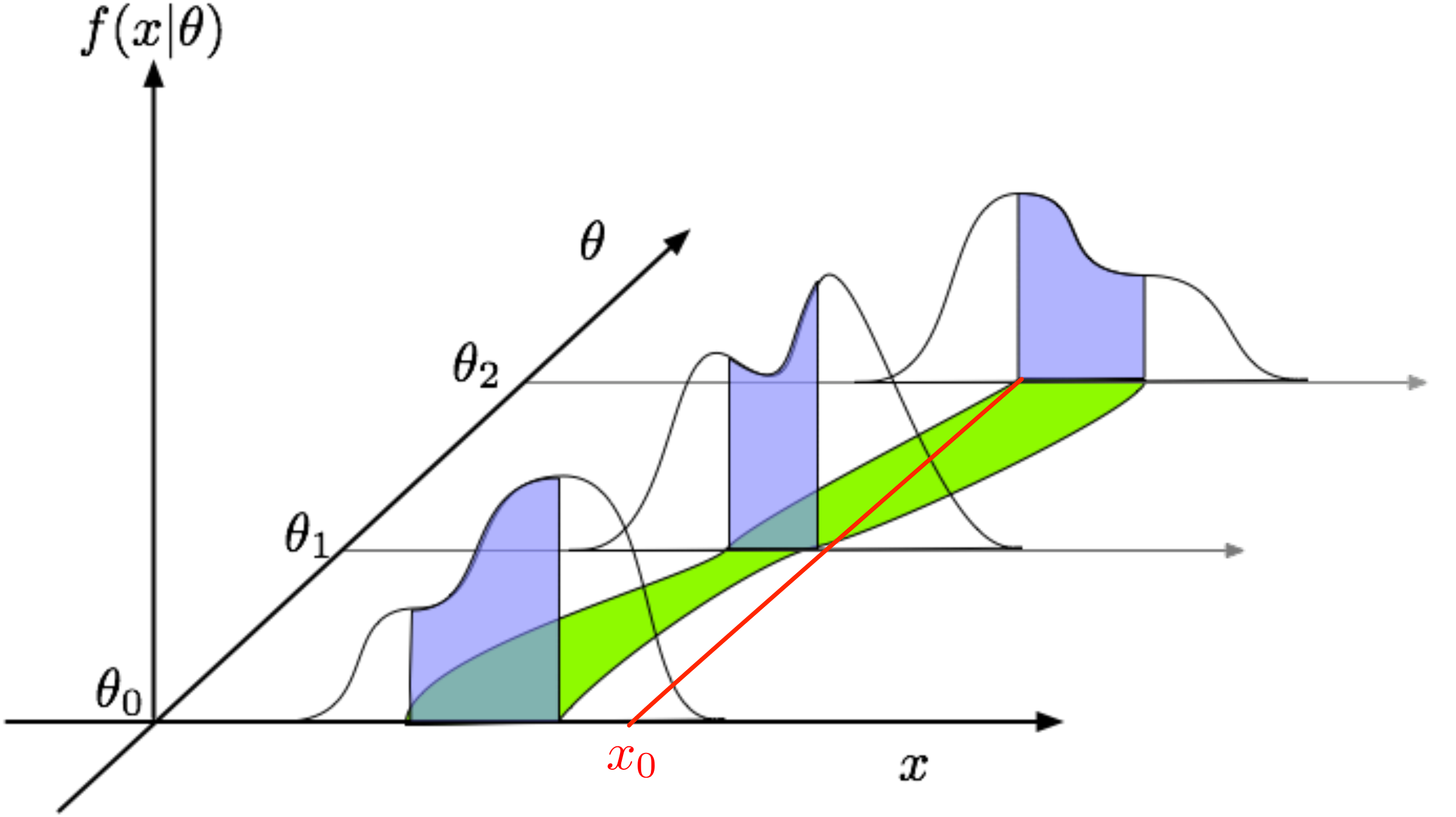}
\caption{A schematic visualization of the Neyman Construction.  For each value of $\theta$ one finds a region in $x$ that satisfies $\int f(x|\theta) dx$ (blue).  Together these regions form a confidence belt (green).  The intersection of the observation $x_0$ (red) with the confidence belt defines the confidence interval $[\theta_1,\theta_2]$.}
\label{fig:NC_schematic}
\end{center}
\end{figure}

\begin{figure}[htbp]
\begin{center}
\includegraphics[width=.9\textwidth]{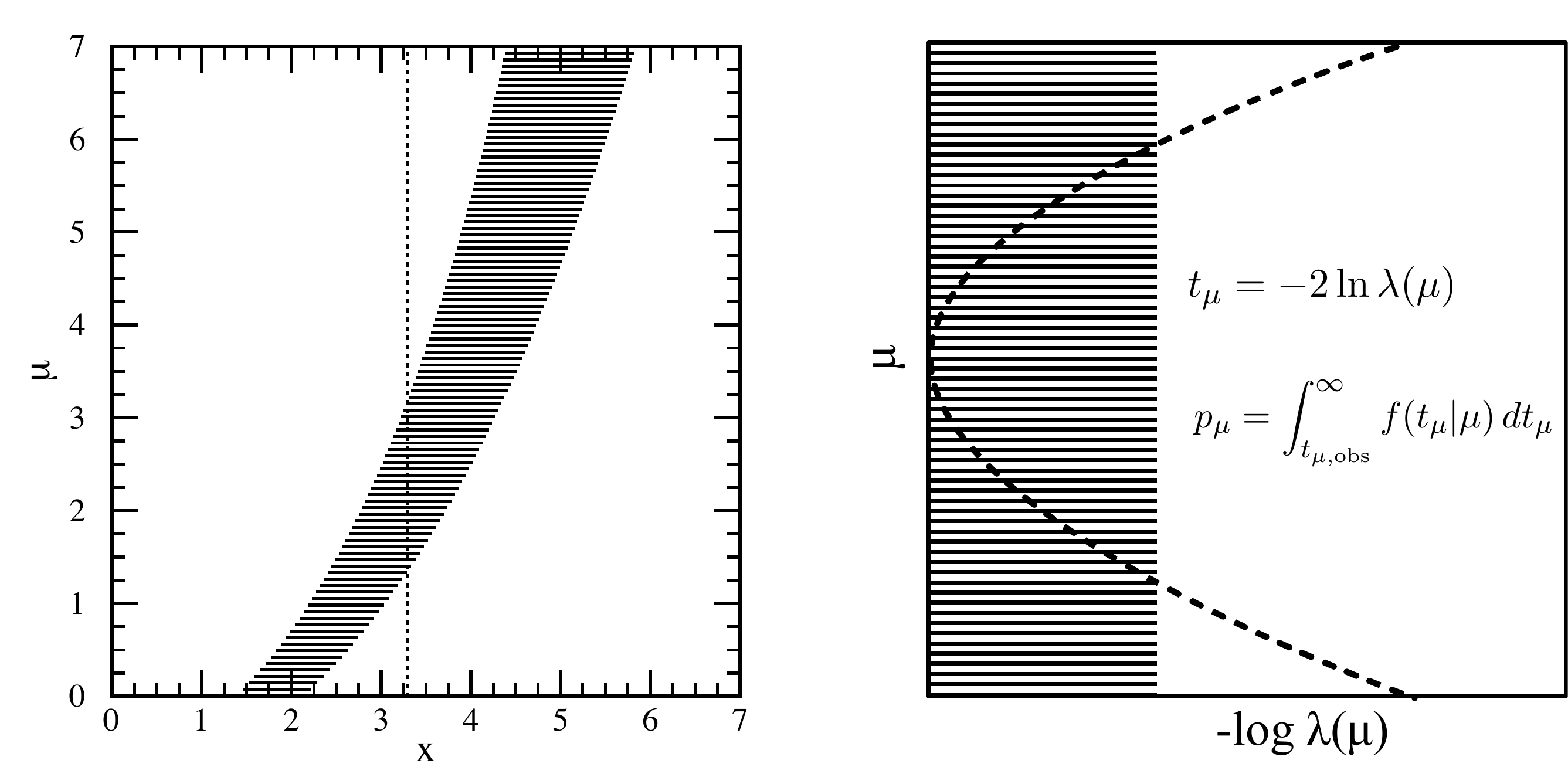}
\caption{Two presentations of a confidence belt (see text).  Left panel taken from Ref.~\cite{Feldman:1997qc}.  Right panel shows a presentation that generalizes to more complicated models.}
\label{fig:neyman}
\end{center}
\end{figure}

Figure~\ref{fig:NC_schematic} provides an overview of the classic Neyman construction corresponding to the left panel of Fig.~\ref{fig:neyman}.  The left panel of  Fig.~\ref{fig:neyman} is taken from the Feldman and Cousins's paper~\cite{Feldman:1997qc} where the parameter of the model is denoted $\mu$ instead of $\theta$.  For each value of the parameter $\mu$, the acceptance region in $x$ is illustrated as a horizontal bar.  Those regions are the ones that satisfy $T(\data)<k_\alpha$, and in the case of Feldman-Cousins the test statistic is the one of Eq.~\ref{eq:tmu}.  This presentation of the confidence belt works well for a simple model in which the data consists of a single measurement $\data=\{x\}$.  Once one has the confidence belt, then one can immediately find the confidence interval for a particular measurement of $x$ simply by taking drawing a vertical line for the measured value of $x$ and finding the intersection with the confidence belt.

Unfortunately, this convenient visualization doesn't generalize to complicated models with many channels or even a single channel marked Poisson model where $\data=\{x_1,\dots,x_n\}$.  In those more complicated cases, the confidence belt can still be visualized where the observable $x$ is replaced with $T$, the test statistic itself.  Thus, the boundary of the belt is given by $k_\alpha$ vs. $\mu$ as in the right panel of Figure~\ref{fig:neyman}. The analog to the vertical line in the left panel is now a curve showing how the observed value of the test statistic depends on $\mu$.  The confidence interval still corresponds to the intersection of the observed test statistic curve and the confidence belt, which clearly satisfies $T(\data)<k_\alpha$.  For more complicated models with many parameters the confidence belt will have one axis for the test statistic and one axis for each model parameter.

Note, a 95\% confidence interval \textit{does not} mean that there is a 95\% chance that the true value of the parameter is inside the interval -- that is a manifestly Bayesian statement.  One can produce a Bayesian \textit{credible interval} with that interpretation; however, that requires a prior probability distribution over the parameters.  Similarly, for any fixed interval $I$ one can compute the Bayesian credibility of the interval 
\begin{equation}
\label{eq:credible}
P(\alpha\in I | \data) = \frac{ \int_{I}  \f(\data | \vec\alpha) \pi(\vec\alpha) d\vec\alpha }{\int \f(\data | \vec\alpha) \pi(\vec\alpha) d\vec\alpha} \;.
\end{equation}

\section{Modeling and the Scientific Narrative}

Now that we have established a general form for a probability model (Eq.~\ref{Eq:simultaneous}) and we have translated the basic questions of measurement, discovery, and exclusion into the statistical language we are ready to address the heart of the statistical challenge -- building the model.  It is difficult to overestimate how important the model building stage is.  So many of the questions that are addressed to the statistical experts in the major particle physics collaborations are not really about statistics \textit{per se}, but about model building.  In fact, the first question that you are likely to be asked by one of the statistical experts is ``what is your model?''

Often people are confused by the question ``what is your model?'' or simply have not written it down.  You simply can't make much progress on any statistical questions if you haven't written down a model.  Of course, people do usually have some idea for what it is that they want to do The process of writing down the model often obviates the answer to the question, reveals some fundamental confusion or assumption in the analysis strategy, or both.  As mentioned in the introduction, writing down the model is intimately related with the analysis strategy and it is a good way to organize an analysis effort.

I like to think of the modeling stage in terms of a  \textit{scientific narrative}.  I find that there are three main narrative elements, though many analyses use a mixture of these elements when building the model.  Below I will discuss these narrative elements, how they are translated into a mathematical formulation, and their relative pros and cons.

\subsection{Simulation Narrative}

The simulation narrative is probably the easiest to explain and produces statistical models with the strongest logical connection to physical theory being tested.  We begin with an relation that every particle physicists should know for the rate of events expected from a specific physical process
\begin{equation}
\textrm{rate} = \textrm{(flux)} \times \textrm{(cross section)}  \times\textrm{(efficiency)}  \times \textrm{(acceptance)} \;,
\end{equation}
where the cross section is predicted from the theory, the flux is controlled by the accelerator\footnote{In some cases, like cosmic rays, the flux must be estimated since the accelerator is quite far away.}, and the efficiency and acceptance are properties of the detector and event selection criteria.  It is worth noting that the equation above is actually a repackaging of a more fundamental relationship. In fact the fundamental quantity that is predicted from first principles in quantum theory is the  \textit{scattering probability}  \mbox{$P(i\to f)=|\langle i|f\rangle|^2/ (\langle i|i\rangle\langle f | f \rangle)$} inside a box of size $V$ over some time interval $T$, which is then repackaged into the Lorentz invariant form above~\cite{Sredniki}.

In the simulation narrative the efficiency and acceptance are estimated with computer simulations of the detector.  Typically, a large sample of events is generated using Monte Carlo techniques~\cite{MonteCarlo}.  The Monte Carlo sampling is performed separately  for the hard (perturbative) interaction (e.g. \texttt{MadGraph}), the parton shower and hadronization process (e.g. \texttt{Pythia} and \texttt{Herwig}), and the interaction of particles with the detector (e.g. \texttt{Geant}).  Note, the efficiency and acceptance depend on the physical process considered, and I will refer to each such process as a \textit{sample} (in reference to the corresponding sample of events generated with Monte Carlo techniques).

To simplify the notation, I will define the effective cross section, $\sigma_{\rm eff.}$ to be the product of the total cross section, efficiency, and acceptance.  Thus, the total number of events expected to be selected for a given scattering process, $\nu$, is the product of the time-integrated flux or time-integrated luminosity, $\lambda$, and the effective cross section
\begin{equation}
\nu = \lambda \sigma_{\rm eff.}\;.
\end{equation}
I use $\lambda$ here instead of the more common $L$ to avoid confusion with the likelihood function and because when we incorporate uncertainty on the time-integrated luminosity it will be a parameter of the model for which I have chosen to use greek letters.  

If we did not need to worry about detector effects and we could measure the final state perfectly, then the distribution for any observable $x$ would be given by
\begin{equation}
\textrm{(idealized)}\hspace{2em}f(x) = \frac{1}{\sigma_{\rm eff.}} \frac{d\sigma_{\rm eff.}}{dx}\;.\hspace{5em}
\end{equation}
Of course, we do need to worry about detector effects and we incorporate them with the detector simulation discussed above.  From the Monte Carlo sample of events\footnote{Here I only consider unweighted Monte Carlo samples, but the discussion below can be generalized for weighted Monte Carlo samples.} $\{x_1, \dots, x_N\}$ we can estimate the underlying distribution $f(x)$ simply by creating a histogram.  If we want we can write the histogram based on $B$ bins centered at $x_b$ with bin width $w_b$ explicitly as
\begin{equation}
\textrm{(histogram)} \hspace{2em}f(x) \approx h(x) =  \sum_{i=1}^N \sum_{b=1}^B  \frac{ \theta(|x_i-x_b|/w_b) }{N} \frac{\theta(|x -x_b|/w_b)}{w_b}\;,  \end{equation}
where the first Heaviside function accumulates simulated events in the bin and the second selects the bin containing the value of $x$ in question.  Histograms are the most common way to estimate a probability density function based on a finite sample, but there are other possibilities.  The downsides of histograms as an estimate for the distribution $f(x)$ is that they are discontinuous and have dependence on the location of the bin boundaries.  A particularly nice alternative is called kernel estimation~\cite{Cranmer:2000du}.  In this approach, one places a kernel of probability $K(x)$ centered around each event in the sample:
\begin{equation}\label{eq:keys}
\textrm{(kernel estimate)}\hspace{2em}f(x) \approx \hat{f}_0(x) =  \frac{1}{N} \sum_{i=1}^N K\left( \frac{x-x_i}{h} \right)\;.\hspace{1em}
\end{equation}
The most common choice of the kernel is a Gaussian distribution, and there are results for the optimal width of the kernel $h$.  Equation~\ref{eq:keys} is referred to as the fixed kernel estimate since $h$ is common for all the events in the sample.  A second order estimate or adaptive kernel estimation provides better performance when the distribution is multimodal or has both narrow and wide features~\cite{Cranmer:2000du}.

\subsubsection{The multi-sample mixture model}
So far we have only considered a single interaction process, or sample.  How do we form a model when there are several scattering processes contributing to the total rate and distribution of $x$?  From first principles of quantum mechanics we must add these different processes together.  Since there is no physical meaning to label individual processes that interfere quantum mechanically, I will consider all such processes as a single sample.  Thus the remaining set of samples that do not interfere simply add incoherently.  The total rate is simply the sum of the individual rates
\begin{equation}
\nu_{\rm tot} = \sum_{s\in\textrm{samples}} \nu_s
\end{equation}
and the total distribution is a weighted sum called a \textit{mixture model}
\begin{equation}
f(x) = \frac{1}{\nu_{\textrm{tot}}} \sum_{s\in\textrm{samples}} \nu_s f_s(x)\;,
\end{equation}
where the subscript $s$ has been added to the equations above for each such sample.  With these two ingredients we can construct our  marked Poisson model of Eq.~\ref{Eq:markedPoisson} for a single channel, and we can simply repeat this for several disjoint event selection requirements to form a multi-channel simultaneous model like Eq.~\ref{Eq:simultaneous}.  In the multi-channel case we will give the additional subscript $c\in\textrm{channels}$ to $\nu_{cs}$, $f_{cs}(x)$, $\nu_{c,\rm{tot}}$, and $f_c(x)$. However, at this point, our model has no free parameters $\vec\alpha$.

\subsubsection{Incorporating physics parameters into the model}

Now we want to parametrize our model interns of some physical parameters $\vec\alpha$, such as those that appear in the Lagrangian of a some theory.  Changing the parameters in the Lagrangian of a theory will in general change both the total rate $\nu$ and the shape of the distributions $f(x)$.  In principle, we can repeat the procedure above for each value of these parameters $\vec\alpha$ to form $\nu_{cs}(\vec\alpha)$ and $f_{cs}(x|\vec\alpha)$ for each sample and selection channel, and, thus, from $\F_\textrm{sim}(\data|\vec\alpha)$.  In practice, we need to resort to some interpolation strategy over the individual parameter points $\{\vec\alpha_i\}$ where we have Monte Carlo samples.  We will return to these interpolation strategies later.

In some case the only effect of the parameter is to scale the rate of some scattering process $\nu_s(\vec\alpha)$ without changing its distribution $f_s(x|\vec\alpha)$.  Furthermore, the scaling is often known analytically, for instance, a coupling constants produce a linear relationship like $\nu(\alpha_p) = \xi \alpha_p  + \nu_0$.  In such cases, interpolation is not necessary and the parametrization of the likelihood function is straightforward.  

Note, not all physics parameters need be considered parameters of interest.  There may be a free physics parameter that is not directly of interest, and as such it would be considered a nuisance parameter.  

\subsubsubsection{An example, the search for the standard model Higgs boson}

In the case of searches for the standard model Higgs boson, the only free parameter in the Lagrangian is $m_H$.  Once $m_H$ is specified the rates and the shapes for each of the scattering processes (combinations of production and decay modes) are specified by the theory.  Of course, as the Higgs boson mass changes the distributions do change so we do need to worry about interpolating the shapes $f(x|m_H)$.  However the results are often presented as a \textit{raster scan} over $m_H$, where one fixes $m_H$ and then asks about the rate of signal events from the Higgs boson scattering process.  With $m_H$ fixed this is really a simple hypothesis test between background-only and signal-plus-background\footnote{Note that  $H\to WW$ interferes with ``background-only'' $WW$ scattering process.  For low Higgs boson masses, the narrow Higgs width means this interference is negligible.  However, at high masses the interference effect is significant and we should really treat these two processes together as a single sample.}, but we usually choose to construct a parametrized model that does not directly correspond to any theory.  In this case the parameter of interest is some scaling of the rate with respect to the standard model prediction, $\mu = \sigma / \sigma_{\rm SM}$, such that $\mu=0$ is the background-only situation and $\mu=1$ is the standard model prediction.  Furthermore, we usually use this global $\mu$ factor for each of the production and decay modes even though essentially all theories of physics beyond the standard model would modify the rates of the various scattering processes differently. Figure~\ref{fig:confidenceIntervals} shows confidence intervals on $\mu$ for fixed values of $m_H$.  Values below the solid black curve are not excluded (since an arbitrarily small signal rate cannot be differentiated from the background-only and this is a one-sided confidence interval).

\subsubsection{Incorporating systematic effects}

The parton shower, hadronization, and detector simulation components of the simulation narrative are based on phenomenological models that have many adjustable parameters.  These parameters are nuisance parameters included in our master list of parameters $\vec\alpha$.  The changes in the rates $\nu(\vec\alpha)$ and shapes $f(x|\vec\alpha)$ due to these parameters lead to systematic uncertainties\footnote{Systematic uncertainty is arguably a better term than systematic \textit{error}.}.  We have already eluded to how one can deal with the presence of nuisance parameters in hypothesis testing and confidence intervals, but here we are focusing on the modeling stage.  In principle, we deal with modeling of these nuisance parameters in the same way as the physics parameters, which is to generate Monte Carlo samples for several choices of the parameters $\{\vec\alpha_i\}$ and then use some interpolation strategy to form a continuous parametrization for $\nu(\vec\alpha), f(x|\vec\alpha)$, and $\F_{\rm sim}(\data|\vec\alpha)$.  In practice, there are many nuisance parameters associated to the parton shower, hadronization, and detector simulation so this becomes a multi-dimensional interpolation problem\footnote{This is sometimes referred to as `template morphing'}.  This is one of the most severe challenges for the simulation narrative.  

Typically, we don't map out the correlated effect of changing multiple $\alpha_p$ simultaneously.  Instead, we have some nominal settings for these parameters $\vec\alpha^0$ and then vary each individual parameter `up' and `down' by some reasonable amount $\alpha_p^\pm$.  So if we have $N_P$ parameters we typically have $1+2N_P$ variations of the Monte Carlo sample from which we try to form  $\F_{\rm sim}(\data|\vec\alpha)$.  This is clearly not an ideal situation and it is not hard to imagine cases where the combined effect on the rate and shapes cannot be factorized in terms of changes from the individual parameters.

What is meant by ``vary each individual parameter `up' and `down' by some reasonable amount'' in the paragraph above?  The nominal choice of the parameters $\vec\alpha^0$ is usually based on experience, test beam studies,  Monte Carlo `tunings', etc..  These studies correspond to auxiliary measurements in the language used in Sec.~\ref{S:AuxMeas} and Sec.~\ref{S:Constraint}.  Similarly, these parameters typically have some maximum likelihood estimates and standard uncertainties from the auxiliary measurements as described in Sec.~\ref{S:estimation}. Thus our complete model $\F_{\rm tot}(\data|\vec\alpha)$ of Eq.~\ref{Eq:ftot} should not only deal with parametrizing the effect of changing each $\alpha_p$ but also include either a constraint term $f_p(a_p | \alpha_p)$ or an additional channel that describes a more complete probability model for the auxiliary measurement.  

Below we will consider a specific interpolation strategy and a few of the most popular conventions for constraint terms.  However, before moving on it is worth emphasizing that while, naively, the matrix element associated to a perturbative scattering amplitude has no free parameters (beyond the physics parameters discussed above), fixed order perturbative calculations do have residual  scale dependence.  This type of \textit{theoretical uncertainty} has no auxiliary measurement associated with it even in principle, thus it really has no frequentist description.  This was discussed briefly in Sec.~\ref{S:Constraint}.  In contrast, the parton density functions are the results of auxiliary measurements and the groups producing the parton density function sets spend time providing sensible multivariate constraint terms for those parameters.  However, those measurements also have uncertainties due to parametrization choices and theoretical uncertainties, which are not statistical in nature.  In short we must take care in ascribing constraint terms to theoretical uncertainties and measurements that have theoretical uncertainties\footnote{``Note that I deliberately called them theory \textit{errors}, not uncertainties.'' -- Tilman Plehn}.  

\subsubsection{Tabulating the effect of varying sources of uncertainty}

The treatment of systematic uncertainties is subtle, particularly when one wishes to take into account the correlated effect of multiple sources of systematic uncertainty across many signal and background samples.   
The most important conceptual issue is that we separate the source of the uncertainty (for instance the uncertainty in the calorimeter's response to jets) from its effect on an individual signal or background sample (eg. the change in the acceptance and shape of a $W$+jets background).  In particular, the same source of uncertainty has a different effect on the various signal and background samples.  The effect of these `up' and `down' variations about the nominal predictions $\nu_s(\vec\alpha^0)$  and $f_{sb}(x|\vec\alpha^0)$ is quantified by dedicated studies.   The result of these studies can be arranged in tables like those below.  The main purpose of the \HF\ XML schema is to represent these tables.  And \HF\ is a tool that can convert these tables into our master model $\F_{\rm tot}(\data |\vec\alpha)$ of Eq.~\ref{Eq:ftot} implemented as a  \texttt{RooAbsPdf} with a  \texttt{ModelConfig} to make it compatible with \roostats\ tools.  The convention used by \HF\ is related to our notation via 
\begin{equation}
\nu_s(\vec\alpha) f_s(x|\vec\alpha) = \eta_{s}(\vec\alpha) \sigma_{s}(x|\vec\alpha) \, 
\end{equation}
where  $\eta_{s}(\vec\alpha)$ represents relative changes in the overall rate $\nu(\vec\alpha)$  and $\sigma_{s}(x|\vec\alpha)$ includes both changes to the rate and the shape $f(x|\vec\alpha)$.  This choice is one of convenience because histograms are often not normalized to unity, but instead in code rate information.  As the name implies, \HF\ works with histograms, so instead of writing $\sigma_{s}(x|\vec\alpha)$ the table is written as $\sigma_{sb}(\vec\alpha)$, where $b$ is a bin index. To compress the notation further, $\eta_{p=1,s=1}^+$ and $\sigma_{psb}^\pm$ represent the value of when $\alpha_p=\alpha_p^\pm$ and all other parameters are fixed to their nominal values.  Thus we arrive at the following tabular form for models built on the simulation narrative based on histograms with individual nuisance parameters varied one at a time:
\begin{table}[h]
\center
\begin{tabular}{l | c c c}
Syst & Sample 1 & $\dots$ & Sample N \\ \hline
Nominal Value & $\eta_{s=1}^0=1$ & \dots & $\eta_{s=N}^0=1$\\\hline
$p$=\OS\ 1 & $\eta_{p=1,s=1}^+$, \;$\eta_{p=1,s=1}^-$ & \dots & $\eta_{p=1,s=N}^+$,\; $\eta_{p=1,s=N}^-$\\ 
$\vdots$ & $\vdots$ & $\ddots$ & $\vdots$ \\
$p$=\OS\ M & $\eta_{p=M,s=1}^+$, $\eta_{p=M,s=1}^-$ & \dots & $\eta_{p=M,s=N}^+$, $\eta_{p=M,s=N}^-$\\ \hline
Net Effect & $\eta_{s=1}(\vec{\alpha})$ & \dots & $\eta_{s=N}(\vec{\alpha})$
\end{tabular}
\caption{Tabular representation of sources of uncertainties that produce a correlated effect in the normalization individual samples (eg. OverallSys).  The $\eta^+_{ps}$ represent histogram when $\alpha_s=1$ and are inserted into the \texttt{High} attribute of the \OS\  XML element.  Similarly, the $\eta^-_{ps}$ represent histogram when $\alpha_s=-1$ and are inserted into the \texttt{Low} attribute of the \OS\  XML element. Note, this does not imply that $\eta^+ > \eta^-$, the $\pm$ superscript correspond to the variation in the source of the systematic, not the resulting effect.}
\end{table}

\begin{table}[h]
\center
\begin{tabular}{l | c c c}
Syst & Sample 1 & $\dots$ & Sample N \\ \hline
Nominal Value & $\sigma_{s=1,b}^0$ & \dots & $\sigma_{s=N,b}^0$\\\hline
$p$=\HS\  1\;\; & $\sigma_{p=1,s=1,b}^+$, $\sigma_{p=1,s=1,b}^-$ & \dots & $\sigma_{p=1,s=N,b}^+$, $\sigma_{p=1,s=N,b}^-$\\ 
$\vdots$ & $\vdots$ & $\ddots$ & $\vdots$ \\
$p$=\HS\  M \;\;\;\;& $\sigma_{p=M,s=1,b}^+$, \;\;$\sigma_{p=M,s=1,b}^-$ \;\;& \dots & $\sigma_{p=M,s=N,b}^+$, \;\; $\sigma_{p=M,s=N,b}^-$\;\;\\ \hline
Net Effect & $\sigma_{s=1,b}(\vec{\alpha})$ & \dots & $\sigma_{s=N,b}(\vec{\alpha})$
\end{tabular}
\caption{Tabular representation of sources of uncertainties that produce a correlated effect in the normalization and shape individual samples (eg. \HS\ ).  The $\sigma^+_{psb}$ represent histogram when $\alpha_s=1$ and are inserted into the \texttt{HighHist} attribute of the \HS\  XML element.  Similarly, the $\sigma^-_{psb}$ represent histogram when $\alpha_s=-1$ and are inserted into the \texttt{LowHist} attribute of the \HS\  XML element.   }
\end{table}

\subsubsection{Interpolation Conventions}
\label{S:Interpolation}

\begin{figure}[h]
\begin{center}
\subfigure[][]{\includegraphics[width=.4\textwidth]{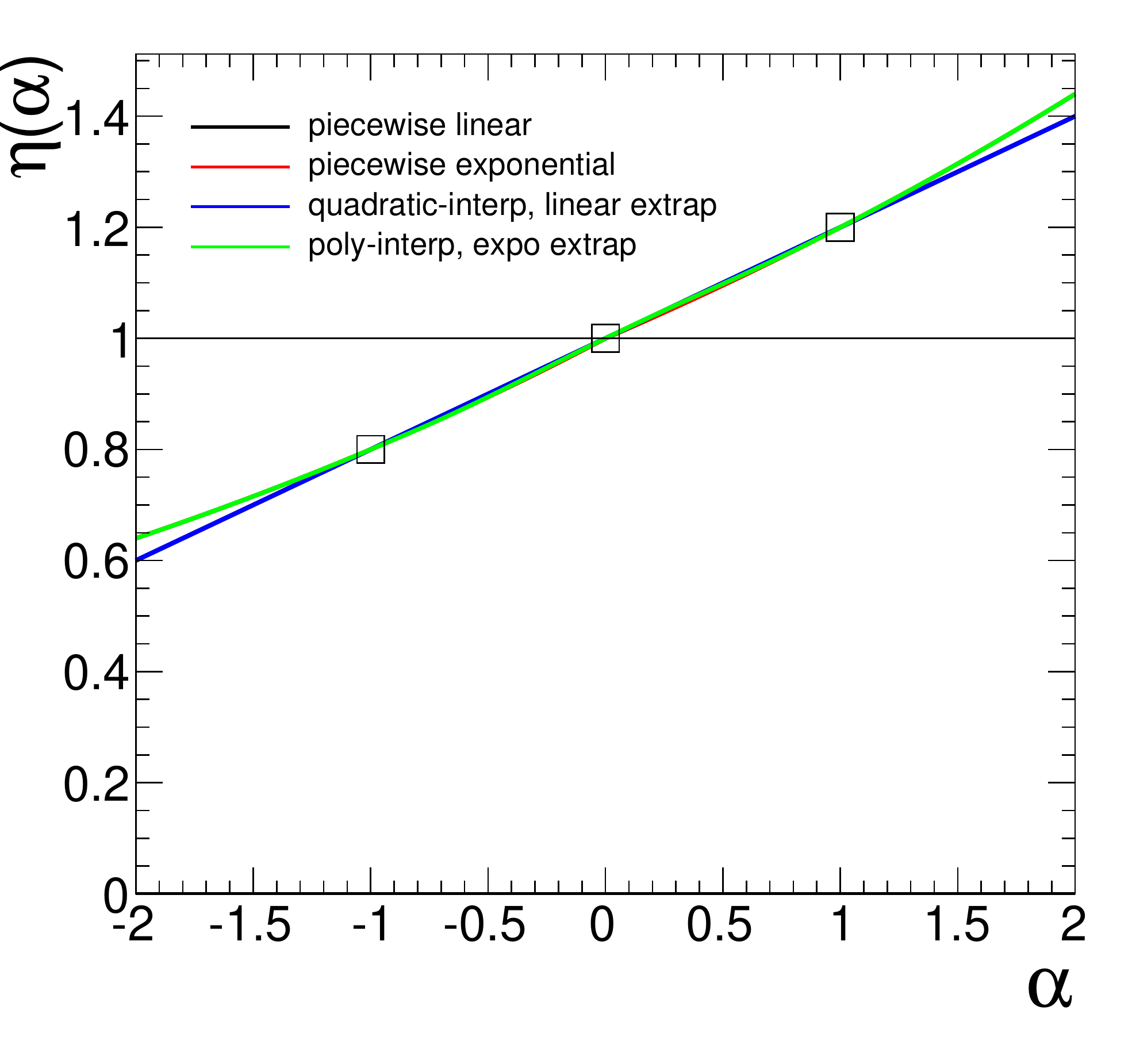}}
\subfigure[][]{\includegraphics[width=.4\textwidth]{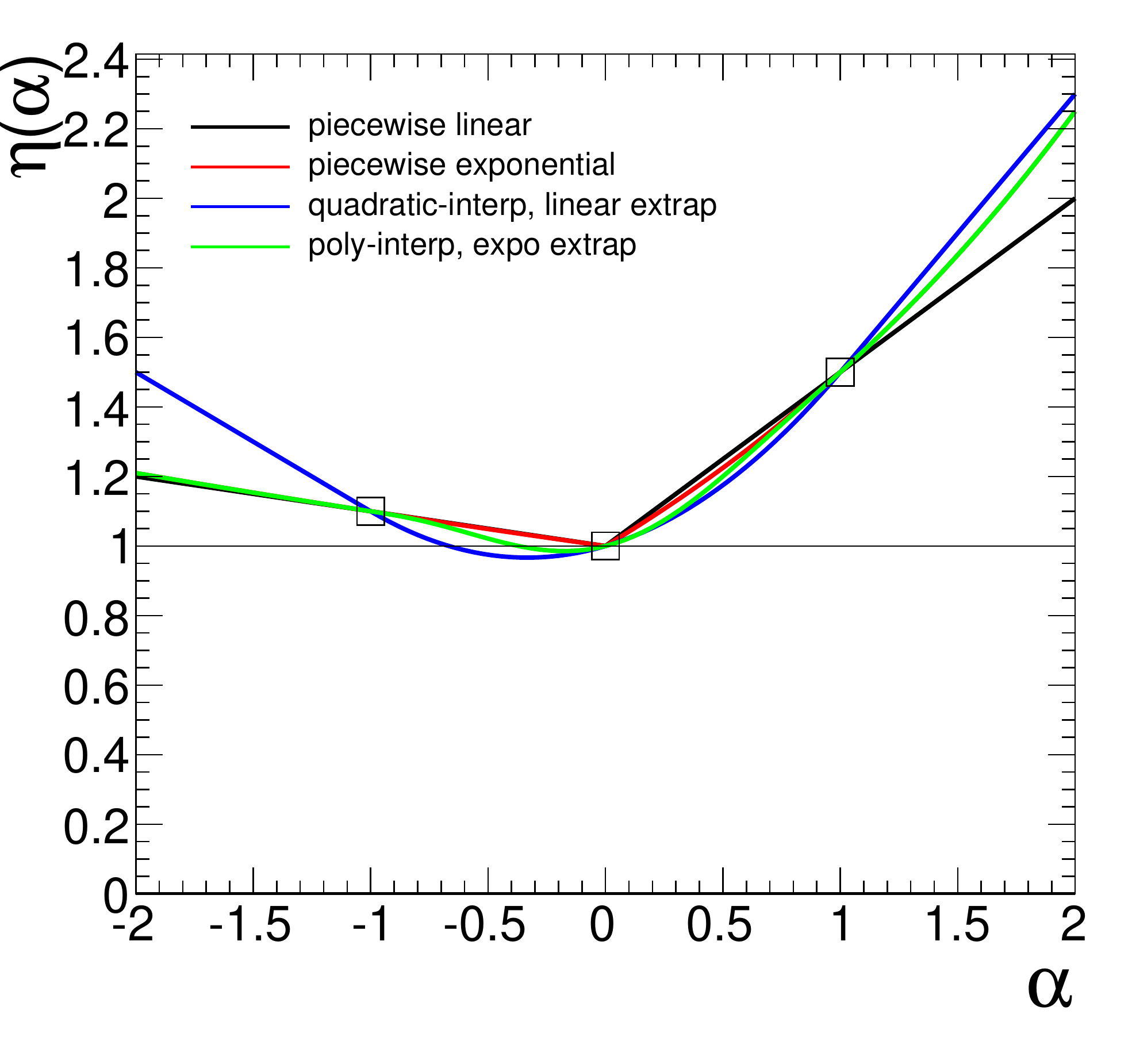}}
\subfigure[][]{\includegraphics[width=.4\textwidth]{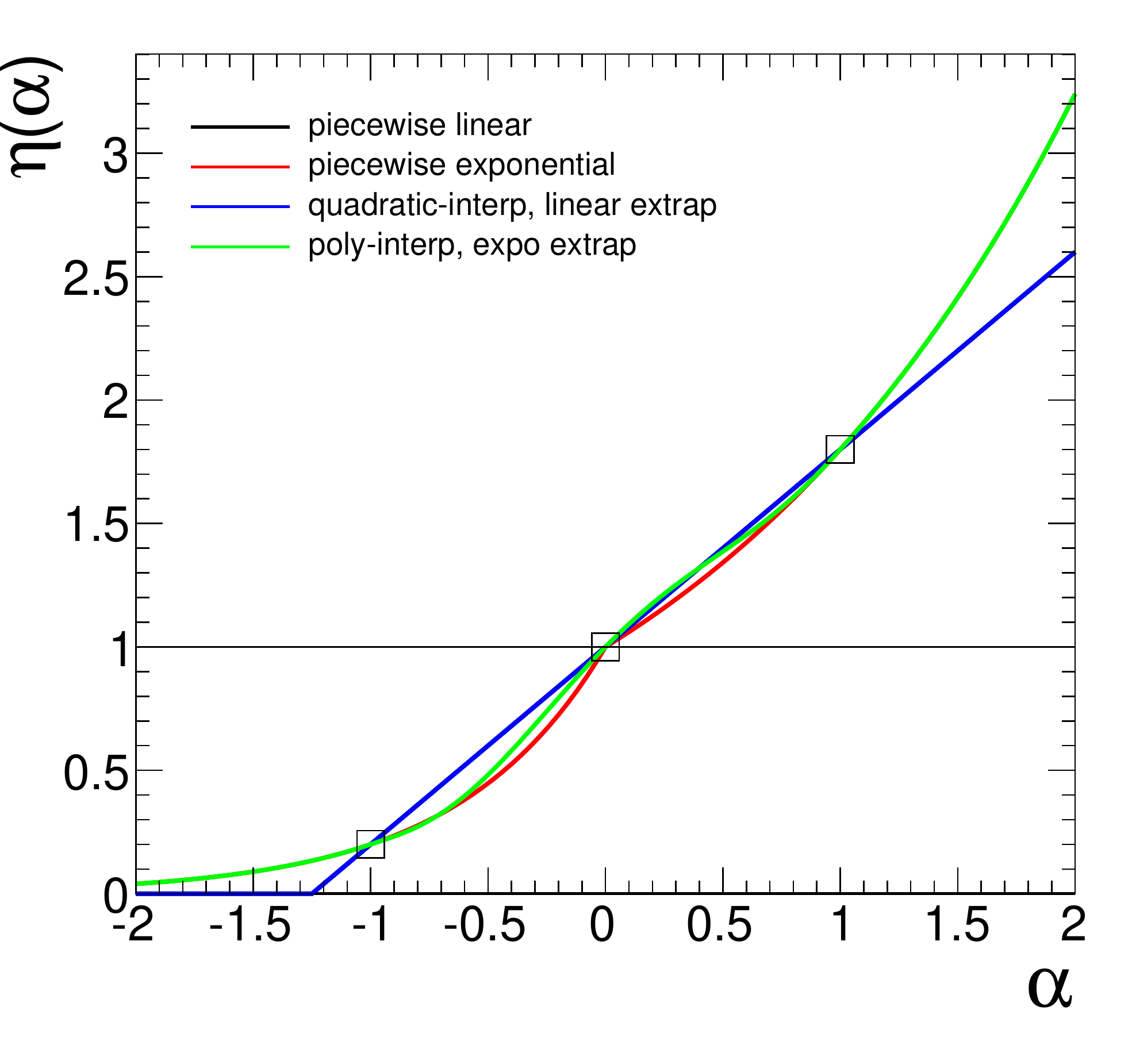}}
\subfigure[][]{\includegraphics[width=.4\textwidth]{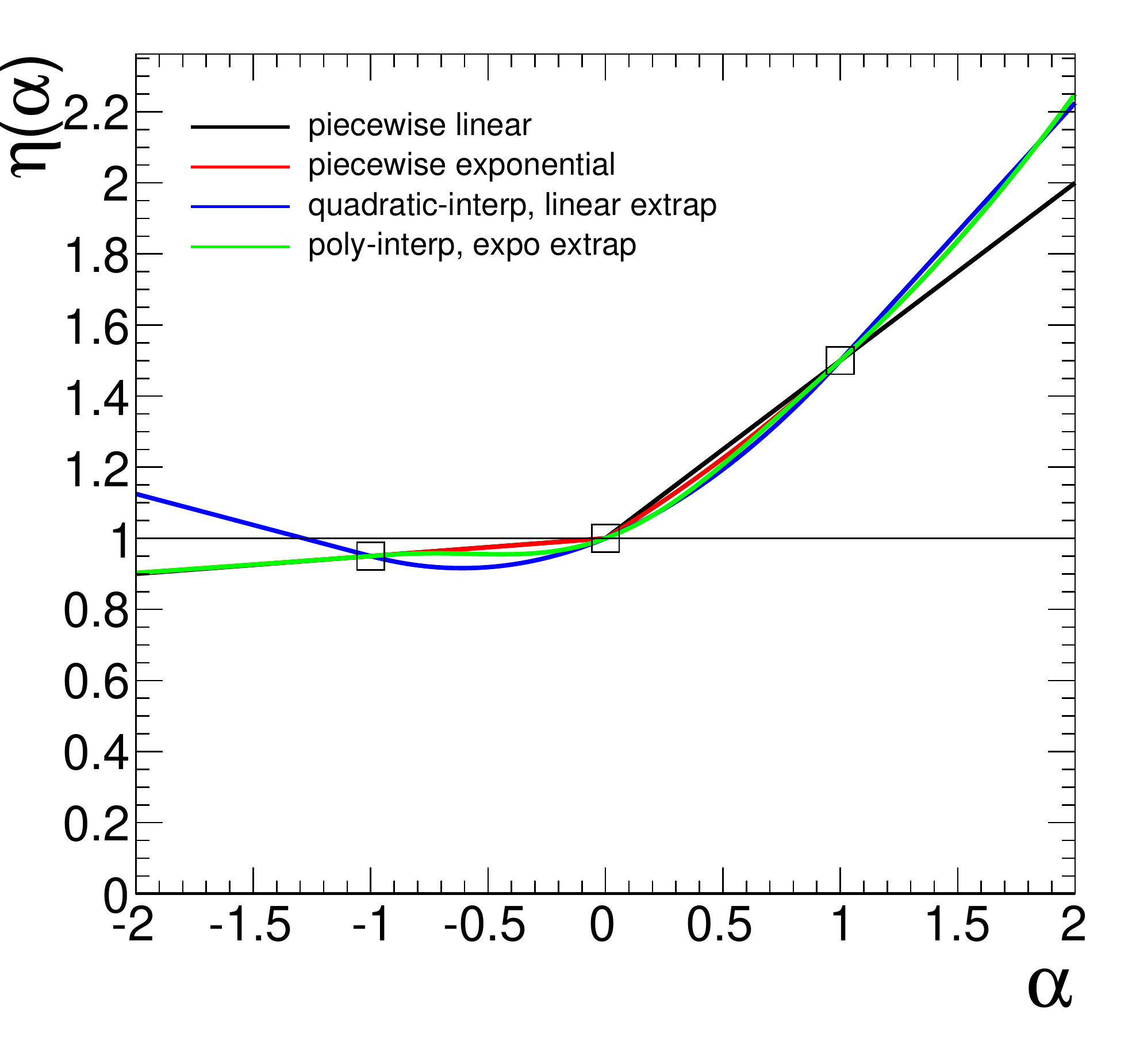}}
\caption{Comparison of the three interpolation options for different $\eta^\pm$.  (a) $\eta^-=0.8$, $\eta^+=1.2$, (b) $\eta^-=1.1$, $\eta^+=1.5$, (c) $\eta^-=0.2$, $\eta^+=1.8$, and (d) $\eta^-=0.95$, $\eta^+=1.5$}
\label{fig:interp1d}
\end{center}
\end{figure}

For each sample, one can interpolate and extrapolate from the nominal prediction $\eta_s^0=1$ and the variations $\eta^\pm_{ps}$ to produce a parametrized $\eta_s(\vec{\alpha})$.   Similarly, one can interpolate and extrapolate from the nominal shape $\sigma_{sb}^0$ and the variations $\sigma^\pm_{psb}$ to produce a parametrized $\sigma_{sb}(\vec{\alpha})$.  We choose to parametrize $\alpha_p$ such that $\alpha_p=0$ is the nominal value of this parameter, $\alpha_p=\pm 1$ are the ``$\pm 1\sigma$ variations''.  Needless to say, there is a significant amount of ambiguity in these interpolation and extrapolation procedures and they must be handled with care.  Bellow are some of the interpolation strategies supported by \texttt{HistFactory}.  These are all  'vertical' style interpolation treated independently per-bin. Four interpolation strategies are described below and can be compared in Fig~\ref{fig:interp1d}.  The interested reader is invited to look at alternative 'horizontal' interpolation strategies,  such as the one developed by Alex Read in Ref.~\cite{Read:1999kh} (the \roofit\ implementation is called \texttt{RooIntegralMorph}) and Max Baak's \texttt{RooMomentMorph}.  These horizontal interpolation strategies are better suited for features moving, such as the location of an invariant mass bump changing with the hypothesized mass of a new particle..
{\flushleft \bf Piecewise Linear (InterpCode=0)}

The piecewise-linear interpolation strategy is defined as
\begin{equation}
\eta_s ({\boldsymbol \alpha}) = 1+\sum_{p\in\textrm{Syst}} I_{\rm lin.}(\alpha_p; 1, \eta_{sp}^+, \; \eta_{sp}^- ) 
\end{equation}
and for shape interpolation it is
\begin{equation}
\sigma_{sb} ({\boldsymbol \alpha}) = \sigma_{sb}^0 + \sum_{p\in\textrm{Syst}} I_{\rm lin.}(\alpha_p;  \sigma_{sb}^0, \sigma_{psb}^+ ,\;
\sigma_{psb}^- )  
\end{equation}
with
\begin{equation}
 I_{\rm lin.}(\alpha;  I^0, I^+,I^- ) =
 \begin{cases}
     \alpha (I^+  -  I^0) &  \text{$\alpha\ge 0$} \\
     \alpha (I^0  -I^-)  &  \text{$\alpha<0$ }
 \end{cases}
\end{equation}

\textsc{Pros:} This approach is the most straightforward of the interpolation strategies.

\textsc{Cons:} It has two negative features.  First, there is a kink (discontinuous first derivative) at $\alpha=0$ (see Fig~\ref{fig:interp1d}(b-d)), which can cause some difficulties for numerical minimization packages such as \texttt{Minuit}.  Second, the interpolation factor can extrapolate to negative values.  For instance, if $\eta^-=0.5$ then  we have $\eta(\alpha)<0$  when $\alpha<-2$  (see Fig~\ref{fig:interp1d}(c)).  

Note that one could have considered the simultaneous variation of $\alpha_{p}$ and $\alpha_{p'}$ in a multiplicative way (see for example, Fig~\ref{fig:interp2d}).  The multiplicative accumulation is not an option currently.

Note that this is the default convention for $\sigma_{sb}(\vec{\alpha})$ (ie. \HS\ ).

{\flushleft\bf Piecewise Exponential (InterpCode=1)}

The piecewise exponential interpolation strategy is defined as
\begin{equation}
\eta_s ({\boldsymbol \alpha}) = \prod_{p\in\textrm{Syst}} I_{\rm exp.}(\alpha_p; 1,\eta_{sp}^+, \; \eta_{sp}^- ) 
\end{equation}
and for shape interpolation it is
\begin{equation}
\sigma_{sb} ({\boldsymbol \alpha}) = \sigma_{sb}^0 \prod_{p\in\textrm{Syst}} I_{\rm exp.}(\alpha_p;  \sigma_{sb}^0, \sigma_{psb}^+ ,\;
\sigma_{psb}^- )  
\end{equation}
with
\begin{equation}
 I_{\rm exp.}(\alpha;  I^0, I^+,I^- ) =
 \begin{cases}
     (I^+/I_0)^ \alpha  &  \text{$\alpha\ge 0$} \\
     (I^-/I_0)^{-\alpha}  &  \text{$\alpha<0$ }
 \end{cases}
\end{equation}

\textsc{Pros:} This approach ensures that $\eta(\alpha)\ge 0$ (see Fig~\ref{fig:interp1d}(c)) and for small response to the uncertainties it has the same linear behavior near $\alpha\sim 0$ as the piecewise linear interpolation (see Fig~\ref{fig:interp1d}(a)).

\textsc{Cons:} It has two negative features.  First, there is a kink (discontinuous first derivative) at $\alpha=0$, which can cause some difficulties for numerical minimization packages such as \texttt{Minuit}.  Second, for large uncertainties it develops a different linear behavior compared to the piecewise linear interpolation.  In particular, even if the systematic has a symmetric response (ie. $\eta^+-1 = 1-\eta^-$) the interpolated response will develop a kink for large response to the uncertainties  (see Fig~\ref{fig:interp1d}(c)).

Note that the one could have considered the simultaneous variation of $\alpha_{p}$ and $\alpha_{p'}$ in an additive way, but this is not an option currently.

Note, that when paired with a Gaussian constraint on $\alpha$ this is equivalent to  linear interpolation and a log-normal constraint in $\ln(\alpha)$.  This is the default strategy for normalization uncertainties $\eta_{s}(\vec{\alpha})$ (ie. \OS\ ) and is the standard convention for normalization uncertainties in the LHC Higgs Combination Group.  In the future, the default may change to the Polynomial Interpolation and Exponential Extrapolation described below.

{\flushleft \bf Polynomial Interpolation and Exponential Extrapolation (InterpCode=4)}

The strategy of this interpolation option is to use the piecewise exponential extrapolation as above with a polynomial interpolation that matches $\eta(\alpha=\pm\alpha_0)$, $d\eta/d\alpha |_{\alpha=\pm\alpha_0}$, and $d^2\eta/d\alpha^2 |_{\alpha=\pm\alpha_0}$ and the boundary $\pm\alpha_0$ is defined by the user (with default $\alpha_0=1$).  
\begin{equation}
\eta_s ({\boldsymbol \alpha}) = \prod_{p\in\textrm{Syst}} I_{\rm poly|exp.}(\alpha_p; 1,\eta_{sp}^+, \; \eta_{sp}^- , \alpha_0) 
\end{equation}
with
\begin{equation}
 I_{\rm poly|exp.}(\alpha;  I^0, I^+,I^- , \alpha_0) =
 \begin{cases}
      (I^+/I_0)^ \alpha  &  \text{$\alpha\ge \alpha_0$} \\
     1+\sum_{i=1}^6 a_i \alpha^i  &  \text{$|\alpha|< \alpha_0$} \\
      (I^-/I_0)^{-\alpha}  &  \text{$\alpha\le-\alpha_0$ }
 \end{cases}
\end{equation}
and the $a_i$ are fixed by the boundary conditions described above.

\textsc{Pros:} This approach avoids the kink (discontinuous first and second derivatives) at $\alpha=0$ (see Fig~\ref{fig:interp1d}(b-d)), which can cause some difficulties for numerical minimization packages such as \texttt{Minuit}.  This approach ensures that $\eta(\alpha)\ge 0$ (see Fig~\ref{fig:interp1d}(c)).

\textbf{Note}: This option is not available in ROOT 5.32.00, but is available for normalization uncertainties (OverallSys) in the subsequent patch releases.  In future releases, this may become the default.

\subsubsection{Consistent Bayesian and Frequentist modeling}
\label{S:ConstraintExamples}

The variational estimates $\eta^\pm$ and $\sigma^\pm$ typically correspond to so called ``$\pm 1\sigma$ variations'' in the source of the uncertainty.  Here we are focusing on the source of the uncertainty, not its affect on rates and shapes.  For instance, we might say that the jet energy scale has a 10\% uncertainty.~\footnote{Without loss of generality, we choose to parametrize $\alpha_p$ such that $\alpha_p=0$ is the nominal value of this parameter, $\alpha_p=\pm 1$ are the ``$\pm 1\sigma$ variations''.}  This is common jargon, but what does it mean?  The most common interpretation of this statement is that the uncertain parameter $\alpha_p$ (eg. the jet energy scale) has a Gaussian distribution.  However, this way of thinking is manifestly Bayesian.  If the parameter was estimated from an auxiliary measurement, then it is the PDF for that measurement that we wish to include into our probability model.  In the frequentist way of thinking, the jet energy scale has an unknown true value and upon repeating the experiment many times the auxiliary measurements estimating the jet energy scale would fluctuate randomly about this true value.  To aid in this subtle distinction, we use greek letters for the parameters (eg. $\alpha_p$) and roman letters for the auxiliary measurements $a_p$.  Furthermore, we interpret the ``$\pm 1\sigma$'' variation in the frequentist sense, which leads to the constraint term $f_p(a_p | \alpha_p)$.  Then, we can pair the resulting likelihood with some prior on $\alpha_p$ to form a Bayesian posterior if we wish according to Eq.~\ref{eq:urprior}.

It is often advocated that a ``log-normal'' or ``gamma'' distribution for $\alpha_p$ is more appropriate than a gaussian constraint~\cite{CousinsLogNormal}.  This is particularly clear in the case of bounded parameters and large uncertainties.    Here we must take some care to build a probability model that can maintain a consistent interpretation in Bayesian a frequentist settings.  Table~\ref{tab:constraints} summarizes a few consistent treatments of the frequentist pdf, the likelihood function, a prior, and the resulting posterior.

\begin{table}[*htb]
\center
\begin{tabular}{llll}
PDF & Likelihood $\propto$ & Prior $\pi_0$ & Posterior $\pi$ \\ \hline
$G(a_p | \alpha_p, \sigma_p)$ & $G(\alpha_p | a_p, \sigma_p)$ & $\pi_0(\alpha_p)\propto$  const & $G(\alpha_p | a_p, \sigma_p)$ \\
$\Pois(n_p | \tau_p \beta_p)$ & $\PGamma(\beta_p | A=\tau_p; B=1+n_p)$ & $\pi_0(\beta_p) \propto$  const & $\PGamma(\beta_p | A=\tau_p; B=1+n_p)$ \\
$\LN(n_p | \beta_p, \sigma_p)$ & $ \beta_p  \cdot \LN(\beta_p | n_p, \sigma_p)$ & $\pi_0(\beta_p) \propto $ const & $\LN(\beta_p | n_p, \sigma_p)$ \\
$\LN(n_p | \beta_p, \sigma_p)$ & $\beta_p  \cdot\LN(\beta_p | n_p, \sigma_p)$ & $\pi_0(\beta_p) \propto 1/\beta_p $  & $\LN(\beta_p | n_p, \sigma_p)$\\
\end{tabular}
\caption{Table relating consistent treatments of PDF, likelihood, prior, and posterior for nuisance parameter constraint terms.}
\label{tab:constraints}
\end{table}

Finally, it is worth mentioning that the uncertainty on some parameters is not the result of an auxiliary measurement -- so the constraint term idealization, it is not just a convenience, but a real  conceptual leap.  This is particularly true for theoretical uncertainties from higher-order corrections or renormalizaiton and factorization scale dependence.  In these cases a formal frequentist analysis would not include a constraint term for these parameters, and the result would simply depend on their assumed values.  As this is not the norm, we can think of reading Table~\ref{tab:constraints} from right-to-left with a subjective Bayesian prior $\pi(\alpha)$ being interpreted as coming from a fictional auxiliary measurement.

\subsubsubsection{Gaussian Constraint}

The Gaussian constraint for $\alpha_p$ corresponds to the familiar situation.  It is a good approximation of the auxiliary measurement when the likelihood function for $\alpha_p$ from that auxiliary measurement has a Gaussian shape.  More formally, it is valid when the maximum likelihood estimate of $\alpha_p$ (eg. the best fit value of $\alpha_p$) has a Gaussian distribution.  Here we can identify the maximum likelihood estimate of $\alpha_p$ with the global observable $a_p$, remembering that it is a number that is extracted from the data and thus its distribution has a frequentist interpretation.  
\begin{equation}
G(a_p | \alpha_p, \sigma_p) = \frac{1}{\sqrt{2\pi \sigma_p^2}} \exp \left[ -\frac{(a_p - \alpha_p)^2}{2\sigma_p^2} \right]
\end{equation}
with $\sigma_p=1$ by default.
Note that the PDF of $a_p$ and the likelihood for $\alpha_p$ are positive for all values. 

\subsubsubsection{Poisson (``Gamma'') constraint}

When the auxiliary measurement is actually based on counting events in a control region (eg. a Poisson process), a more accurate to describe the auxiliary measurement with a Poisson distribution.  It has been shown that the truncated Gaussian constraint can lead to undercoverage (overly optimistic) results, which makes this issue practically relevant~\cite{Cousins:2008zz}.  Table~\ref{tab:constraints} shows that a Poisson PDF together with a uniform prior leads to a gamma posterior, thus this type of constraint is often called a ``gamma'' constraint.  This is a bit unfortunate since the gamma distribution is manifestly Bayesian and with a different choice of prior, one might not arrive at a gamma posterior.  When dealing with the Poisson constraint, it is no longer convenient to work with  our conventional scaling for $\alpha_p$ which can be negative.  Instead, it is more natural to think of the number of events measured in the auxiliary measurement $n_p$ and the mean of the Poisson parameter.  This information is not usually available, instead one usually has some notion of the relative uncertainty in the parameter $\sigma_p^{\rm rel}$ (eg. a the jet energy scale is known to 10\%).  In order to give some uniformity to the different uncertainties of this type and think of relative uncertainty, the nominal rate is factored out into a constant $\tau_p$ and the mean of the Poisson is given by $\tau_p \alpha_p$.  
\begin{equation}
\Pois(n_p | \tau_p \alpha_p) =\frac{ (\tau_p \alpha_p)^{n_p} \; e^{-\tau_p \alpha_p} } {n_p!}
\end{equation}
Here we can use the fact that Var$[n_p]=\sqrt{\tau_p\alpha_p}$ and reverse engineer the nominal auxiliary measurement 
\begin{equation}
n_p^0 =  \tau_p = (1/\sigma_{p}^{\rm rel})^2\; .
\end{equation}
where the superscript $0$ is to remind us that $n_p$ will fluctuate in repeated experiments but $n_p^0$ is the value of our measured estimate of the parameter.

One important thing to keep in mind is that there is only one constraint term per nuisance parameter, so there must be only one $\sigma_p^{rel}$ per nuisance parameter.  This $\sigma_p^{rel}$ is related to the fundamental uncertainty in the source and we cannot infer this from the various response terms $\eta_{ps}^\pm$ or $\sigma_{pub}^\pm$. 

Another technical difficulty is that the Poisson distribution is discrete. So if one were to say the relative uncertainty was 30\%, then we would find $n_p^0=11.11...$, which is not an integer.  Rounding $n_p$ to the nearest integer while maintaining $\tau_p= (1/\sigma_{p}^{\rm rel})^2$ will bias the maximum likelihood estimate of $\alpha_p$ away from 1.  To avoid this, one can use the gamma distribution, which generalizes more continuously with 
\begin{equation}
\PGamma(\alpha_p | A=\tau_p, B=n_p-1) = A (A \alpha_p)^{B} e^{-A \alpha_p} / \Gamma(B)\;.
\end{equation}
This approach works fine for likelihood fits, Bayesian calculations, and frequentist techniques based on asymptotic approximations, but it does not offer a consistent treatment of the pdf for the global observable $n_p$ that is needed for techniques based on Monte Carlo sampling. 


\subsubsubsection{Log-normal constraint}

\begin{figure}[htbp]
\begin{center}
\includegraphics[width=.6\textwidth]{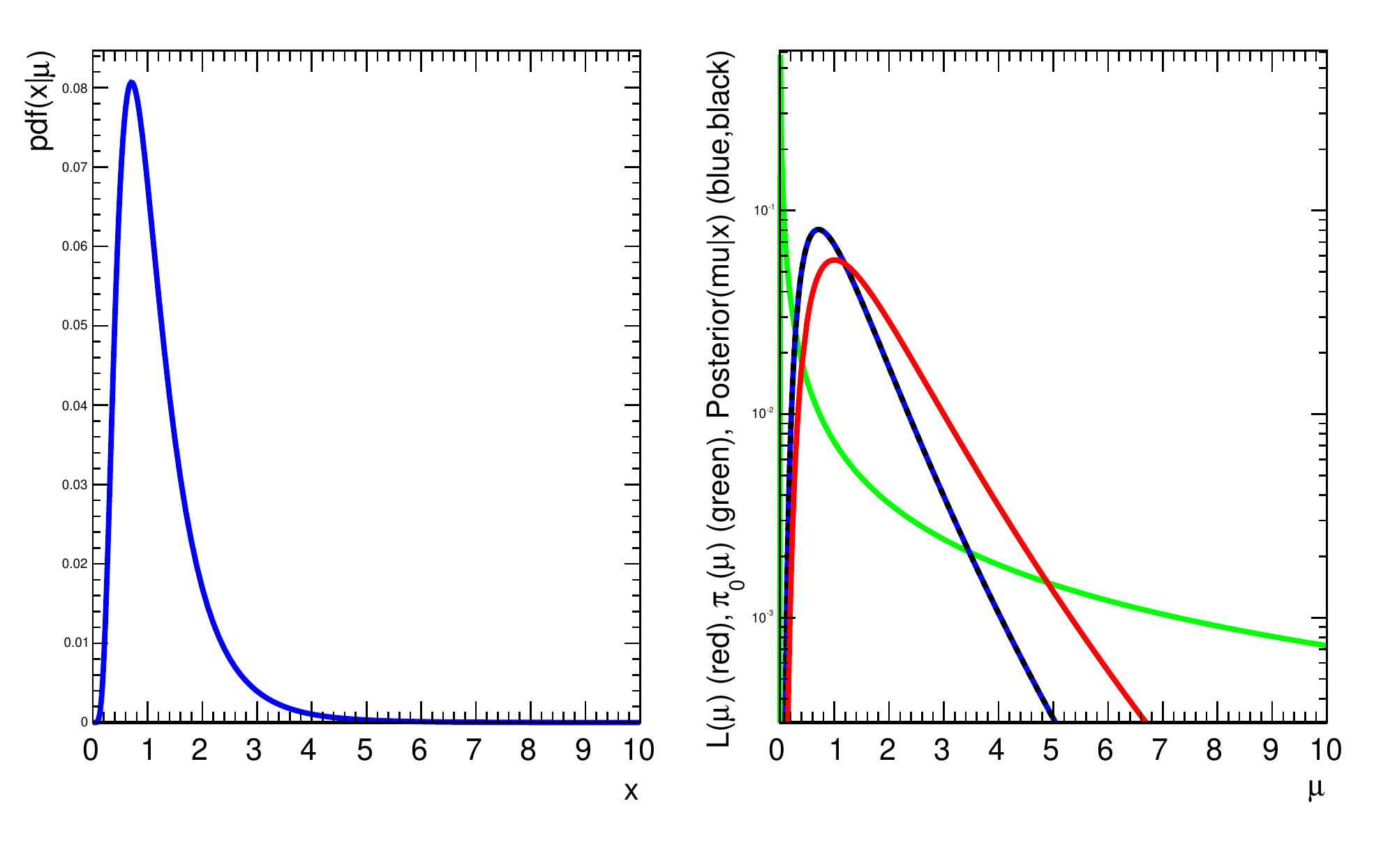}
\caption{The lognormal constraint term: (left) the pdf for the global observable $a_p$ and (right) the likelihood function, the posterior based on a flat prior on $\alpha_p$, and the posterior based on a $1/\alpha_p$ prior.}
\label{fig:lognormal}
\end{center}
\end{figure}

From Eadie et al., ``The log-normal distribution represents a random variable whose logarithm follows a normal distribution. It provides a model for the error of a process involving many small multiplicative errors (from the Central Limit Theorem). It is also appropriate when the value of an observed variable is a random proportion of the previous observation.''~\cite{Eadie:qy,CousinsLogNormal}.  This logic of multiplicative errors applies to the the measured value, not the parameter.  Thus, it is natural to say that there is some auxiliary measurement (global observable) with a log-normal distribution.  As in the gamma/Poisson case above, let us again say that the global observable is $n_p$ with a nominal value
\begin{equation}
n_p^0 =  \tau_p = (1/\sigma_{p}^{\rm rel})^2\; .
\end{equation}
Then the conventional choice for the corresponding log-normal distribution is
\begin{equation}
\LN(n_p | \alpha_p, \kappa_p) = \frac{1}{\sqrt{2\pi}\ln \kappa}\frac{1}{n_p} \exp \left[ -\frac{\ln(n_p/ \alpha_p)^2}{2(\ln \kappa_p)^2} \right]
\end{equation}
while the likelihood function is (blue curve in Fig.~\ref{fig:lognormal}(a)).
\begin{equation}
L( \alpha_p) = \frac{1}{\sqrt{2\pi}\ln \kappa}\frac{1}{n_p} \exp \left[ -\frac{\ln(n_p/ \alpha_p)^2}{2(\ln \kappa_p)^2} \right];.
\end{equation}
To get to the posterior for $\alpha_p$ given $n_p$ we need an ur-prior $\eta(\alpha_p$)
\begin{equation}
\pi( \alpha_p) \propto \eta(\alpha_p)  \; \frac{1}{\sqrt{2\pi}\ln \kappa}\frac{1}{n_p} \exp \left[ -\frac{\ln(n_p/ \alpha_p)^2}{2(\ln \kappa_p)^2} \right]
\end{equation}
If $\eta(\alpha_p)$ is uniform, then the posterior looks like the red curve in Fig.~\ref{fig:lognormal}(b).  However, when paired with an ``ur-prior'' $\eta(\alpha_p) \propto 1/\alpha_p$ (green curve in Fig.~\ref{fig:lognormal}(b)), this results in a posterior distribution that is also of a log-normal form for $\alpha_p$ (blue curve in Fig.~\ref{fig:lognormal}(b)).

\subsubsection{Incorporating Monte Carlo statistical uncertainty on the histogram templates}

The histogram based approach described above are based Monte Carlo simulations of full detector simulation.  These simulations are very computationally intensive and often the histograms are sparsely populated.  In this case the histograms are not good descriptions of the underlying distribution, but are estimates of that distribution with some statistical uncertainty.  Barlow and Beeston outlined a treatment of this situation in which each bin of each sample is given a nuisance parameter for the true rate, which is then fit using both the data measurement and the Monte Carlo estimate~\cite{Barlow:1993dm}.  This approach would lead to several hundred nuisance parameters in the current analysis.  Instead, the \texttt{HistFactory} employs a lighter weight version in which there is only one nuisance parameter per bin associated with the total Monte Carlo estimate  and the total statistical uncertainty in that bin.  If we focus on an individual bin with index $b$ the contribution to the full statistical model is the factor
\begin{equation}
\Pois(n_b | \nu_b(\vec\alpha) + \gamma_b \nu_b^{\rm MC}(\vec\alpha)) \,  \Pois(m_b | \gamma_b \tau_b) \;,
\end{equation}
where $n_b$ is the number of events observed in the bin, $\nu_b(\vec\alpha)$ is the number of events expected in the bin where Monte Carlo statistical uncertainties need not be included (either because the estimate is data driven or because the Monte Carlo sample is sufficiently large), $\nu_b^{\rm MC}(\vec\alpha)$ is the number of events estimated using Monte Carlo techniques where the statistical uncertainty needs to be taken into account.  Both expectations include the dependence on the parameters $\vec\alpha$.  The factor $\gamma_b$ is the nuisance parameter reflecting that the true rate may differ from the Monte Carlo estimate $\nu_b^{\rm MC}(\vec\alpha) $ by some amount.  If the total statistical uncertainty is $\delta_b$, then the relative statistical uncertainty is given by $\nu_b^{\rm MC}/\delta_b$.  This corresponds to a total Monte Carlo sample in that bin of size $m_b =  (\delta_b/\nu_b^{\rm MC})^2$.  Treating the Monte Carlo estimate as an auxiliary measurement, we arrive at a Poisson constraint term $ \Pois(m_b | \gamma_b \tau_b)$, where $m_b$ would fluctuate about $\gamma_b \tau_b$ if we generated a new Monte Carlo sample.  Since we have scaled $\gamma$ to be a factor about 1, then we also have $\tau_b=(\nu_b^{\rm MC}/\delta_b)^2$; however, $\tau_b$ is treated as a fixed constant and does not fluctuate when generating ensembles of pseudo-experiments.

It is worth noting that the conditional maximum likelihood estimate $\hat{\hat{\gamma_b}}(\vec\alpha)$ can be solved analytically with a simple quadratic expression.
\begin{equation}
\hat{\hat{\gamma_b}}(\vec\alpha) = \frac{-B + \sqrt{B^2 - 4 AC}}{2A} \;,
\end{equation}
with
\begin{equation}
A =  \nu_b^{\rm MC}(\vec\alpha)^2 + \tau_b  \nu_b^{\rm MC}(\vec\alpha)
\end{equation}
\begin{equation}
B= \nu_b(\vec\alpha) \tau +  \nu_b(\vec\alpha) \nu_b^{\rm MC}(\vec\alpha) - n_b  \nu_b^{\rm MC}(\vec\alpha) - m_b  \nu_b^{\rm MC}(\vec\alpha)
\end{equation}
\begin{equation}
C= m_b \nu_b(\vec\alpha) \;.
\end{equation}

In a Bayesian technique with a flat prior on $\gamma_b$, the posterior distribution is a gamma distribution.  Similarly, the distribution of $\hat\gamma_b$ will take on a skew distribution with an envelope similar to the gamma distribution, but with features reflecting the discrete values of $m_b$.  Because the maximum likelihood estimate of $\gamma_b$ will also depend on $n_b$ and $\hat{\vec\alpha}$, the  features from the discrete values of $m_b$ will be smeared.  This effect will be more noticeable for large statistical uncertainties where $\tau_b$ is small and the distribution  of $\hat\gamma_b$  will have several small peaks.  For smaller statistical uncertainties where $\tau_b$ is large the distribution of $\hat\gamma_b$ will be approximately Gaussian.

\subsection{Data-Driven Narrative}

The strength of the simulation narrative lies in its direct logical link from the underlying theory to the modeling of the experimental observations.  The weakness of the simulation narrative derives from the weaknesses in the simulation itself.  Data-driven approaches are more motivated when they address specific deficiencies in the simulation.  Before moving to a more abstract or general discussion of the data-driven narrative, let us first consider a few examples.

The first example we have already considered in Sec.~\ref{S:AuxMeas} in the context of the ``on/off'' problem.  There we introduced an auxiliary measurement that counted $n_{CR}$ events in a  control region to estimate the background $\nu_B$ in the signal region.  In order to do this we needed to understand the ratio of the number of events from the background process in the control and signal regions, $\tau$.  This ratio $\tau$ either comes from some reasonable assumption or simulation.  For example, if one wanted to estimate the background due to jets faking muons $j\to\mu$ for a search selecting $\mu^+\mu^-$ , then one might use a sample of $\mu^\pm\mu^\pm$ events as a control region.  Here the motivation for using a data-driven approach  is that modeling the processes that lead to $j\to\mu$ rely heavily on the tails of fragmentation functions and detector response, which one might reasonably have some skepticism.  If one assumes that control region is expected to have negligible signal in it, that backgrounds that produce $\mu^+\mu^-$ other than the jets faking muons, and that the rate for $j\to\mu^-$ is the same\footnote{Given that the LHC collides $pp$ and not $p\bar{p}$, there is clearly a reason to worry if this assumption is valid.} as the rate for $j\to\mu^+$, then one can assume $\tau=1$.  Thus, this background estimate is as trustworthy as the assumptions that went into it.    In practice, several of these assumptions may be violated.  Another approach is to use simulation of these background processes to estimate the ratio $\tau$; a hybrid of the data-driven and simulation narratives.

\begin{figure}[htbp]
\begin{center}
\includegraphics[width=.4\textwidth]{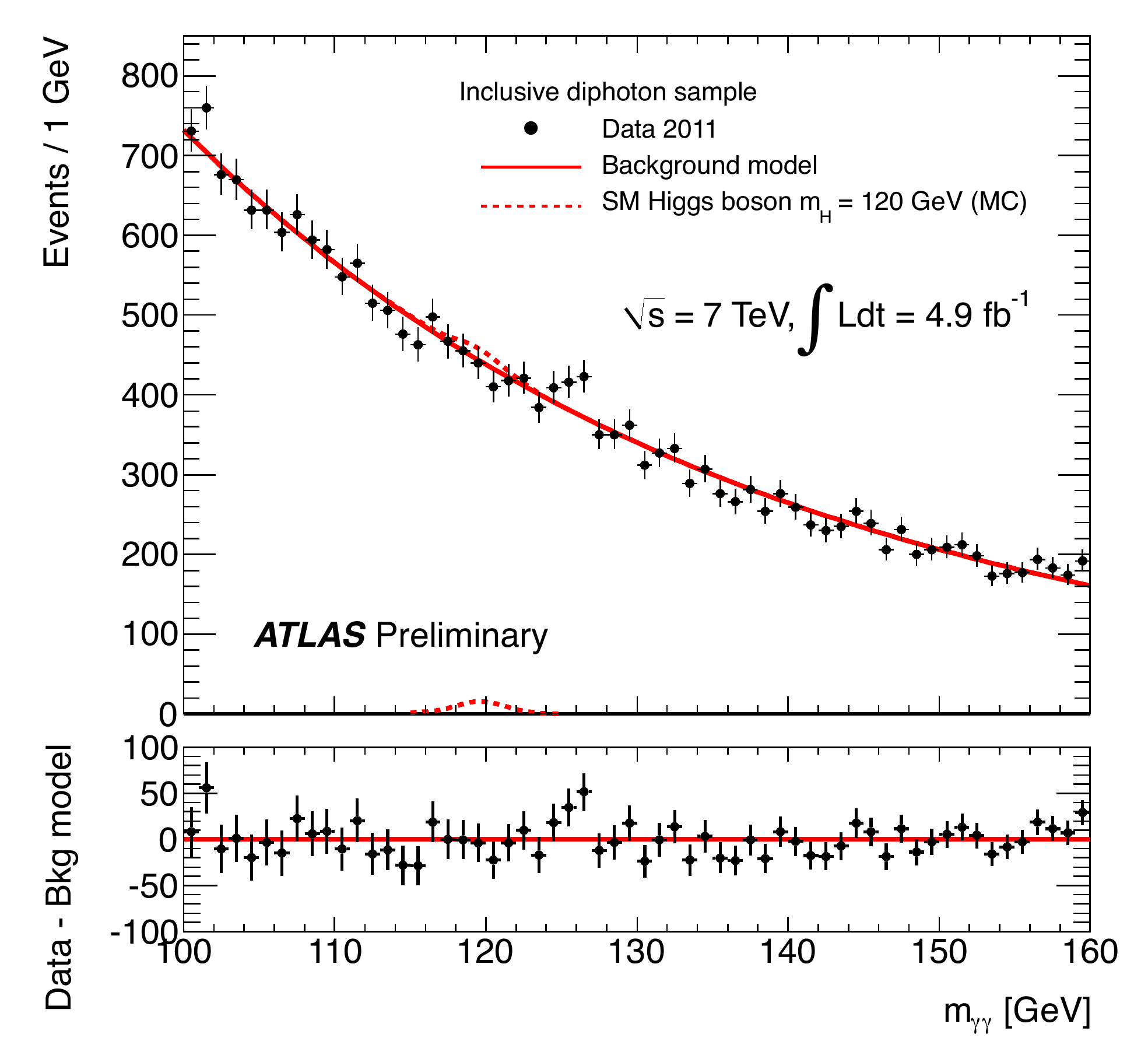}
\includegraphics[width=.5\textwidth]{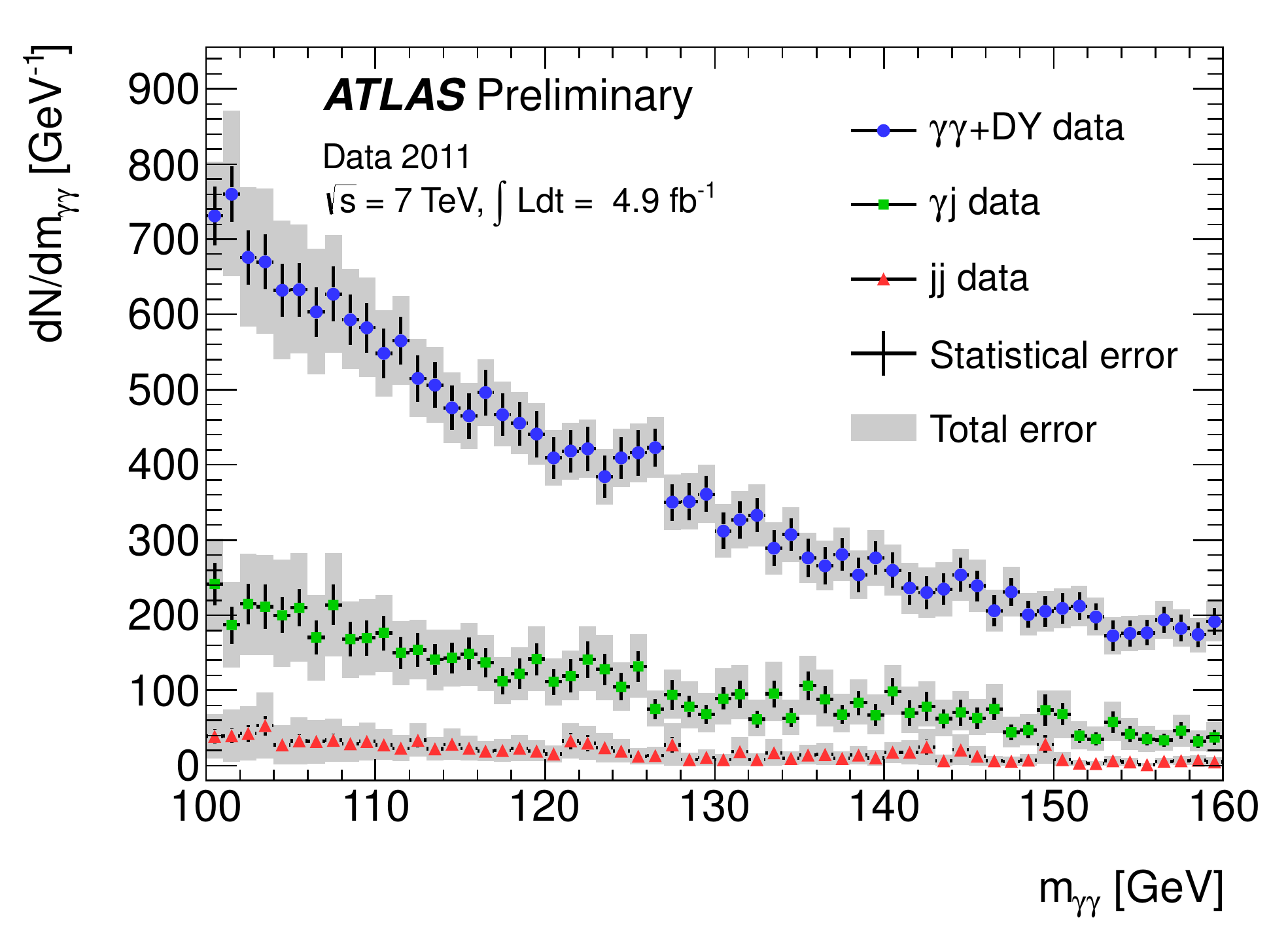}
\caption{Distribution of diphoton invariant mass distributions in the ATLAS $H\to\gamma\gamma$ search.  The left plot shows a fit of a an effective model to the data and the right plot shows an estimate of the $\gamma\gamma$, $\gamma$+jet, and diet contributions.}
\label{fig:H2photons}
\end{center}
\end{figure}

Let us now consider the search for $H\to\gamma\gamma$ shown in Fig~\ref{fig:H2photons}~\cite{ATLAS-CONF-2011-161,ATLAS:2012ad}.  The right plot of  Fig~\ref{fig:H2photons} shows the composition of the backgrounds in this search, including the continuum production of $pp\to\gamma\gamma$, the $\gamma$+jets process with a jet faking a photon $j\to\gamma$, and the multi jet process with two jets faking photons.  The continuum production of $\gamma\gamma$ has a  theoretical uncertainty that is much larger than the statistical fluctuations one would expect in the data.  Similarly, the rate of jets faking photons is sensitive to fragmentation and the detector simulation.  These uncertainties are large compared to the statistical fluctuations in the data itself.  Thus we can use the distribution in  Fig~\ref{fig:H2photons} to measure the total background rate.  Of course, the signal would also be in this distribution, so one either needs to apply a mass window around the signal and consider the region outside of the window as a sideband control sample or model the signal and background contributions to the distribution.  In the case of the   $H\to\gamma\gamma$ shown in Fig~\ref{fig:H2photons}~\cite{ATLAS-CONF-2011-161,ATLAS:2012ad} the modeling of the distribution signal and background distributions is not based on histograms from simulation, but instead a continuous function is used as an effective model.  I will discuss this effective modeling narrative below, but point out that here this is another example of a hybrid narrative.

\begin{figure}[htbp]
\begin{center}
\includegraphics[width=.47\textwidth]{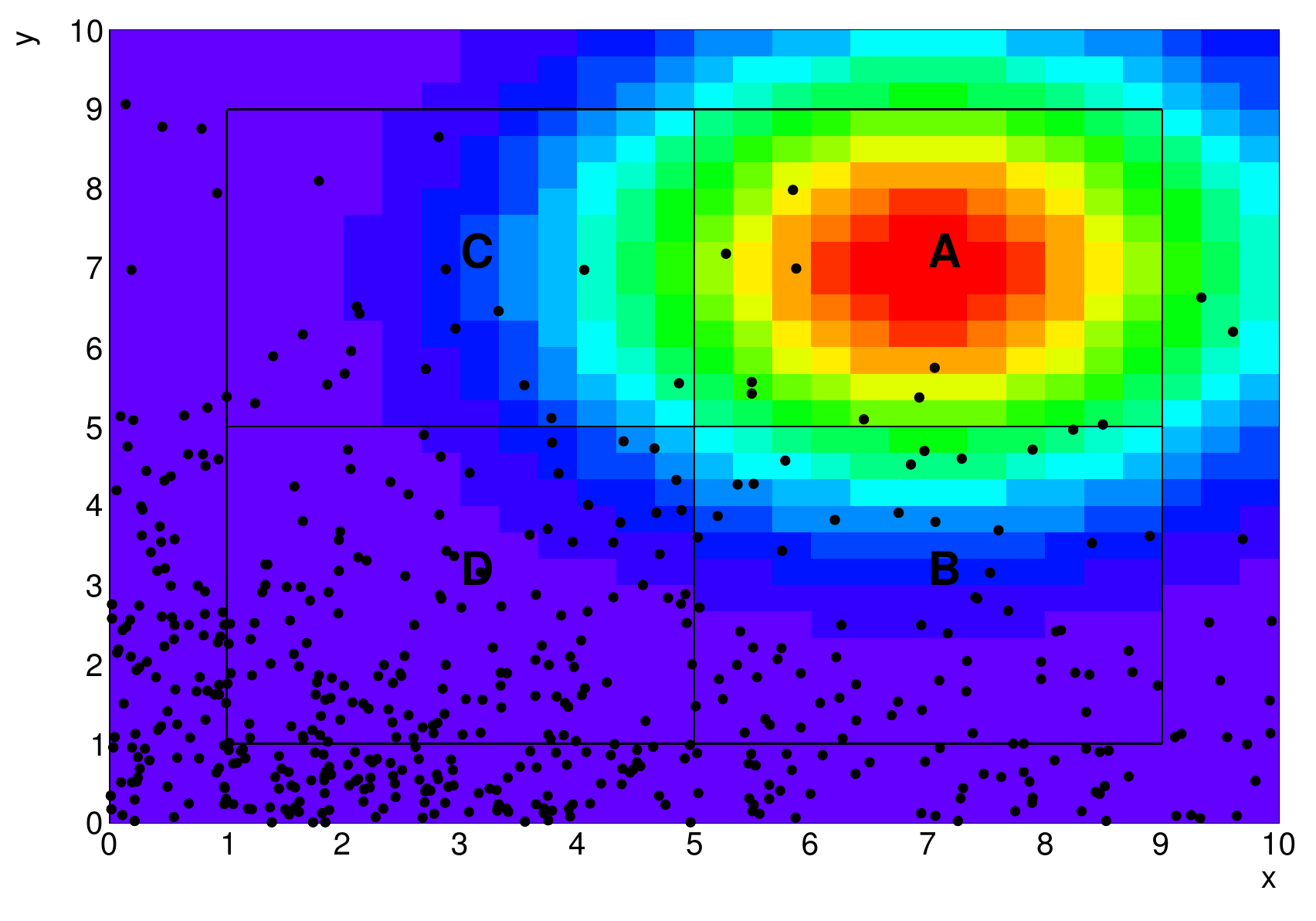}%
\includegraphics[width=.5\textwidth]{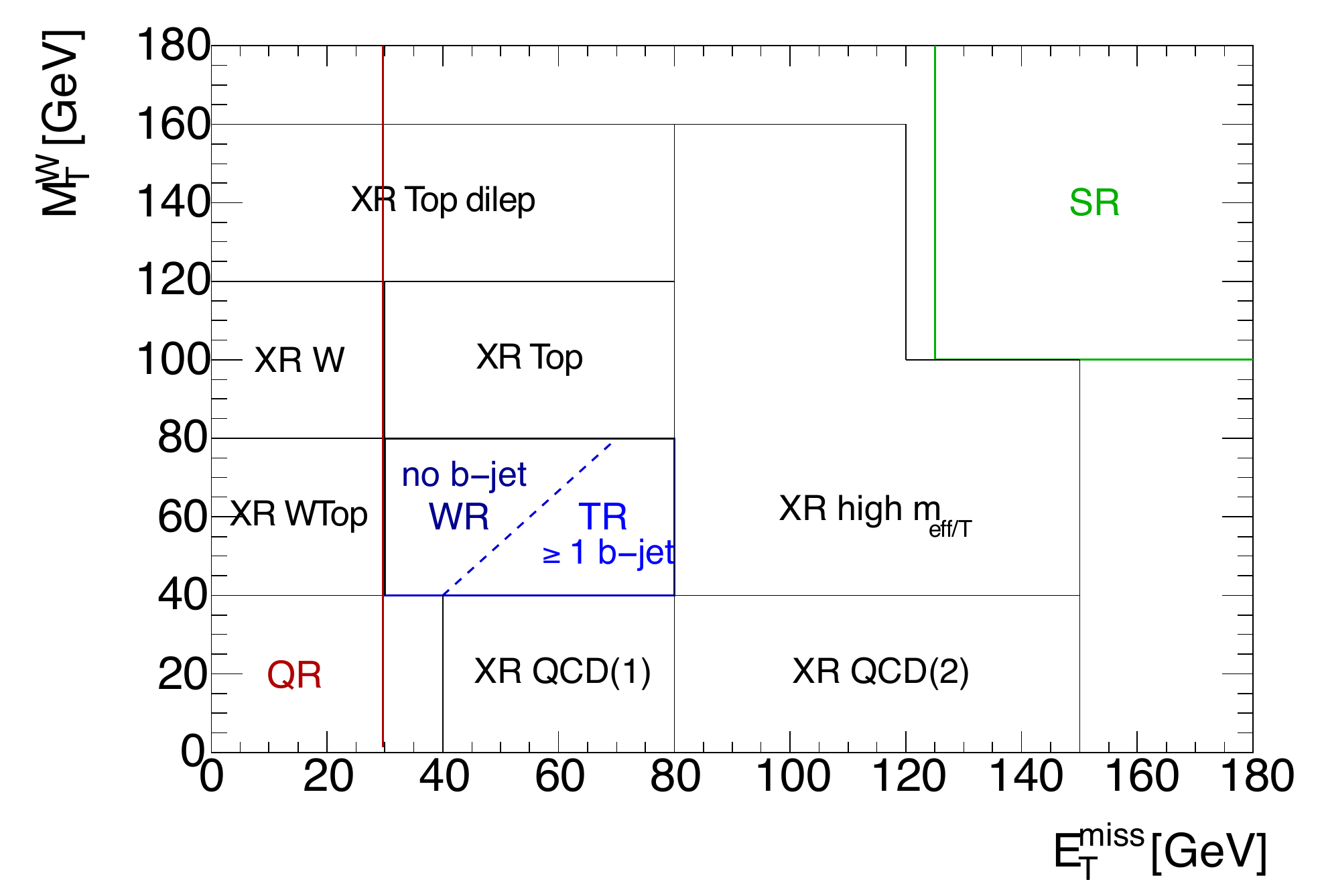}
\caption{An example of ABCD (from Alex Read) in the $x-y$ plane of two observables $x$ and $y$ (left).  A more complex example with several regions in the $M_T^W-E_T^{\rm miss}$ plane~\cite{ATLAS:2011ad}.}
\label{fig:ABCD}
\end{center}
\end{figure}

The final example to consider is an extension of the `on/off' model, often referred to as the `ABCD' method.  Let us start with the `on/off' model:
\mbox{ $\Pois(n_{\rm SR} | \nu_S + \nu_B)\cdot \Pois(n_{\rm CR}|\tau\nu_B)$}.  As mentioned above, this requires that one estimate $\tau$ either from simulation or through some assumptions. The ABCD method aims to estimate introduce two new control regions that can be used to measure  $\tau$.  To see this, let us imagine that the signal and control regions correspond to requiring some continuous variable $x$ being greater than or less than some threshold value $x_c$.  If we could introduce a second discriminating variable $y$ such that the distribution for background factorizes $f_B(x,y)=f_B(x)f_B(y)$, then we have a handle to measure the factor $\tau$.  Typically, one introduces a threshold $y_c$  so that the signal contribution is small below this threshold\footnote{The relative sign of the cut is not important, but has been chosen for consistency with Fig~\ref{fig:ABCD}.}.  Figure~\ref{fig:ABCD} shows an example where $x_c=y_c=5$.  With this we these two thresholds we have four regions that we can schematically refer to as A, B, C, and D.  In the case of simply counting events in these regions we can write the total expectation as
\begin{eqnarray}
\nu_A &=& 1\cdot\mu + \nu_A^{MC} + 1\cdot\nu_A \\\nonumber
\nu_B &=& \epsilon_B\mu \,+ \nu_B^{MC} + \tau_B\nu_A \\\nonumber
\nu_C &=& \epsilon_C\mu \,+ \nu_C^{MC} + \tau_C\nu_A \\\nonumber
\nu_D &=& \epsilon_D\mu \,+ \nu_D^{MC} + \tau_B\tau_C\nu_A 
\end{eqnarray}
where $\mu$ is the signal rate in region A, $\epsilon_i$ is the ratio of the signal in the regions B, C, D with respect to the signal in region A, $\nu_i^{MC}$ is the rate of background in each of the regions being estimated from simulation,  $\nu_i$ is the rate of the background being estimated with the data driven technique in the signal region, and $\tau_i$ are the ratios of the background rates in the regions B, C, and D with respect to the background in region A.  The key is that we have used the factorization $f_B(x,y)=f_B(x)f_B(y)$ to write $\tau_D=\tau_B\tau_C$.  The right panel of Fig.~\ref{fig:ABCD} shows a more complicated extension of the ABCD method from a recent ATLAS SUSY analysis~\cite{ATLAS:2011ad}.

d

An alternative parametrization, which can be more numerically stable is\\
\begin{eqnarray}
\nu_A &=& 1\cdot\mu + \nu_A^{MC} + \eta_C\eta_B\nu_D \\\nonumber
\nu_B &=& \epsilon_B\mu \,+ \nu_B^{MC} + \eta_B\nu_D \\\nonumber
\nu_C &=& \epsilon_C\mu \,+ \nu_C^{MC} + \eta_C\nu_D \\\nonumber
\nu_D &=& \epsilon_D\mu \,+ \nu_D^{MC} + 1\cdot\nu_D 
\end{eqnarray}

\subsection{Effective Model Narrative}

In the simulation narrative the model of discriminating variable distributions $f(x|\vec\alpha)$ is derived from discrete samples of simulated events $\{x_1,\dots,x_N\}$.   We discussed above how one can use histograms or kernel estimation to approximate the underlying distribution and interpolation strategies to incorporate systematic effects.  Another approach is to assume some parametric form for the distribution to serve as an effective model.  For example, in the $H\to\gamma\gamma$ analysis shown in Fig.~\ref{fig:H2photons} a simple exponential distribution was used to model the background.  The state-of-the-art theoretical predictions for the continuum $\gamma\gamma$ background process do not predict exactly an exponentially falling distribution, and the analysis must (and does) incorporate the systematic associated to the effective model.  Similarly, it is common to use a polynomial in some limited sideband region to estimate backgrounds under a peak.  These effective models can range from very ad hoc~\footnote{For instance, the modeling of $H\to ZZ^{(*)}\to 4l$  described in \cite{Aad:2009wy} (see Eq. 2 of the corresponding section)  } to more motivated.  For instance, one might use knowledge of kinematics and phase space and/or detector resolution to construct an effective model that captures the relevant physics.  The advantage of a well motivated effective model is that few nuisance parameters may describe well the relevant family of probability densities, which is the challenge for generic (and relatively unsophisticated) interpolation strategies usually employed in the simulation narrative.


\subsection{The Matrix Element Method}
Ideally, one would not use a single discriminating variable to distinguish the process of interest from the other background processes, but instead would use as much discriminating power as possible.  This implies forming a probability model over a multi-dimensional discriminating variable (ie. a multivariate analysis technique).  In principle, both the histogram-based and kernel-based approach generalize to distributions of multi-dimensional discriminating variables; however, in practice, they are limited to only a few dimensions. In the case of histograms this is particularly severe unless one employs clever binning choices, while in the kernel-based approach one can model up to about 5-dimensional distributions with reasonable Monte Carlo sample sizes.   In practice, one often uses multivariate algorithms like Neural Networks or boosted decision trees\footnote{A useful toolkit for high-energy physics is TMVA, which is packaged with ROOT~\cite{tmva}.} to map the multiple variables into a single discriminating variable.  Often these multivariate techniques are seen as somewhat of a black-box.  If we restrict ourselves to discriminating variables associated with the kinematics of final state particles (as opposed to the more detailed signature of particles in the detector), then we can often approximate he detailed simulation of the detector with a parametrized detector response.  If we denote the kinematic configuration of all the final state particles in the Lorentz invariant phase space as $\Phi$, the initial state as $i$,  the matrix element (potentially averaged over unmeasured spin configurations) as $\mathcal{M}(i,\Phi)$, and the probability due to parton density functions for the initial state $i$ going into the hard scattering  as $f(i)$, then we can write that the distribution of the, possibly multi-dimensional, discriminating variable $x$ as
\begin{equation}
f(x) \propto \int d\Phi \, f(i) |\mathcal{M}(i,\Phi)|^2 \, W(x | \Phi) \;,
\end{equation}
where $W(x|\Phi)$ is referred to as the transfer function of $x$ given the final state configuration $\Phi$.  It is natural to think of $W(x|\Phi)$ as a conditional distribution, but here I let $W$ encode the efficiency and acceptance so that we have
\begin{equation}
\frac{\sigma_{\rm eff.}}{\sigma} = \frac{\int dx \int d\Phi \, |\mathcal{M}(i,\Phi)|^2 \, W(x | \Phi) }{\int d\Phi \, |\mathcal{M}(i,\Phi)|^2 }\;.
\end{equation}
Otherwise, the equation above looks like another application one Bayes's theorem where $W(x|\Phi)$ plays the role of the pdf/likelihood function and $\mathcal{M}(i,\Phi)$ plays the role of the prior over the $\Phi$.  It is worth pointing out that this is a frequentist use of Bayes's theorem since $d\Phi$ is the Lorentz invariant phase space which explicitly has a measure associated with it.

\subsection{Event-by-event resolution, conditional modeling, and Punzi factors}

In some cases one would like to provide a distribution for the discriminating variable $x$ based conditional on some other observable in the event $y$: $f(x|\vec\alpha,y)$.  For instance, one might want to say that the energy resolution for electrons depends on the energy itself through a well-known calorimeter resolution parametrization like $\sigma(E)/E = A/\sqrt{E}\oplus B$.  These types of conditional distributions can be built in \roofit.  A subtle point studied by Punzi is that if $f(y|\vec\alpha)$ depends on $\vec\alpha$ the inference on $\vec\alpha$ can be biased~\cite{Punzi:2004wh}.  In particular, if one is trying to estimate the amount of signal in a sample and the distribution of $y$ for the signal is different than for the background, the estimate of the signal fraction will be biased.  This can be remedied by including terms related to $f(y|\vec\alpha)$, colloquially called `Punzi Factors'.  Importantly, this means one cannot build conditional models like this without knowing or assuming something about $f(y|\vec\alpha)$.

\section{Frequentist Statistical Procedures}
\label{UE:Inputs}

Here I summarize the procedure used by the LHC Higgs combination group  for computing frequentist  $p$-values uses for 
quantifying the agreement with the background-only hypothesis and for determining exclusion limits.  
The procedures are based on the profile likelihood ratio test statistic.  

The parameter of interest is the overall signal strength factor $\mu$, which acts as a scaling to the total rate of signal events.  We often write $\mu=\sigma/\sigma_{SM}$, where $\sigma_{SM}$ is the standard model production cross-section; however, it should be clarified that the same $\mu$ factor is used for all production modes and could also be seen as a scaling on the branching ratios.  The signal strength is called so that $\mu=0$ corresponds to the background-only model and $\mu=1$ is the standard model signal.  It is convenient to separate the full list of parameters $\vec\alpha$ into the parameter of interest $\mu$ and the nuisance parameters $\vec\theta$: $\vec\alpha=(\mu,\vec\theta)$.

For a given data set $\datasim$ and values for the global observables $\globs$ there is an associated likelihood function over $\mu$ and $\theta$ derived from combined model over all the channels including all the constraint terms in Eq.~\ref{Eq:ftot}
\begin{equation}
L(\mu,\vec\theta;\datasim,\globs) = \F_{\rm tot}(\datasim,\globs|\mu,\vec\theta) \;.
\end{equation}
The notation $L(\mu,\vec\theta)$ leaves the dependence on the data implicit, which can lead to confusion.  Thus, we will explicitly write the dependence on the data when the identity of the dataset is important and only suppress $\datasim,\globs$ when the statements about the likelihood are generic.

We begin with the definition of the procedure in the abstract and then describe three implementations of the method based on asymptotic distributions, ensemble tests (Toy Monte Carlo),  and importance sampling.

\subsection{The test statistics and estimators of $\mu$ and $\vec\theta$}

This definitions in this section are all relative to a given dataset $\datasim$ and value of the global observables $\globs$, thus we will suppress their appearance.  The nomenclature follows from Ref.~\cite{asimov}.

The maximum likelihood estimates (MLEs) $\hat\mu$ and $\hat{\vec\theta}$ and the values of the parameters that maximize the likelihood function $L(\mu,\vec\theta)$ or, equivalently, minimize $-\ln L(\mu,\vec\theta)$.  The dependence of the likelihood function on the data propagates to the values of the MLEs, so when needed the MLEs will be given subscripts to indicate the data set used.  For instance, $\hat{\vec\theta}_{\rm obs}$ is the MLE of $\vec\theta$ derived from the observed data and global observables. 

The conditional maximum likelihood estimate (CMLEs) $\hathatthetamu$ is the value of $\vec\theta$ that maximizes the likelihood function with $\mu$ fixed; it can be seen as a multidimensional function of the single variable $\mu$.  Again, the dependence on $\datasim$ and $\globs$ is implicit. This procedure for choosing specific values of the nuisance parameters for a given value of $\mu$, $\datasim$, and $\globs$ is often referred to as ``profiling''.  Similarly, $\hathatthetamu$ is often called ``the profiled value of $\vec\theta$''.

Given these definitions, we can construct the profile likelihood ratio
\begin{equation}
\label{eq:lambdatilde}  
{\lambda}({\mu}) =  \frac{ L(\mu, \hat{\hat{\vec{\theta}}}(\mu)) }
{L(\hat{\mu}, \hat{\vec{\theta}}) } \;,
\end{equation}
which depends explicitly on the parameter of interest $\mu$, implicitly on the data $\datasim$ and global observables $\globs$, and is independent of the nuisance parameters $\vec\theta$ (which have been eliminated via ``profiling'').

\begin{figure}[h]
\begin{center}
\subfigure[][]{\includegraphics[width=.3\textwidth]{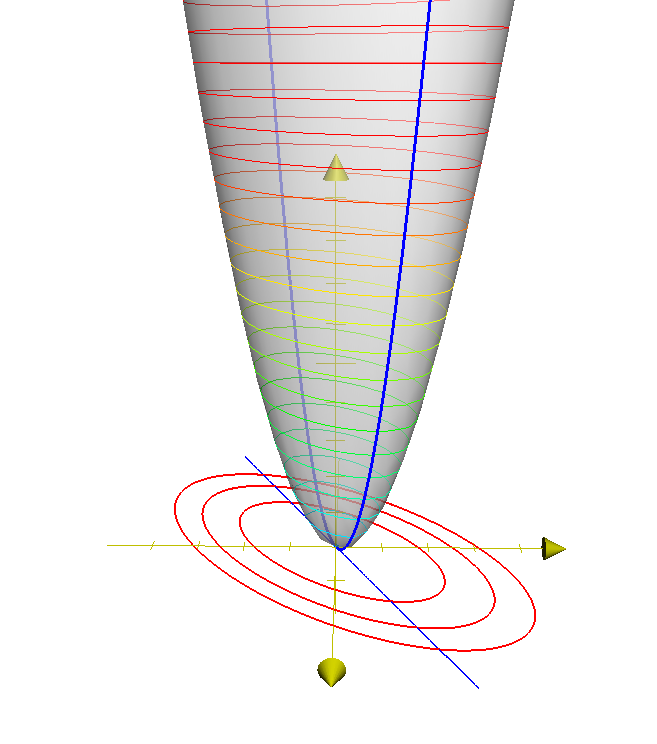}}
\subfigure[][]{\includegraphics[width=.3\textwidth]{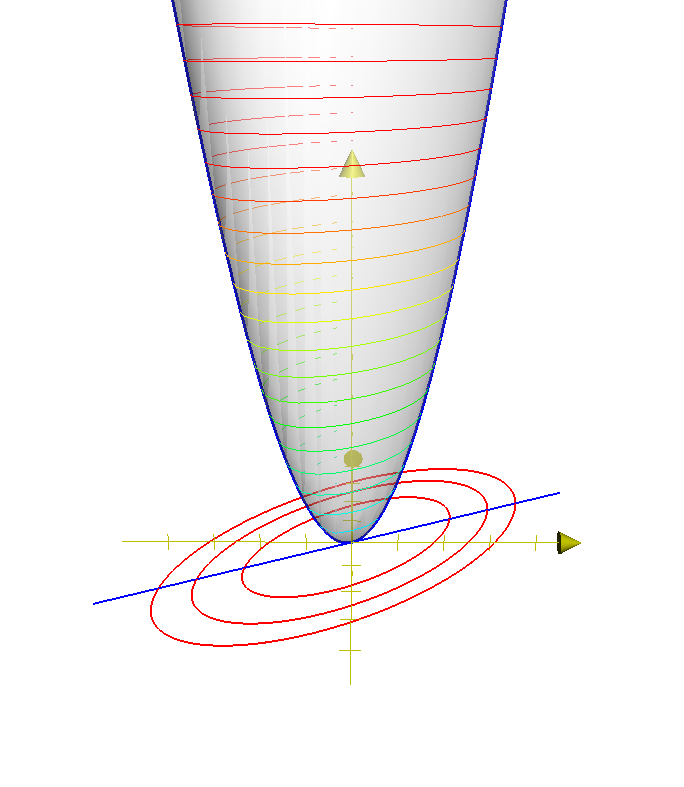}}\\
\subfigure[][]{\includegraphics[width=.3\textwidth]{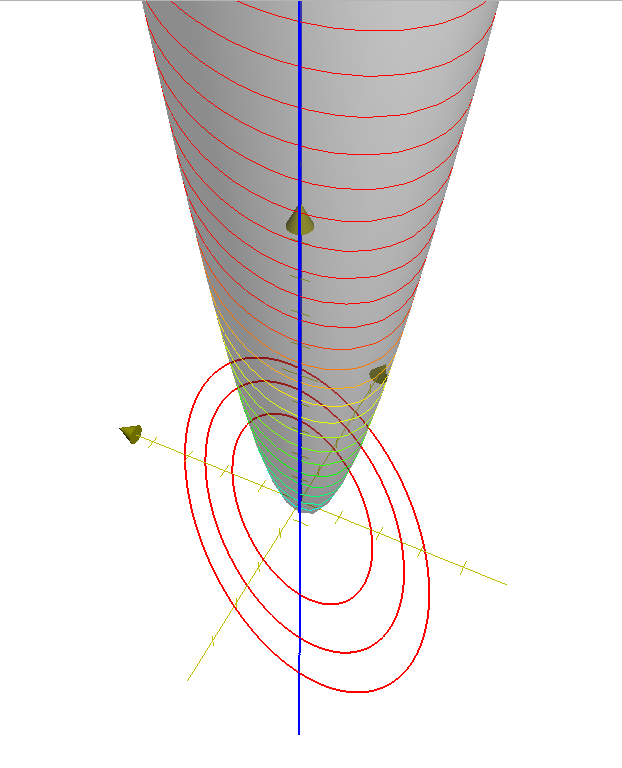}}
\subfigure[][]{\includegraphics[width=.3\textwidth]{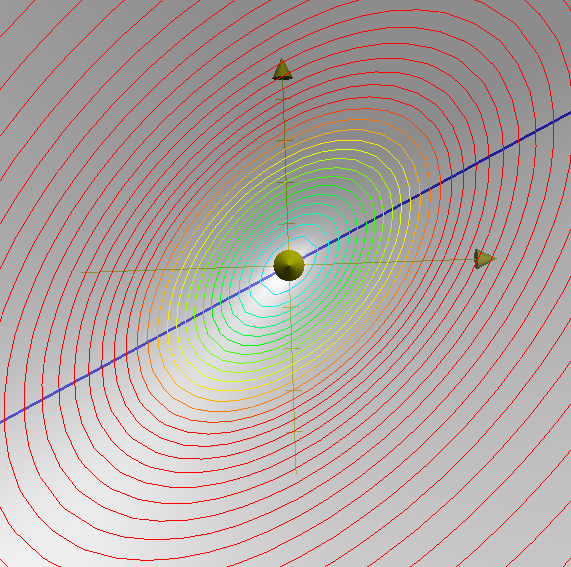}}
\caption{Visualization of a two dimensional likelihood function $-2\ln L(\mu,\theta)$.  The blue line in the plane represents the profiling operation $\hat{\hat{\theta}}(\mu)$ and the blue curve along the likelihood surface represents $-2\ln \lambda(\mu)$.  Note it is was to show that the blue line exits the contours of $-2\ln L(\mu,\theta)$ when they are perpendicular to the $\mu$ axis, which provides the correspondence between the profile likelihood ratio and the description of the \texttt{Minos} algorithm.}
\label{fig:profileLR}
\end{center}
\end{figure}

In any physical theory the rate of signal events is non-negative, thus $\mu\ge 0$.  However, it is often convenient to allow $\mu<0$ (as long as the pdf $f_c(x_c | \mu,\vec\theta)\ge 0$ everywhere).  In particular, $\hat\mu<0$ indicates a deficit of events signal-like with respect to the background only and the boundary at $\mu=0$ complicates the asymptotic distributions.  Ref.~\cite{asimov} uses a  trick that is equivalent to requiring $\mu\ge 0$ while avoiding the formal complications of a boundary, which is to allow $\mu< 0$ and   impose the constraint in the test statistic itself.  In particular, one defines $\tilde \lambda(\mu)$
\begin{equation}
\label{eq:lambdatilde}  \tilde{\lambda}({\mu}) =  \left\{ 
\begin{array}{ll} \frac{ L(\mu, \hat{\hat{\vec{\theta}}}(\mu)) }
{L(\hat{\mu}, \hat{\vec{\theta}}) } & \hat{\mu} \ge 0 , \\*[0.3 cm]
\frac{ L(\mu, \hat{\hat{\vec{\theta}}}(\mu)) }  {L(0,
\hat{\hat{\vec{\theta}}}(0)) } & \hat{\mu} < 0 
              \end{array} \right.
\end{equation}
This is not necessary when ensembles of pseudo-experiments are generated with ``Toy'' Monte Carlo techniques, but since they are equivalent we will write $\tilde\lambda$ to emphasize the boundary at $\mu=0$.

For discovery the test statistic $\tilde{q}_0$  is used to differentiate the background-only hypothesis $\mu=0$ from the alternative hypothesis $\mu>0$:
\begin{equation}
\tilde{q}_{0} =  \left\{ \!
\! \begin{array}{ll} - 2 \ln \tilde{\lambda}(\mu)  & \hat{\mu} > 0
\\*[0.2 cm] 0 & \hat{\mu} \le 0 
              \end{array} \right.  
\end{equation}
Note that $\tilde{q}_0$ is test statistic for a one-sided alternative.  Note also that if we consider the parameter of interest $\mu\ge 0$, then it is equivalent to the two-sided test (because there are no values of $\mu$ less than $\mu=0$. 

For limit setting the test statistic $\tilde{q}_{\mu}$ is used to differentiate the hypothesis of signal being produced at a rate $\mu$ from the alternative hypothesis of signal events being produced at a lesser rate $\mu'<\mu$:
\begin{equation}
\label{eq:qmutilde}  
\tilde{q}_{\mu} =  \left\{ \!
\! \begin{array}{ll} - 2 \ln \tilde{\lambda}(\mu)  & \hat{\mu} \le \mu
\\*[0.2 cm] 0 & \hat{\mu} > \mu 
              \end{array} \right.  \quad = \quad \: \left\{ \!
\! \begin{array}{lll} - 2 \ln \frac{L(\mu,
\hat{\hat{\vec{\theta}}}(\mu))} {L(0, \hat{\hat{\theta}}(0))}   &
\hat{\mu} < 0 \;, \\*[0.2 cm] -2 \ln \frac{L(\mu,
\hat{\hat{\vec{\theta}}}(\mu))} {L(\hat{\mu}, \hat{\vec{\theta}})} &
0 \le \hat{\mu} \le \mu  \;, \\*[0.2 cm] 0 & \hat{\mu} > \mu \;.
              \end{array} \right.
\end{equation}
Note that $\tilde{q}_{\mu}$ is a test statistic for a one-sided alternative; it is a test statistic for a one-sided upper limit. 

The test statistic $\tilde{t}_\mu$ is used to differentiate signal being produced at a rate $\mu$ from the alternative hypothesis of signal events being produced at a lesser or greater rate $\mu' \ne\mu$.
\begin{equation}
\label{eq:tmu}
\tilde{t}_{\mu} =  - 2 \ln \tilde{\lambda}(\mu)   \; .  
\end{equation}
Note that $\tilde{t}_\mu$ is a test statistic for a two-sided alternative (as in the case of the Feldman-Cousins technique, though this is more general as it incorporates nuisance parameters).  Note  that if we consider the parameter of interest $\mu\ge 0$ and we the test at $\mu=0$ then there is no ``other side'' and we have $\tilde{t}_{\mu=0} = \tilde{q}_0$.  Finally, if one relaxes the constraint $\mu\ge0$ then the two-sided test statistic is written $t_\mu$ or, simply, $-2\ln\lambda(\mu)$.

\subsection{The distribution of the test statistic and $p$-values}

The test statistic should be interpreted as a single real-valued number that represents the outcome of the experiment.  More formally, it is a mapping of the data to a single real-valued number:  \mbox{$\tilde{q}_\mu: \datasim,\globs \rightarrow \mathbb{R}$}.  For the observed data the test statistic has a given value, eg. $\tilde{q}_{\mu,\rm obs}$.  If one were to repeat the experiment many times the test statistic would take on different values, thus, conceptually, the test statistic has a distribution.  Similarly, we can use our model to generate pseudo-experiments using Monte Carlo techniques or more abstractly consider the distribution.  Since the number of expected events $\nu(\mu,\vec\theta)$ and the distributions of the discriminating variables $f_c(x_c|\mu,\vec\theta)$ explicitly depend on $\vec\theta$ the distribution of the test statistic will also depend on $\vec\theta$.  Let us denote this distribution 
\begin{equation}
f(\tilde{q}_\mu | \mu, \vec\theta) \;,
\end{equation}
and we have analogous expressions for each of the test statistics described above.

The $p$-value for a given observation under a particular hypothesis ($\mu,\vec\theta$) is the probability for an equally or more `extreme' outcome than observed assuming that hypothesis 
\begin{equation}
p_{\mu,\vec\theta} = \int_{\tilde{q}_{\mu,\rm obs}}^\infty f(\tilde{q}_\mu | \mu, \vec\theta) \, d\tilde{q}_\mu\;.
\end{equation}
The logic is that small $p$-values are evidence against the corresponding hypothesis.  In Toy Monte Carlo approaches, the integral above is really carried out in the space of the data $\int d\datasim  d\globs$.

The immediate difficulty is that we are interested in $\mu$ but the $p$-values depend on both $\mu$ and $\vec\theta$.  In the frequentist approach the hypothesis $\mu=\mu_0$ would not be rejected unless the $p$-value is sufficiently small \textit{for all} values of $\vec\theta$.  Equivalently, one can use the supremum $p$-value for over all $\vec\theta$ to base the decision to accept or reject the hypothesis at $\mu=\mu_0$.
\begin{equation}
p^{\rm sup}_{\mu} = \sup_{\vec\theta}\; p_{\mu,\vec\theta} 
\end{equation}

The key conceptual reason for choosing the test statistics based on the profile likelihood ratio is that asymptotically (ie. when there are many events) the distribution of the profile likelihood ratio \mbox{$\lambda(\mu=\mu_{\rm true})$} is  independent of the values of the nuisance parameters.  This follows from Wilks's theorem.  In that limit $p^{\rm sup}_{\mu} =  p_{\mu,\vec\theta}$ for all $\vec\theta$.  

The asymptotic distributions \mbox{$f(\lambda(\mu) | \mu, \vec\theta)$} and \mbox{$f(\lambda(\mu) | \mu', \vec\theta)$} are known and described in Sec.~\ref{sec:asymptotic}.  For results based on generating ensembles of pseudo-experiements using Toy Monte Carlo techniques does not assume the form of the distribution $f(\tilde{q}_\mu | \mu, \vec\theta)$, but knowing that it is approximately independent of $\vec\theta$ means that one does not need to calculate $p$-values for all $\vec\theta$ (which is not computationally feasible).  Since there may still be some residual dependence of the $p$-values on the choice of $\vec\theta$ we would like to know the specific value of $\vec\theta^{\rm sup}$ that produces the supremum $p$-value over $\vec\theta$.  Since larger $p$-values indicate better agreement of the data with the model, it is not surprising that choosing $\vec\theta^{\rm sup}=\hathatthetamu$ is a good estimate of $\vec\theta^{\rm sup}$.  This has been studied in detail by statisticians, and is called the Hybrid Resampling method and is referred to in physics as the `profile construction'~\cite{Feldman,Cranmer,hybridResampling,Bodhi}.

Based on the discussion above, the following $p$-value is used to quantify consistency with the hypothesis of a signal strength of $\mu$:
\begin{equation}
p_{\mu}=\int_{\tilde q_{\mu,\rm obs}}^{\infty} f(\tilde q_\mu|\mu,\hat{\hat{\vec{\theta}}}(\mu,\textrm{obs})) \,d\tilde q_\mu \;.
\end{equation}
A standard 95\% confidence-level, one-sided frequentist confidence interval (upper limit) is obtained by  solving for $p'_{\mu_{up}}=5\%$.  For downward fluctuations the upper limit of the confidence interval can be arbitrarily small, though it will always include $\mu=0$.  This feature is considered undesirable since a physicist would not claim sensitivity to an arbitrarily small signal rate.  The feature was the motivation for the modified frequentist method called $CL_s$~\cite{Read2,Read1,CLsWikipedia}. and the alternative approach called power-constrained limits~\cite{2011arXiv1105.3166C}. 

To calculate the $CL_s$ upper limit, we define $p'_\mu$ as a ratio of p-values,
\begin{equation}
p'_\mu=\frac{p_\mu}{1-p_b} \; ,
\end{equation}
where  $p_b$ is the $p$-value derived from the same test statistic under the background-only hypothesis
\begin{equation}
\label{eq:pb}
p_b=1-\int_{\tilde q_{\mu,obs}}^\infty f(\tilde q_\mu|0,\hat{\hat{\vec{\theta}}}(\mu=0,\textrm{obs}))d\tilde q_\mu \;.
\end{equation}
The $CL_s$ upper-limit on $\mu$ is denoted $\mu_{up}$ and obtained by solving for $p'_{\mu_{up}}=5\%$.  
It is worth noting that while confidence intervals produced with the ``CLs'' method over cover, a value of $\mu$ is regarded as excluded at the 95\% confidence level if $\mu<\mu_{up}$.  The amount of over coverage is not immediately obvious; however, for small values of $\mu$ the coverage approaches 100\% and for large values of $\mu$ the coverage is near the nominal 95\% (due to $\langle p_b\rangle\approx0$).

For the purposes discovery one is interested in compatibility of the data with the background-only hypothesis.  Statistically, a discovery corresponds to rejecting the background-only hypothesis.  This compatibility is based on the following $p$-value
\begin{equation}
p_0=\int_{\tilde q_{0,obs}}^\infty f(\tilde q_0|0,\hat{\hat{\vec{\theta}}}(\mu=0,\textrm{obs}))d\tilde q_0 \;.
\end{equation}
This $p$-value is also based on the background-only hypothesis, but the test statistic $\tilde q_0$ is suited for testing the background-only while the test statistic $\tilde{q}_\mu$ in Eq.~\ref{eq:pb} is suited for testing a hypothesis with signal.

It is customary to convert the background-only $p$-value into the quantile (or ``sigma'') of a unit Gaussian.  This conversion is purely conventional and makes no assumption that the test statistic $q_0$ is Gaussian distributed.  The conversion is defined as:
\begin{equation}
Z = \Phi^{-1}(1-p_0) ;\,
\end{equation}
where $\Phi^{-1}$ is the inverse of the cumulative distribution for a unit Gaussian.  One says the significance of the result is $Z\sigma$ and the standard discovery convention is $5\sigma$, corresponding to $p_0=2.87  \cdot 10^{-7}$.

\subsection{Expected sensitivity and bands}

The expected sensitivity for limits and discovery are useful quantities, though subject to some degree of ambiguity.  Intuitively, the expected upper limit is the upper limit one would expect to obtain if the background-only hypothesis is true.  Similarly, the expected significance is the significance of the observation assuming the standard model signal rate (at some $\mh$).  To find the expected limit one needs a distribution $f(\mu_up | \mu=0,\vec\theta)$.  To find the expected significance one needs the distribution $f(Z | \mu=1,\vec\theta)$ or, equivalently, $f(p_0 | \mu=1,\vec\theta)$.  We use the median instead of the mean, as it is invariant to the choice of $Z$ or $p_0$.  More importantly, is that the expected limit and significance depend on the value of the nuisance parameters $\vec\theta$, for which we do not know the true values.  Thus, the expected limit and significance will depend on some convention for choosing $\vec\theta$.  While many nuisance parameters have a nominal estimate (i.e. the global observables in the constraint terms), others do not (eg. the exponent in the $H\to\gamma\gamma$ background model).  Thus, we choose a convention that treats all of the nuisance parameters consistently, which is the profiled value based on the observed data.  Thus for  the expected limit we use $ f(\mu_{\rm up}|0,\hat{\hat{\vec{\theta}}}(\mu=0,\textrm{obs}))$ and for the expected significance we use $f(p_0 | \mu=1,\hat{\hat{\vec\theta}}(\mu=1, \rm obs))$.  An unintuitive and possibly undesirable feature of this choice is that the expected limit and significance depend on the observed data through the conventional choice for $\vec\theta$.

With these distributions we can also define bands around the median upper limit.  Our standard limit plot shows a dark green band corresponding to $\mu_{\pm 1}$ defined by 
\begin{equation}
\int_{0}^{\mu_{\pm 1}}  f(\mu_{\rm up}|0,\hat{\hat{\vec{\theta}}}(\mu=0,\textrm{obs})) d\mu_{\rm up} = \Phi^{-1}(\pm 1) 
\end{equation}
and a light yellow band corresponding to $\mu_{\pm 2}$ defined by 
\begin{equation}
\int_{0}^{\mu_{\pm 2}}  f(\mu_{\rm up}|0,\hat{\hat{\vec{\theta}}}(\mu=0,\textrm{obs})) d\mu_{\rm up} = \Phi^{-1}(\pm 2) 
\end{equation}

\subsection{Ensemble of pseudo-experiments generated with ``Toy'' Monte Carlo}

The $p$-values in the procedure described above require performing several integrals.  In the case of the asymptotic approach, the distributions for $\tilde q_\mu$ and $\tilde q_0$ are known and the integral is performed directly.  When the distributions are not assumed to take on their asymptotic form, then they must be constructed using Monte Carlo methods.  In the ``toy Monte Carlo'' approach one generates pseudo-experiments in which the number of events in each channel $n_c$, the values of the discriminating variables $\{x_{ec}\}$ for each of those events, and the auxiliary measurements (global observables) $a_p$ are all randomized according to $\F_{\rm tot}$.    We  denote the resulting data $\data_{\rm toy}$ and global observables $\globs_{\rm toy}$.  By doing this several times one can build an ensemble of pseudo-experiments and evaluate the necessary integrals.  Recall that Monte Carlo techniques can be viewed as a form of numerical integration.

The fact that the auxiliary measurements $a_p$ are randomized is unfamiliar in particle physics.  The more familiar approach for toy Monte Carlo is that the nuisance parameters are randomized.  This requires a distribution for the nuisance parameters, and thus corresponds to a Bayesian treatment of the nuisance parameters.  The resulting $p$-values are a hybrid Bayesian-Frequentist quantity with no consistent definition of probability.  To maintain a strictly frequentist procedure, the corresponding operation is to randomize the auxiliary measurements.  

While formally this procedure is well motivated, as physicists we also know that our models can have deficiencies and we should check that the distribution of the auxiliary measurements does not deviate too far from our expectations.  In Section~\ref{Sec:crossChecks} we show the distribution of the auxiliary measurements and the corresponding $\hat{\vec\theta}$ from the toy Monte Carlo technique.  

Technically, the pseudo-experiments are generated with the \texttt{RooStats} \texttt{ToyMCSampler}, which is used by the higher-level tool \texttt{FrequentistCalculator}, which is in turn used by \texttt{HypoTestInverter}.

 \subsection{Asymptotic Formulas }

The following has been extracted from Ref.~\cite{asimov} and has been reproduced here for convenience.  The primary message of Ref.~\cite{asimov} is that for a sufficiently large data sample the distributions of the likelihood ratio based test statistics above converge to a specific form.  In particular, Wilks's theorem~\cite{Wilks} can be used to obtain the distribution $f(\lambda(\mu)|\mu)$, that is the distribution of the test statistic $\lambda(\mu)$ when $\mu$ is true.  Note that the asymptotic distribution is independent of the value of the nuisance parameters. Wald's theorem~\cite{Wald} provides the generalization to $f(\lambda(\mu)|\mu',\vec\theta)$, that is when the true value is not the same as the tested value.  The various formulae listed below are corollaries of Wilks's and Wald's theorems for the likelihood ratio test statistics described above.  The Asimov data described immediately below was a novel result of Ref.~\cite{asimov}.

 \subsubsection{The Asimov data and  $\sigma=\textrm{var}$($\hat\mu$)}
  \label{S:Asimov}

 The asymptotic formulae below require knowing the variance of the maximum likelihood estimate of $\mu$
 \begin{equation}
 \sigma=\textrm{var}[\hat\mu]\;.
 \end{equation}
 One result of Ref.~\cite{asimov} is that $\sigma$ can be
estimated with an artificial dataset referred to as the \textit{ Asimov} dataset.  The Asimov dataset is defined as a binned dataset, where the number of events in bin $b$ is exactly the number of events expected in bin $b$.  Note, this means that the dataset generally has non-integer number of events in each bin.  For our general model one can write
\begin{equation}
\label{eq:asimovData}
n_{b,A} = \int_{x \in \textrm{bin}~b} \nu(\vec\alpha) f(x|\vec\alpha) dx \;
\end{equation}
where the subscript $A$ denotes that this is the Asimov data.  Note, that the dataset depends on the value of $\vec\alpha$ implicitly.  For an model of unbinned data, one can simply take the limit of narrow bin widths for the Asimov data.    We denote the likelihood evaluated with the Asimov data as $L_{\rm A}(\mu)$. 
The important result is that one can calculate the expected Fisher information of Eq.~\ref{Eq:expfisher} by computing the observed Fisher information on the likelihood function based on this special Asimov dataset.  

A related and convenient way to calculate the variance of $\hat\mu$ is 
\begin{equation}
\label{eq:sigmaofmu}
\sigma \sim \frac{\mu}{\sqrt {\tilde q_{\mu,A}}} \;.
\end{equation}
where $\tilde q_{\mu,A}$ is the to use the $\tilde q_\mu$ test statistic based on a background-only Asimov data (ie. the one with$\mu=0$ in Eq.~\ref{eq:asimovData}).  It is worth noting that higher-order corrections to the formulae below are being developed to address the case when the variance of $\hat\mu$ depends strongly on $\mu$.

 \subsubsection{Asymptotic Formulas for $\tilde q_{0}$}
For a sufficiently large data sample,  the pdf $f(\tilde{q}_{0} | \mu')$ is found to approach
\begin{equation}
\label{eq:fq0muprimewald}
f(q_0 | \mu^{\prime}) = \left( 1 - 
\Phi \left( \frac{ \mu^{\prime}}{\sigma} \right) \right) \delta(q_0)  + 
\frac{1}{2}
\frac{1}{\sqrt{2 \pi}} \frac{1}{\sqrt{q_0}} \exp 
\left[ - \frac{1}{2} \left( \sqrt{q_0} - \frac{\mu^{\prime}}{\sigma} 
\right)^2 \right] 
\;.
\end{equation}
For the special case of $\mu^{\prime} = 0$, this reduces to
\begin{equation}
\label{eq:fq00}
f(q_0 | 0) = \frac{1}{2} \delta(q_0) + 
\frac{1}{2} \frac{1}{\sqrt{2 \pi}} \frac{1}{\sqrt{q_0}} e^{-q_0/2} \;.
\end{equation}
That is, one finds a mixture of a delta function at zero and
a chi-square distribution for one degree of freedom, with each term
having a weight of $1/2$.  In the following we will refer to this
mixture as a half chi-square distribution or $\half \chi^2_1$.

From Eq.~(\ref{eq:fq0muprimewald}) the corresponding cumulative
distribution is found to be
\begin{equation}
\label{cdfq0muprimewald}
F(q_0 | \mu^{\prime}) = \Phi \left( \sqrt{q_0} - \frac{\mu^{\prime}}{\sigma} 
\right) \;.
\end{equation}

The important special case $\mu^{\prime} = 0$ is therefore simply
\begin{equation}
\label{cdfq00wald}
F(q_0 | 0) = \Phi \Big( \sqrt{q_0} \Big)
\;.
\end{equation}
The $p$-value of the $\mu=0$ hypothesis is 
\begin{equation}
\label{eq:pval0}
p_0 = 1 - F(q_0 | 0) \;, 
\end{equation}
and therefore for the significance gives the simple formula
\begin{equation}
\label{eq:Z0}
Z = \Phi^{-1}(1 - p_0) = \sqrt{q_0} \;.
\end{equation}

 \subsubsection{Asymptotic Formulas for $\tilde q_{\mu}$}
 \label{sec:tildeqmu}
 
For a sufficiently large data sample,  the pdf $f(\tilde{q}_{\mu} | \mu)$ is found to approach
\begin{eqnarray}
\label{eq:ftildeqmmp} 
f(\tilde{q}_{\mu}|\mu^{\prime}) & = & 
\Phi \left( \frac{\mu^{\prime} - \mu}{\sigma} \right) 
\delta (\tilde{q}_{\mu}) \nonumber \\*[0.3 cm] 
& + &
 \: \left\{ \! \! \begin{array}{lll}
\frac{1}{2} \frac{1}{\sqrt{2 \pi}} \frac{1}{\sqrt{\tilde{q}_{\mu}}}
\exp \left[ -\frac{1}{2} \left( \sqrt{\tilde{q}_{\mu}} -
\frac{\mu - \mu^{\prime}}{\sigma} \right)^2 \right]
                 & 0 < \tilde{q}_{\mu} \le \mu^2/\sigma^{2}  \\*[0.5 cm]
\frac{1}{\sqrt{2 \pi} \sigma} \exp \left[
-\frac{1}{2} \frac{ (\tilde{q}_{\mu} - 
(\mu^2 - 2 \mu \mu^{\prime})/\sigma^{2} )^2 }
{(2 \mu/\sigma)^2} \right] 
                 &  \quad \tilde{q}_{\mu} > \mu^2/\sigma^{2} 
              \end{array}
       \right.
\;.
\end{eqnarray}
The special case $\mu = \mu^{\prime}$ is therefore
\begin{equation}
\label{eq:ftildeqmm} 
f(\tilde{q}_{\mu}|\mu) = 
\frac{1}{2} \delta (\tilde{q}_{\mu}) +
 \: \left\{ \! \! \begin{array}{lll}
\frac{1}{2} \frac{1}{\sqrt{2 \pi}} \frac{1}{\sqrt{\tilde{q}_{\mu}}}
e^{- \tilde{q}_{\mu}/2}
                 & 0 < \tilde{q}_{\mu} \le \mu^2/\sigma^2  \\*[0.5 cm]
\frac{1}{\sqrt{2 \pi} \sigma} \exp \left[
-\frac{1}{2} \frac{ (\tilde{q}_{\mu} + \mu^2/\sigma^2 )^2 }
{(2 \mu/\sigma)^2} \right] 
                 &  \quad \tilde{q}_{\mu} > \mu^2/\sigma^2 \;.
              \end{array}
       \right.
\end{equation}
The corresponding cumulative distribution is
\begin{equation}
\label{eq:tildeqmmpcdf} 
F(\tilde{q}_{\mu}|\mu^{\prime}) = 
 \: \left\{ \! \! \begin{array}{lll}
\Phi\left( \sqrt{\tilde{q}_{\mu}} - 
\frac{\mu - \mu^{\prime}}{\sigma} \right)
                 & \quad 0 < \tilde{q}_{\mu} \le \mu^2/\sigma^{2}  
\;, \\*[0.5 cm]
\Phi \left( \frac{ \tilde{q}_{\mu} - 
(\mu^2 - 2 \mu \mu^{\prime})/\sigma^{2}}
{2\mu/\sigma} \right)
                 &  \quad \tilde{q}_{\mu} > \mu^2/\sigma^{2} \;.
              \end{array}
       \right.
\end{equation}

 The special case $\mu = \mu^{\prime}$ is
\begin{equation}
\label{eq:tildeqmmcdf} 
F(\tilde{q}_{\mu}|\mu) = 
 \: \left\{ \! \! \begin{array}{lll}
\Phi\Big( \sqrt{\tilde{q}_{\mu}} \Big)
                 & \quad 0 < \tilde{q}_{\mu} \le \mu^2/\sigma^2  
\;, \\*[0.5 cm]
\Phi \left( \frac{ \tilde{q}_{\mu} + \mu^2/\sigma^2}
{2\mu/\sigma} \right)
                 &  \quad \tilde{q}_{\mu} > \mu^2/\sigma^2 \;.
              \end{array}
       \right.
\end{equation}

 The $p$-value of the hypothesized $\mu$ is as before
given by one minus the cumulative distribution,

\begin{equation}
\label{eq:pvalmutilde}
p_{\mu} = 1 - F(\tilde{q}_{\mu} | \mu) \;.
\end{equation}

As when using $q_{\mu}$, the upper limit on $\mu$ at confidence level
$1 - \alpha$ is found by setting $p_{\mu} = \alpha$ and solving for
$\mu$, which reduces to the same result as found when using $q_{\mu}$,
namely,

\begin{equation}
\label{eq:muuptilde}
\mu_{\rm up} =  \hat{\mu} + \sigma \Phi^{-1}(1 - \alpha) \;.
\end{equation}

 Note that because $\sigma$ depends in general on $\mu$,
Eq.~(\ref{eq:muuptilde}) must be solved numerically.

 \subsubsection{Expected $\mathrm{CL}_s$ Limit and Bands}
For the $CL_s$ method we need distributions for $\tilde{q}_\mu$ for the hypothesis at $\mu$ and $\mu=0$.   We find
 \begin{equation}
 p'_{\mu}=\frac{1-\Phi(\sqrt{q_\mu})}{\Phi(\sqrt{q_{\mu,A}}-\sqrt{q_{\mu}})}
 \end{equation}
The median and expected error bands will therefore be
   \begin{equation}
   \mu_{{up}+N}=\sigma(\Phi^{-1}(1-\alpha \Phi(N))+N)
   \end{equation}   
\noindent    with 
   \begin{equation}
   \sigma^2=\frac{\mu^2}{q_{\mu,A}}
      \end{equation}      
  \noindent     $\alpha=0.05$, $\mu$ can be taken as $\mu_{up}^{med}$ in the calculation of $\sigma$.
        Note that for $N=0$ we find the median limit    
        \begin{equation}
   \mu_{up}^{med}=\sigma \Phi^{-1}(1-0.5\alpha)
      \end{equation}

The fact that $\sigma$ (the variance of $\hat{\mu}$) defined in Eq.~\ref{eq:sigmaofmu} in general depends on $\mu$ complicates situations and can lead to some discrepancies between the correct value of the bands and those obtained with the equation above.  The bands tend to be too narrow.  A modified treatment of the bands taking into account the $\mu$ dependence of $\sigma$ is under development.

%
%
%

\subsection{Importance Sampling}

[The following section has been adapted from text written primarily by Sven Kreiss, Alex Read, and myself for the ATLAS Higgs combination.  It is reproduced here for convenience. ]

To claim a discovery, it is necessary to populate a small tail of a test statistic distribution. Toy Monte-Carlo techniques use the model $\F_{\rm tot}$ to generate toy data $\data_{toy}$. For every pseudo-experiment (toy), the test statistic is calculated and added to the test statistic distribution. Building this distribution from toys is independent of the assumptions that go into the asymptotic calculation that describes this distribution with an analytic expression.   Recently progress has been made using Importance Sampling to populate the extreme tails of the test statistic distribution, which is much more computationally intensive with standard methods. The presented algorithms are implemented in \roostats\ \texttt{ToyMCSampler}.

\subsubsection{Naive Importance Sampling}


An ensemble of "standard toys" is generated from a model representing the Null hypothesis with $\mu=0$ and the nuisance parameters $\vec\theta$ fixed at their profiled values to the observed data $\nuisObs$, written\\ \mbox{$\F_{\rm tot}(\datasim,\globs|\mu=0,\nuisObs)$}. With importance sampling however, the underlying idea is to generate toys from a different model, called the importance density. A valid importance density is for example the same model with a non-zero value of $\mu$. The simple Likelihood ratio is calculated for each toy and used as a weight.
\[
{\rm weight} = \frac{\F_{\rm tot}( \data_{\rm toy}, \globs_{\rm toy} | \mu=0,\nuisObs)} {\F_{\rm tot}( \data_{\rm toy}, \globs_{\rm toy} | \mu=\mu',\nuisObs)}
\]

The weighted distribution is equal to a distribution of unweighted toys generated from the Null. The choice of the importance density is a delicate issue. Michael Woodroofe presented a prescription for creating a well behaved importance density~\cite{Woodroofe}. Unfortunately, this method is impractical for models as large as the combined Higgs models. An alternative approach is shown below.

\subsubsection{Phase Space Slicing}
The first improvement from naive importance sampling is the idea of taking toys from both, the null density and the importance density. There are various ways to do that. Simply stitching two test statistic distributions together at an arbitrary point has the disadvantage that the normalizations of both distributions have to be known.

Instead, it is possible to select toys according to their weights. First, toys are generated from the Null and the simple Likelihood ratio is calculated. If it is larger than one, the toy is kept and otherwise rejected. Next, toys from the importance density are generated. Here again, the simple Likelihood ratio is calculated but this time the toy is rejected when the Likelihood ratio is larger than one and kept when it is smaller than one. If kept, the toy's weight is the simple Likelihood ratio which is smaller than one by this prescription.

In the following section, this idea is restated such that it generalizes to multiple importance densities.

\subsubsection{Multiple Importance Densities}

The above procedure for selecting and reweighting toys that were generated from both densities can be phrased in the following way:
\begin{itemize}
   \item A toy is generated from a density with $\mu=\mu'$ and the Likelihoods $\F_{\rm tot}( \data_{\rm toy}, \globs_{\rm toy} | \mu=0,\nuisObs)$ and $\F_{\rm tot}( \data_{\rm toy}, \globs_{\rm toy} | \mu=\mu',\nuisObs)$ are calculated.
   \item The toy is veto-ed when the Likelihood with $\mu=\mu'$ is not the largest. Otherwise, the toy is used with a weight that is the ratio of the Likelihoods.
\end{itemize}
This can be generalized to any number of densities with $\mu_i=\{0, \mu', \mu'', \ldots\}$. For the toys generated from model $i$:
\begin{align}
\textrm{veto:}& \textrm{ if } \F_{\rm tot}( \data_{\rm toy}, \globs_{\rm toy} | \mu=\mu_i,\nuisObs) \neq \max\left\{ \F_{\rm tot}( \data_{\rm toy}, \globs_{\rm toy} | \mu=\mu_j,\nuisObs) : \mu_j = \{0, \mu', \mu'', \ldots\}\right\} \\
\textrm{weight} &= \frac{\F_{\rm tot}( \data_{\rm toy}, \globs_{\rm toy} | \mu=0,\nuisObs)}{\F_{\rm tot}( \data_{\rm toy}, \globs_{\rm toy} | \mu=\mu_i,\nuisObs)}
\end{align}

The number of importance densities has to be known when applying the vetos. It should not be too small to cover the parameter space appropriately and it should not be too large, because too many importance densities lead to too many vetoed toys which decreases overall efficiency. The value and error of $\hat{\mu}$ from a fit to data can be used to estimate the required number of importance densities for a given target overlap of the distributions.

The sampling efficiency in the tail can be further improved by generating a larger number of toys for densities with larger values of $\mu$. For example, for $n$ densities, one can generate $2^k / 2^n = 2^{k-n}$ of the overall toys per density $k$ with $k=0, \ldots, n-1$. The toys have to be re-weighted for example by $2^{n-1} / 2^k$ resulting in a  minimum re-weight factor of one. The current implementation of the error calculation for the p-value is independent of an overall scale in the weights.

The method using multiple importance densities is similar to Michael Woodroofe's \cite{Woodroofe} prescription of creating a suitable importance density with an integral over $\mu$. In the method presented here, the integral is approximated by a sum over discrete values of $\mu$. Instead of taking the sum, a mechanism that allows for multiple importance densities is introduced.

\begin{figure}[htb]
  \begin{center}
    \includegraphics[width=.5\textwidth]{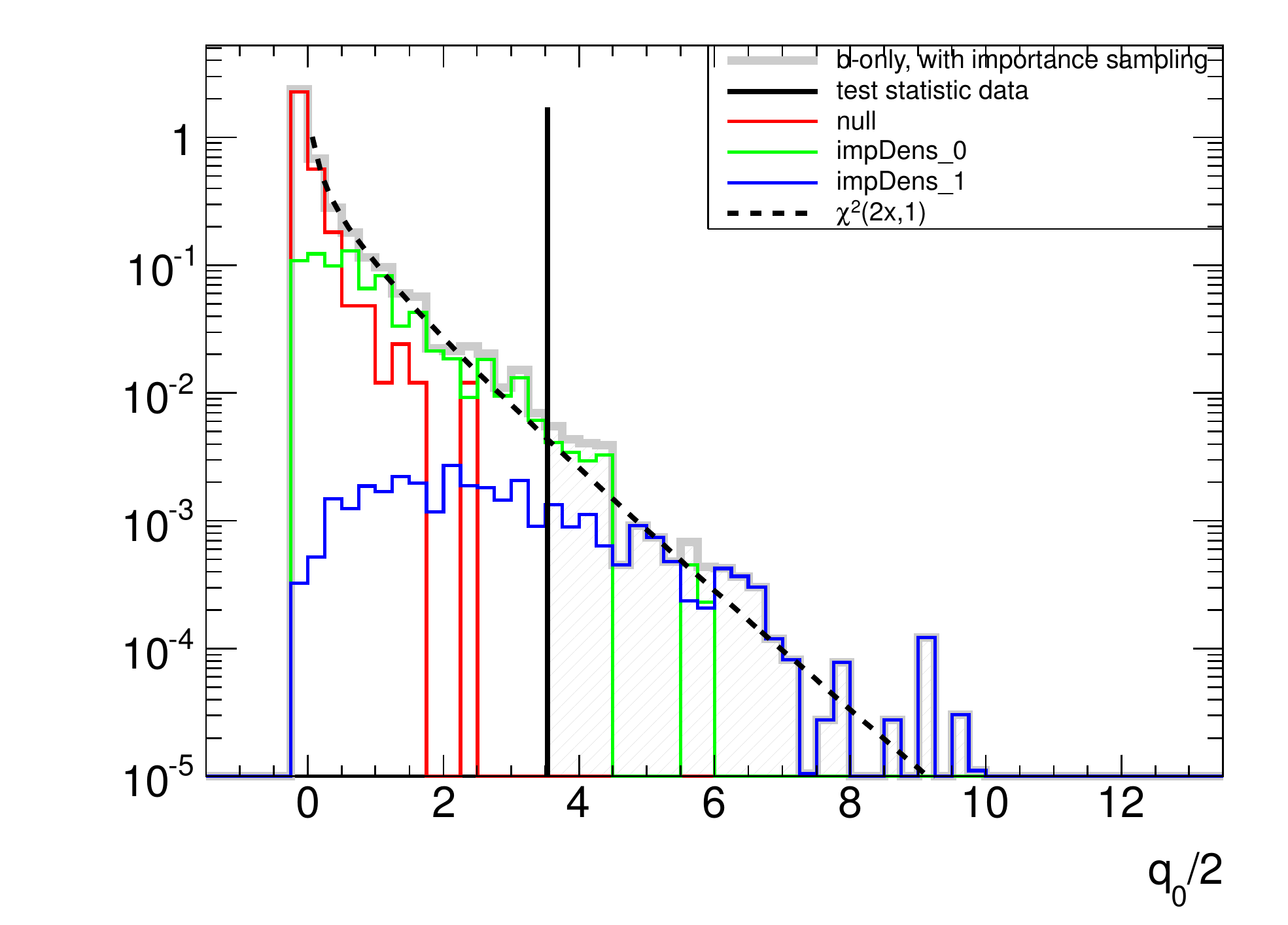}
  \end{center}
\vspace*{-0.5cm}
\caption{An example sampling of a test statistic distribution using three densities, the original null density and two importance densities.}
  \label{fig:ImpSampl}
\end{figure}

\subsection{Look-elsewhere effect, trials factor, Bonferoni}

Future versions of this document will discuss the so-called look-elsewhere effect in more detail.  Here we point to the primary development recently: \cite{LEE,ATL-PHYS-PUB-2011-011}.

\subsection{One-sided intervals, CLs, power-constraints, and Negatively Biased Relevant Subsets}

Particle physicists regularly set upper-limits on cross sections and other parameters that are bounded to be non-negative.  Standard frequentist confidence intervals should nominally cover at the stated value.  The implication that a 95\% confidence level upper-limit covers the true value 95\% of the time is that it doesn't cover the true value 5\% of the time.  This is true no matter how small the cross section is.  That means that if there is no signal present, 5\% of the time we would be excluding any positive value of the cross-section.  Experimentalists do not like this since we would not consider ourselves sensitive to arbitrarily small signals.  

\begin{figure}[htb]
  \begin{center}
	\subfigure[]{\includegraphics[width=.45\textwidth]{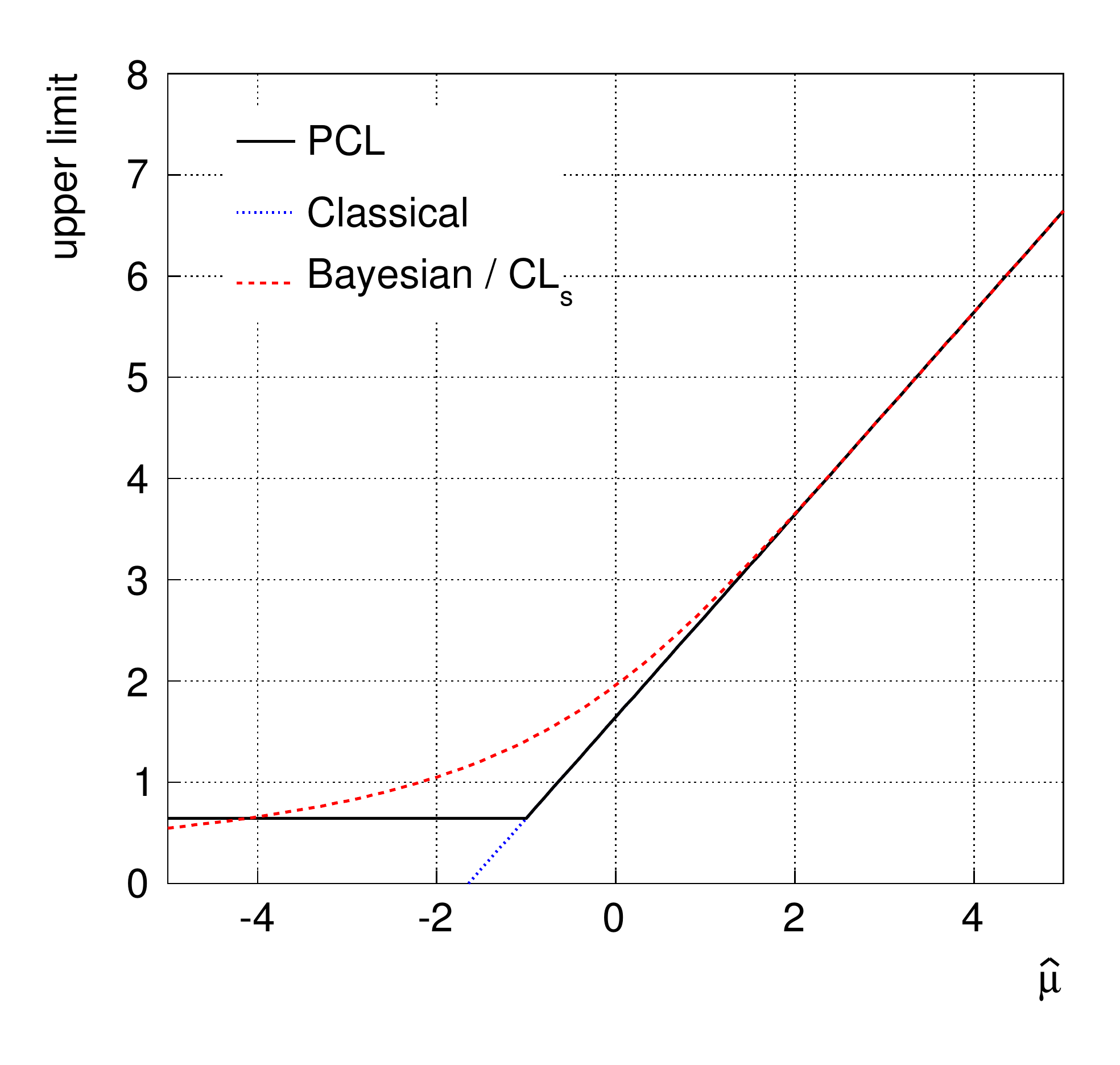}}
	\subfigure[]{\includegraphics[width=.45\textwidth]{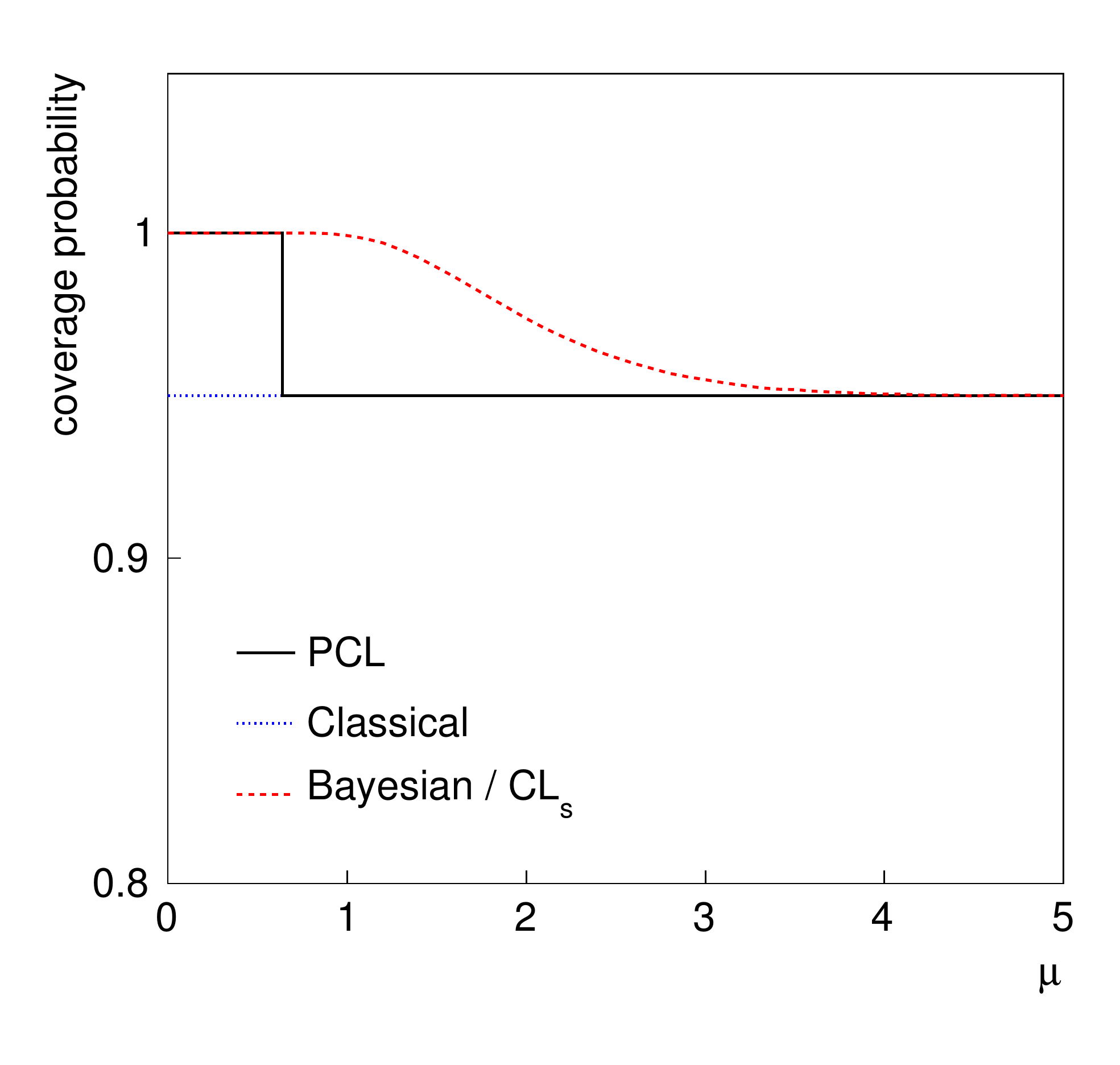}}
  \end{center}
\vspace*{-0.5cm}
\caption{Taken from Fig.3 of \cite{2011arXiv1105.3166C}: (a) Upper limits from the PCL (solid), CLs and Bayesian (dashed), and classical (dotted) procedures as a function of $\mu$), which is assumed to follow a Gaussian distribution with unit standard deviation. (b) The corresponding coverage probabilities as a function of $\mu$.}
  \label{fig:CLscoverage}
\end{figure}

Two main approaches have been proposed to protect from excluding signals to which we do not consider ourselves sensitive.  The first is the CLs procedure introduced by Read and described above~\cite{Read2,Read1,CLsWikipedia}.  The CLs procedure produce intervals that over-cover -- meaning that the intervals cover the true value more than the desired level.  The  coverage for small values of the cross-section approaches 100\%, while for large values of the cross section, where the experiment does have sensitivity, the coverage converges to the nominal level  (see Fig.~\ref{fig:CLscoverage}).  Unfortunately, the coverage for intermediate values is not immediately accessible without more detailed studies.  Interestingly, the modified frequentist CLs procedure reproduces the one-sided upper limit from a Bayesian procedure with a uniform prior on the cross section for simple models like number counting analyses.  Even in very complicated models we see very good numerical agreement between CLs and the Bayesian approach, even though the interpretation of the numbers is  different.

An alternate approach called power-constrained limits (PCL) is to leave the standard frequentist procedure unchanged while adding an additional requirement for a parameter point to be considered `excluded'.  The additional requirement is directly a measure of the sensitivity of to that parameter point based on the notion of power (or Type II error).  This approach makes the coverage of the procedure manifest~\cite{2011arXiv1105.3166C}.

Surprisingly, one-sided upper limits on a bounded parameter are a subtle topic that has led to debates among the experts of statistics in the collaborations and a string of interesting articles from statisticians.  The discussion is beyond the scope of the current version of these notes, but the interested reader is invited and encouraged to read~\cite{Mandelkern2002} and the responses from notable statisticians on the topic.  More recently Cousins tried to formalize the sensitivity problem in terms of a concept called Negatively Biased Relevant Subsets (NBRS)~\cite{2011arXiv1109.2023C}.  While the power-constrained limits do not formally emit NBRS, it is an interesting insight.  Even more recently, Vitells has  found interesting connections with CLs and the work of Birnbaum~\cite{Birnbaum:1962,CLsWikipedia}. This connection is significant since statisticians have primarily seen CLs as an ad hoc procedure mixing the notion of size and power with no satisfying properties.

%
%

%

%



\section{Bayesian Procedures}

[This section is far from complete.  Some key practical issues and references to other literature are given.]

Unsurprisingly, Bayesian procedures are based on Bayes's theorem as in Eq.~\ref{Eq:Bayes} and Eq.~\ref{eq:urprior}.  The Bayesian approach requires one to provide a prior over the parameters, which can be seen either as an advantage or a disadvantage~\cite{DAgostiniInference,Cousins:1994yw}.  In practical terms, one typically wants to build the posterior distribution for the parameter of interest.  This typically requires integrating, or \textit{marginalizing}, over all the nuisance parameters as in Eq.~\ref{eq:credible}.  These integrals can be over very high dimensional posteriors with complicated structure.  One of the most powerful algorithms for this integration is Markov Chain Monte Carlo, described below.  In terms of the prior one can either embrace the subjective Bayesian approach~\cite{Jaynes:2003fk} or take a more 'objective' approach in which the prior is derived from formal rules.  For instance, Jeffreys's Prior~\cite{JeffreysPrior} or their generalization in terms of Reference Priors~\cite{Demortier:2010sn}.  

Given the logical importance of the choice of prior, it is generally recommended to try a few options to see how the result numerically depends on the choice of priors (i.e.. sensitivity analysis).  This leads me to a few great quotes from prominent statisticians:

``Sensitivity analysis is at the heart of scientific Bayesianism'' --Michael Goldstein

``Perhaps the most important general lesson is that the facile use of what appear to be uninformative priors is a dangerous practice in high dimensions'' -Brad Efron

``Meaningful prior specification of beliefs in probabilistic form over very large possibility spaces is very difficult and may lead to a lot of arbitrariness in the specification'' -- Michael Goldstein

``Objective Bayesian analysis is the best frequentist tool around'' --Jim Berger

\subsection{Hybrid Bayesian-Frequentist methods}

It is worth mentioning that in particle physics there has been widespread use of a hybrid Bayesian-Frequentist approach in which one marginalizes nuisance parameters.  Perhaps the most well known example is due to a paper by Cousins and Highland~\cite{CousinsHighland:1991qz}.  In some instances one obtains a Bayesian-averaged model that depends only on the parameters of interest
\begin{equation}
\bar{\F}(\data | \vec\alpha_{\rm poi}) = \int  \F_{\rm tot}(\data | \vec\alpha) \eta(\vec\alpha_{\rm nuis}) \; d\vec\alpha_{\rm nuis}
\end{equation}
and then proceeds with the typical frequentist methodology for calculating p-values and constructing confidence intervals. Note, in this approach the constraint terms that are appended to $\F_{\rm sim}$ of Eq.~\ref{Eq:simultaneous} to obtain $\F_{\rm tot}$ of Eq.~\ref{Eq:ftot} are interpreted as in Eq.~\ref{eq:urprior} and $\eta(\vec\alpha_{\rm nuts})$ is usually a uniform prior.  Furthermore, the global observables or auxiliary measurements $a_p$ are typically left fixed to their nominal or observed values and not randomized.
  In other variants the full model without constraints $\F_{\rm sim}(\data | \vec\alpha)$ is used to define the test statistic but the distribution of the test statistic is obtained by marginalizing (or randomizing) the nuisance parameters as in Eq.~\ref{eq:urprior}.  See the following references for more details  \cite{Conrad:2005zm,Tegenfeldt:2004dk,Conrad:2002ur,Conrad:2002kn,Rolke:2004mj,PhysRevD.67.118101,Demortier:2007zz,Cousins:2008zz}.  
  
The shortcomings of this approach are that the coverage is not guaranteed and the method uses an inconsistent notion of probability.  Thus it is hard to define exactly what the p-values and intervals mean in a formal sense.

\subsection{Markov Chain Monte Carlo and the Metropolis-Hastings Algorithm}

	The Metropolis-Hastings algorithm is used to construct a Markov chain $\{\vec\alpha_i\}$, where the samples $\vec\alpha_i$ are proportional to the target posterior density or likelihood function.  The algorithm requires a proposal function $Q(\vec\alpha | \vec\alpha')$ that gives the probability density to propose the point $\vec\alpha$ given that the last point in the chain is $\vec\alpha'$.  Note, the density only depends on the last step in the chain, thus it is considered a Markov process.  At each step in the algorithm, a new point in parameter space is proposed and possibly appended to the chain based on its likelihood relative to the current point in the chain.  Even when the proposal density function is not symmetric, Metropolis Hastings maintains `detailed balance' when constructing the Markov chain by counterbalancing the relative likelihood between the two points with the relative proposal density.  That is, given the current point $\vec\alpha$, proposed point $\vec\alpha'$, likelihood function $L$, and proposal density function $Q$, we visit $\vec\alpha'$ if and only if
\begin{equation}
\displaystyle \frac{L(\vec\alpha')}{L(\vec\alpha)} \frac{Q(\vec\alpha | \vec\alpha')}{Q(\vec\alpha' | \vec\alpha)} \geq Rand[0,1]
\end{equation}
Note, if the proposal density is symmetric, $Q(\vec\alpha | \vec\alpha')=Q(\vec\alpha' | \vec\alpha)$, then the ratio of the proposal densities can be neglected (which can be computationally expensive).  Above we have written the algorithm to sample the likelihood function $L(\vec\alpha)$, but typically one would use the posterior $\pi(\vec\alpha)$. Within \roostats\ the Metropolis-Hastings algorithm is implemented with the \texttt{MetropolisHastings} class, which returns a \texttt{MarkovChain}.  Another powerful tool is the Bayesian Analysis Toolkit (BAT)~\cite{Caldwell:2009ve}.  Note, one can use a \roofit\ / \roostats\ model in the BAT environment.

Note, an alternative to Markov Chain Monte Carlo is the nested sampling approach of Skilling~\cite{skilling:395} and the \texttt{MultiNest} implementation~\cite{Feroz:2008xx}.

Lastly, we mention that sampling algorithms associated to Bayesian belief networks and graphical models may offer enormous advantages to both MCMC and nested sampling due to the fact that they can take advantage of the conditional dependencies in the model.

\subsection{Jeffreys's and Reference Prior}

\begin{figure}[htbp]
\begin{center}
\includegraphics[width=.8\textwidth]{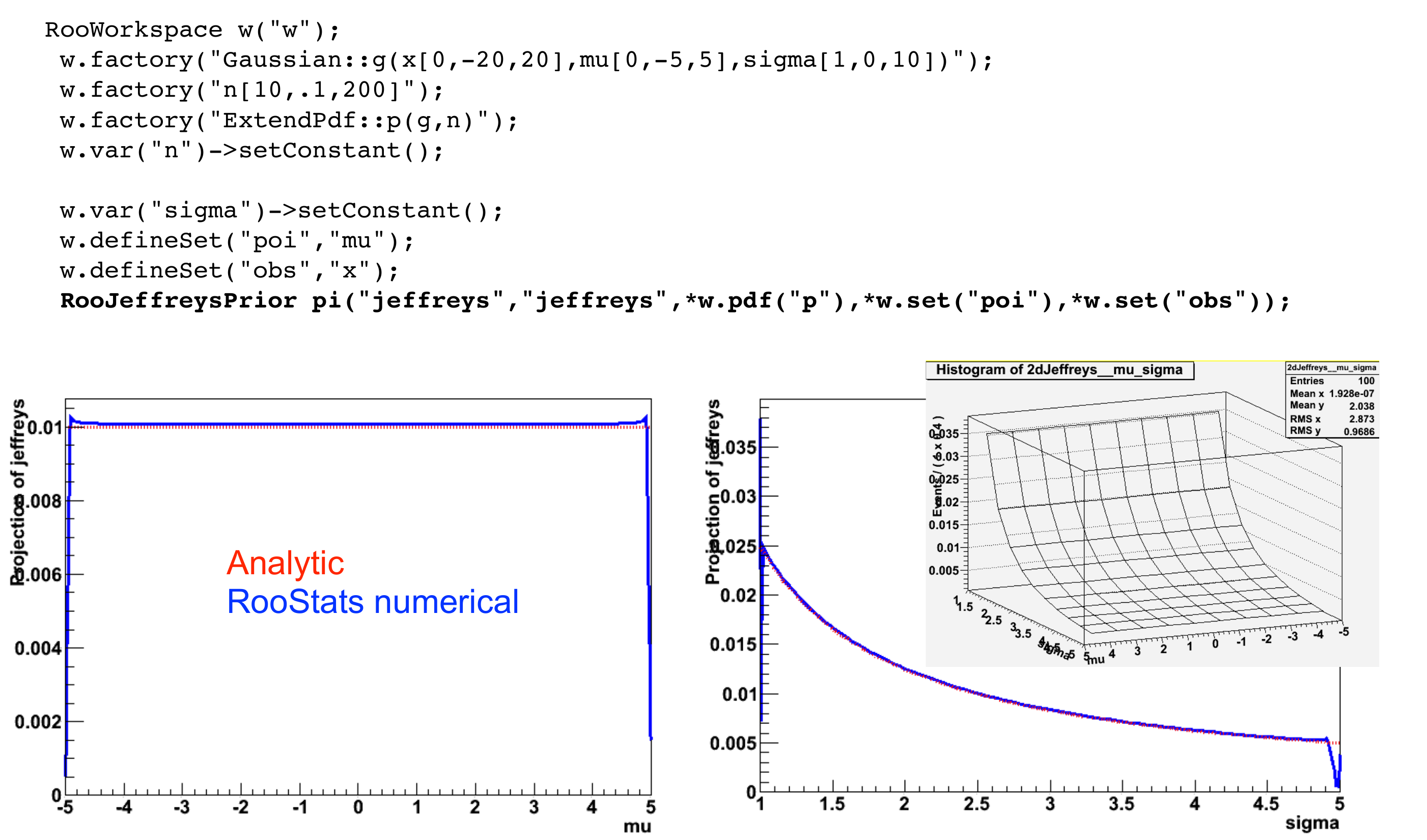}
\caption{Example code making a Gaussian distribution (with 10 events expected) and the Jeffreys Prior for $\mu$ and $\sigma$ calculated numerically in \texttt{RooStats}  and compared to the analytic result.  }
\label{fig:JeffreysPriorGaussian}
\end{center}
\end{figure}

\begin{figure}[htbp]
\begin{center}
\includegraphics[width=.8\textwidth]{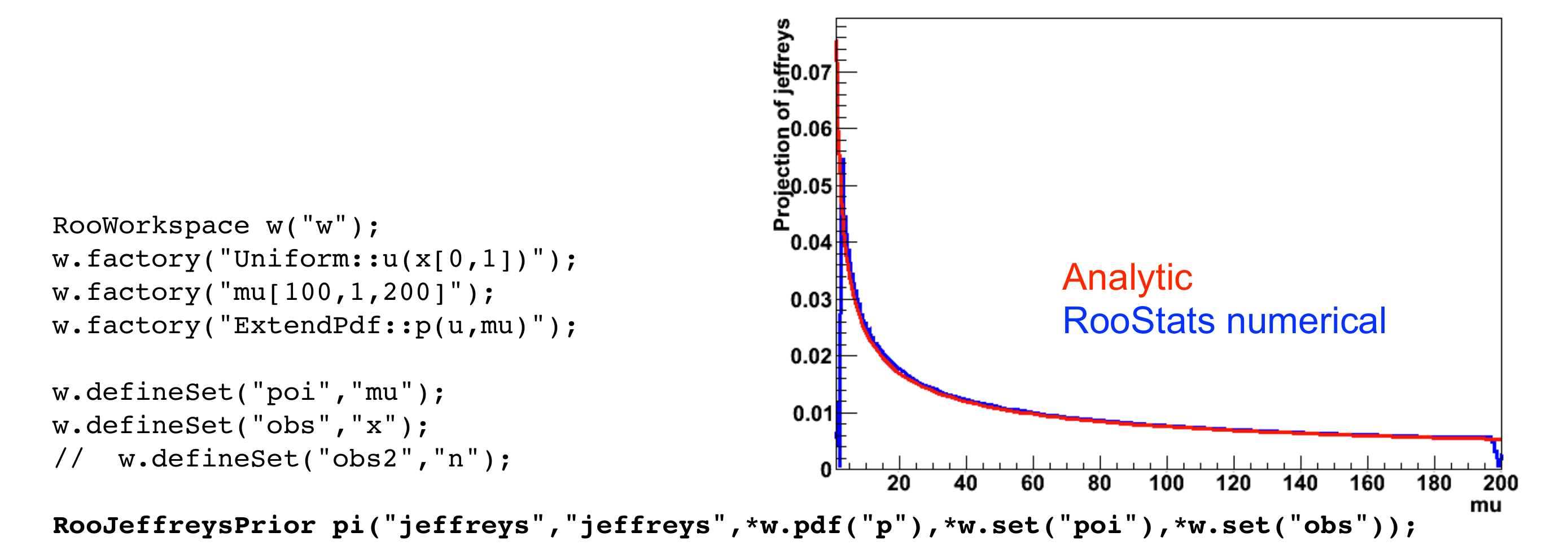}
\caption{Example code making a Poisson distribution (with 100 replications expected) and the Jeffreys Prior for $\mu$ calculated numerically in \texttt{RooStats} and compared to the analytic result.  }
\label{fig:JeffreysPriorPoisson}
\end{center}
\end{figure}

One of the great advances in Bayesian methodology was the introduction of Jeffreys's rule for selecting a prior based on a formal rule~\cite{JeffreysPrior}.  The rule selects a prior that is invariant under reparametrization of the observables and covariant with reparametrization of the parameters.  The rule is based on information theoretic arguments and the prior is given by the square root of the determinant of the Fisher information matrix, which we first encountered in Eq.~\ref{Eq:expfisher}.
\begin{equation}
\pi(\vec\alpha)  =  \sqrt{\det \Sigma^{-1}_{pp'}(\vec\alpha)} = \sqrt{ \det \left[ \int  \F_{\rm tot}(\data | \vec\alpha) \; \frac{-\partial^2 \log \F_{\rm tot}(\data | \vec\alpha)}{\partial\alpha_p\alpha_{p'}} \; d\data  \right]}\
\end{equation}
While the right-most form of the prior looks daunting with complex integrals over partial derivatives, the Asimov data described in Sec.~\ref{S:Asimov} and Ref.~\cite{asimov} provide a convenient way to calculate the Fisher information.  Fig.~\ref{fig:JeffreysPriorGaussian} and \ref{fig:JeffreysPriorPoisson} show examples of \roostats\ numerical algorithm for calculating Jeffreys's prior compared to analytic results on a simple Gaussian and a Poisson model.

Unfortunately, Jeffreys's prior does not behave well in multidimensional problems.  Based on a similar information theoretic approach, Bernardo and Berger have developed the Reference priors~\cite{Berger:1992ys,Berger:1992vn,Berger:1989kx,Bernardo:1979uq} and the associated Reference analysis.  While attractive in many ways, the approach is fairly difficult to implement.  Recently,  there has been some progress within the particle physics context in deriving the reference prior for problems relevant to particle physics~\cite{Demortier:2010sn,Casadei:2011hx}.

%
%
%

\subsection{Likelihood Principle}

For those interested in the deeper and more philosophical aspects of statistical inference, the likelihood principle is incredibly interesting.  This section will be expanded in the future, but for now I simply suggest searching on the internet, the Wikipedia article, and Ref.~\cite{Birnbaum:1962}.  In short the principle says that all inference should be based on the likelihood function of the observed data.  Frequentist procedures violate the likelihood principle since p-values are tail probabilities associated to hypothetical outcomes (not the observed data).  Generally, Bayesian procedures and those based on the asymptotic properties of likelihood tests do obey the likelihood principle.  Somewhat ironically, the objective Bayesian procedures such as Reference priors and Jeffreys's prior can violate the likelihood principle since the prior is based on expectations over hypothetical outcomes.



%

\section{Unfolding}
Another topic for the future.  The basic aim of unfolding is to try to correct distributions back to the true underlying distribution before  detector 'smearing'.  For now, see \cite{Prosper:1306523,DAgostini1995487,Adye:2011gm,Malaescu:2011yg,Blobel:2002pu,Hocker:1995kb,Choudalakis2012,Tikhonov}.

\section{Conclusions}

It was a pleasure to lecture at the 2011 ESHEP school in Cheile Gradistei  and the 2013 CLASHEP school in Peru.  Quite a bit of progress has been made in the last few years in terms of statistical methodology, in particular the formalization of a fully frequentist approach to incorporating systematics, a deeper understanding of the look-elsewhere effect, the development of asymptotic approximations of the distributions important for particle physics, and in roads to Bayesian reference analysis. Furthermore, most of these developments are general purpose and can be applied across diverse models.   While those developments are interesting, the most important area for most physicists to devote their attention in terms of statistics is to improve the modeling of the data for his or her individual analysis.

\end{document}